\documentclass[twoside,12pt]{article}
\usepackage{epsfig}
\usepackage{braket}
\usepackage{xcolor}
\usepackage{bm}
\usepackage{amsmath}

\newcommand{\be}{\begin{equation}}
\newcommand{\ee}{\end{equation}}
\newcommand{\bea}{\begin{eqnarray}}
\newcommand{\eea}{\end{eqnarray}}

\begin{document}

\title{ \vspace{1cm}Recent developments for the optical model of nuclei}
\author{W. H. Dickhoff,$^{1}$ R. J. Charity$^2$
\\
${}^1$Department of Physics, Washington University, St. Louis, MO 63130, USA\\
${}^2$Department of Chemistry, Washington University, St. Louis, MO 63130, USA
}
\maketitle
\begin{abstract} 
A brief overview of various approaches to the optical-model description of nuclei is presented.
A survey of some of the formal aspects is given which links the Feshbach formulation for either the hole or particle Green's function to the time-ordered quantity of many-body theory.
The link between the reducible self-energy and the elastic nucleon-nucleus scattering amplitude is also presented using the development of Villars.
A brief summary of the essential elements of the multiple-scattering approach is also included.
Several ingredients contained in the time-ordered Green's function are summarized for the formal framework of the dispersive optical model (DOM).
Empirical approaches to the optical potential are reviewed with emphasis on the latest global parametrizations for nucleons and composites.
Various calculations that start from an underlying realistic nucleon-nucleon interaction are discussed with emphasis on more recent work. 
The efficacy of the DOM is illustrated in relating nuclear structure and nuclear reaction information.
Its use as an intermediate between experimental data and theoretical calculations is advocated.
Applications of the use of optical models are pointed out in the context of the description of nuclear reactions other than elastic nucleon-nucleus scattering.
\end{abstract}
\section{Introduction}

The concept of an optical potential with both real and imaginary components was introduced in 1949 to describe the elastic scattering of neutrons. This quantum-mechanical approach allows the incident particle waves to be scattered, transmitted, and absorbed by the potential. The reference to ``optical'' comes from the use of complex refractive indices to explain similar phenomenona for light rays.
The optical potential is an effective interaction which is used not just for elastic scattering, but as an ingredient in predicting the cross sections and angular distributions of many direct reaction processes and therefore plays an important role in nuclear physics. Reviews of the history and early use of the optical model by Feshbach and Hodgson can be found in Refs.~\cite{Feshbach:1958,Hodgson:1967,Hodgson:1971,Hodgson:1984,Hodgson:2002}. However, since the last of these reviews was in 2002, it is timely to survey some of the more recent advances. 

The new era of radioactive-beam experiments has clearly brought into focus the need to synthesize our approach to nuclear reactions and nuclear structure.
Strongly-interaction probes of such exotic nuclei involve interactions that are not yet well constrained by experimental data.
THerefore, there is an ongoing need to better understand the potentials that protons and neutrons in such systems experience.
One important aspect is covered by the positive-energy optical potentials that govern elastic nucleon scattering while there is a clear need to establish the link with bound-state potentials which are hardly distinct from the positive-energy ones when the drip lines are approached.
 
The number of works that are encompassed by the title of ``optical model'' is immense and any review must be selective. The topics and studies present in this review are biased by the interests of the authors.   We start with several theoretical derivations of the optical potential and its connection to the self-energy in many-body formulations in Sec.~\ref{sec:formal}.
In Sec.~\ref{sec:Feshbach} the analysis of Capuzzi and Mahaux~\cite{CapMah:00} will be employed to link the Feshbach approach to the many-body perspective provided by the time-ordered Green's function.
We complement this in Sec.~\ref{sec:viilars} with a derivation by Villars~\cite{vill67} that formalizes the work of Ref.~\cite{Bell:59} that demonstrates the equivalence of the nucleon-nucleus $\mathcal{T}$-matrix to the on-shell reducible nucleon self-energy $\Sigma_{red}$ associated with the time-ordered Green's function.
This approach is important as it can be extended to the description of more complicated reactions involving composite projectiles or ejectiles.
In Sec.~\ref{sec:MSc} we briefly summarize some elements of the multiple-scattering approach but refrain from a detailed presentation which is available in Ref.~\cite{Amos:00}.
We will take particular interest in the empirical dispersive optical model (DOM) where the real and imaginary potentials are connected by dispersion relations thereby enforcing causality.
This approach allows one to exploit the formal properties of the time-ordered Green's function including its link to the nucleon self-energy through the Dyson equation~\cite{Dickhoff:08,Dickhoff:17}. 
As the dispersion relations require the optical potentials to be defined at both positive and negative energy, the formalism makes connections between nuclear reactions and nuclear structure needed for a better understanding of rare isotopes.  
Some ingredients relevant for the discussion of the DOM are summarized in Sec.~\ref{sec:Mah}.

Since the last review there are significantly more data available on elastic scattering, total and absorption cross sections. Of particular importance is the availability of more neutron data with separated isotopes and, in addition,  a few elastic-scattering experiments in inverse kinematics with radioactive beams.  With the new data have come better global empirical parametrizations. These will be discussed in Sec.~\ref{sec:empirical} for scattering of nucleons and light projectiles.  

Since the concept of the optical model was first formulated, there has been considerable interest in calculating it microscopically from the underlying nucleon-nucleon (NN) interaction. Some recent attempts of such studies are presented in Sec.~\ref{sec:calc} with examples obtained from a number of approaches.
The nuclear-matter approach pioneered in Ref.~\cite{PhysRevC.16.80} is summarized and illustrated in Sec.~\ref{sec:JLM} as it has played an important role in practical applications.
The method accounts for the short-range repulsion of NN interactions in nuclear matter including the effect of the Pauli principle and utilizes suitable local-density approximations.
An example of a recent nuclear-matter study relying on modern chiral NN interactions emphasizing the isovector properties of optical potentials, is discussed in Sec.~\ref{sec:Holt}.
Recent applications of the multiple-scattering approach are summarized in Sec.~\ref{sec:Trho}.

Calculations based on the Green's-function method are presented in Sec.~\ref{sec:GF}.
We emphasize and illustrate the influence of short-range correlations in Sec.~\ref{sec:dussanCD}.
Such calculations attempt to properly account for the short-range repulsion of the NN interaction by summing ladder diagrams in the nucleus under consideration and are therefore quite distinct from nuclear-matter approaches.
We devote attention to the effect of long-range correlations in Sec.~\ref{sec:LRC}.
The emphasis in this analysis is on the role of low-lying collective states in determining the nucleon self-energy in an energy domain that includes the physics of giant resonances.
In Sec.~\ref{sec:BarLap} a recent analysis is presented that combines the structure calculations of rare isotopes employing \textit{ab initio} many-body methods with optical potentials in an attempt to extract matter radii.
Recent work employing the coupled-cluster method to determine the Green's function and thereby the optical potential, is presented in Sec.~\ref{sec:CC}
 
Applications of the DOM provide an important tool to establish a connection between experimental data and theory.
We discuss various incarnations of the developments of the DOM in Sec.~\ref{sec:DOM}.
A brief section on applications of optical potentials is presented in Sec.~\ref{sec:applic}.
This includes transfer reactions (Sec.~\ref{sec:transfer}), heavy-ion knockout reactions (Sec.~\ref{sec:knock}), electron and proton induced knockout reactions (Sec.~\ref{sec:pek}), and in Sec.~\ref{sec:DOMknockout} some other applications of DOM potentials to these reactions.
Finally, we draw some conclusions in Sec.~\ref{sec:conc}.

\section{Formal background}
\label{sec:formal}
In this section we illustrate several different approaches to the optical potential.

\subsection{Hole and particle Green' functions and the Feshbach formulation and its relation to the time-ordered Green's function}
\label{sec:Feshbach}
We start by taking a slightly different route in introducing the optical potential by following some of the development discussed in Ref.~\cite{CapMah:00}, which provides a more complete formal discussion.
This paper starts by introducing the hole Green's function for complex $z$
\be
G^{(h)}(\bm{r}, \bm{r}'; z) =\bra{\Psi_0^A} a^\dagger_{\bm{r}'} \frac{1}{z-(E^A_0 - \hat{H})} a_{\bm{r}} \ket{\Psi^A_0} ,
\label{eq:hgf}
\ee
where $\ket{\Psi^A_0}$ represents the nondegenerate ground state of the $A$-particle system belonging to the Hamiltonian $\hat{H}$.
We employ a coordinate-space notation but implicitly refer to a complete set of single-particle (sp) quantum numbers relevant for the system under study.
By inserting a complete set of states for the $A-1$-particle system on can write
\begin{equation}
G^{(h)}(\bm{r} ,\bm{r}' ; z)  =  
\sum_n  \frac{ \chi^{n-}(\bm{r}') \left[\chi^{n-}(\bm{r})\right]^*
}{ z - E^{-}_n}  
 + 
\sum_c \int^{T^{(-)}_c}_{-\infty} dE'\  \frac{\chi^{cE'}(\bm{r}') \left[\chi^{cE'}(\bm{r})\right]^* }{
z - E'} .
 \label{eq:proph} 
\end{equation}
Overlap functions for bound $A-1$ states are given by $ \chi^{n-}(\bm{r}')=\bra{\Psi^A_0} a^\dagger_{\bm{r}'}
\ket{\Psi^{A-1}_n}$, whereas those in the continuum are given by $ \chi^{cE}(\bm{r}')=\bra{\Psi^A_0} a^\dagger_{\bm{r}'} \ket{\Psi^{A-1}_{cE}}$ indicating the relevant channel by $c$ and the energy by $E$.
Excitation energies in the $A-1$ system are given with respect to the $A$-body ground state $E^{-}_n = E^A_0-E^{A-1}_n$.
Each channel $c$ has an appropriate threshold indicated by $T^{(-)}_c$, which is the experimental threshold with respect to the ground-state energy of the $A$-body system.
The hole spectral density is given by
\begin{equation}
S^{(h)}(\bm{r} ,\bm{r}' ; E)  =  
\sum_n   \chi^{n-}(\bm{r}') \left[\chi^{n-}(\bm{r})\right]^*
\delta( E - E^{-}_n)    +  
\sum_c \chi^{cE}(\bm{r}') \left[\chi^{cE}(\bm{r})\right]^*  ,
\label{eq:sfunch}
\end{equation}
which allows one to write the hole Green's function as
\begin{equation}
G^{(h)}(\bm{r} ,\bm{r}' ; z)  =  
\int_{-\infty}^{\infty} dE'\  \frac{S^{(h)}(\bm{r}, \bm{r}' ; E') }{
z - E'} .
 \label{eq:prophL} 
\end{equation}

For complex $z$, one may define the ``hole Hamiltonian'' $h^{(h)}(z)$ operator by using the operator form of $G^{(h)}$ as follows
\be
\left[ z - h^{(h)}(z) \right] G^{(h)}(z) = 1.
\label{eq:hham}
\ee
For the propagator it is useful to adopt the limit
\be
h^{(h)}(E) = \lim_{\eta \rightarrow +0} h^{(h)}(E-i\eta) 
\label{eq:hbc}
\ee
for $E$ real. 
This hole Hamiltonian has eigenstates corresponding to the discrete overlap states
\be
h^{(h)}(E^{-}_n) \ket{\chi^{n-}} = E^{-}_n\  \ket{\chi^{n-}} .
\label{eq:evh}
\ee
As was introduced in Refs.~\cite{Feshbach:58,FESHBACH1962287} for nucleon scattering, one can decompose any vector in the Hilbert space for $A-1$ particles with two hermitian projection operators
\be
P^{(h)} + Q^{(h)} = 1
\label{eq:projh}
\ee
with $P^{(h)}  Q^{(h)} = 0$.
It is then required that the hole projection operator has the property
\be
P^{(h)} \ket{\Psi^{A-1}} = a_{u^-} \ket{\Psi_0^A} 
\label{eq:u}
\ee
for all $\ket{\Psi^{A-1}}$.
The one-body density matrix $\rho(\bm{r},\bm{r}')=\bra{\Psi^A_0} a^\dagger_{\bm{r}'} a_{\bm{r}} \ket{\Psi^A_0}$ provides the following link between overlap states and $\ket{u^-}$
\be
\ket{\chi^-} = \rho \ket{u^-} .
\label{eq:dm}
\ee
It is possible to show that~\cite{CapMah:00} 
\be
P^{(h)} = \int d\bm{r}\ \int d\bm{r}'\ a_{\bm{r}} \ket{\Psi^A_0} \frac{1}{\rho(\bm{r}, \bm{r}')} \bra{\Psi^A_0} a^\dagger_{\bm{r}'} .
\label{eq:prodm}
\ee
Using standard manipulation, the hole Hamiltonian can be written in terms of the projection operators given in Eq.~(\ref{eq:projh}) according to
\be
h^{(h)}(E)= E(1-\rho^{-1}) + \rho^{-1} R^{(h)} \rho^{-1} +\rho^{-1} D^{(h)}(E-i\eta ) \rho^{-1}
\label{eq:hhamp}
\ee
which leads to the corresponding eigenvalue equation
\be
\left[E^-_n \rho^{-1} - \rho^{-1} R^{(h)} \rho^{-1} -\rho^{-1} D^{(h)}(E^-_n) \rho^{-1} \right]
\ket{\chi^{(-)}_n} = 0,
\label{eq:hammp}
\ee
where in the coordinate representation
\be
R^{(h)}_{\bm{r} \bm{r}'} = \bra{\Psi^A_0} a^\dagger_{\bm{r}'} \left[ a_{\bm{r}} , \hat{H} \right] \ket{\Psi^A_0} 
\label{eq:rrr}
\ee
and
\be
D^{(h)}_{\bm{r} \bm{r}'}(E-i\eta) = \bra{\Psi^A_0} a^\dagger_{\bm{r}'} \hat{H} Q^{(h)} \frac{1}{E + Q^{(h)} \hat{H} Q^{(h)} -E^A_0 -i \eta} Q^{(h)} \hat{H} a_{\bm{r}} \ket{\Psi^A_0} .
\label{eq:drr}
\ee
Note that Eq.~(\ref{eq:hammp}) is equivalent to Eq.~(\ref{eq:evh}).

It is possible to employ the original Hamiltonian to separate the kinetic and interaction contributions to $R^{(h)}$ in Eq.~(\ref{eq:rrr}) written as
\be
R^{(h)}= \mathcal{T}^{(h)} + \mathcal{V}^{(h)} .
\label{eq:RTV}
\ee
It is then possible to introduce the hole self-energy operator in the one-body space by subtracting the corresponding kinetic-energy operator as in
\be
\Sigma^{(h)}(E) = h^{(h)}(E) - T .
\label{eq:hse}
\ee

The first term in Eq.~(\ref{eq:hhamp}) diverges at large $E$. 
This can be avoided by deriving other, related, Schr\"{o}dinger-type equations.
Indeed, by multiplying Eq.~(\ref{eq:hammp}) on the left by any energy-independent, but possibly nonlocal, operator $\mathcal{N}$ one obtains
\be
\left[ E^-_n - h^{(h)}_{\mathcal{N}} (E^{-}_n) \right] \ket{\chi^{n-}} = 0 ,
\label{eq:evhN}
\ee
where
\be
h^{(h)}_{\mathcal{N}}(E)= E(1-\mathcal{N} \rho^{-1}) + \mathcal{N} \rho^{-1} R^{(h)} \rho^{-1} +\rho^{-1} D^{(h)}(E) \rho^{-1} .
\label{eq:hhampp}
\ee
The divergence can now be eliminated by setting $\mathcal{N} = \rho$ in Eq.~(\ref{eq:hhampp}) which then becomes
\bea
h^{(h)}_{\mathcal{F}}(E) & = & \rho h^{(h)}(E) +E(1-\rho)  \nonumber \\ 
& = & \left[ R^{(h)}  + D^{(h)}(E) \right] \rho^{-1} \nonumber \\
 & = & T + \left( \mathcal{V}^{(h)} +D^{(h)} \right) \rho^{-1} .
\label{eq:hhamF}
\eea
Note that this Hamiltonian does not have a scattering eigenstate because it reflects the action of a removal operator on the localized ground state.
This form of the Hamiltonian is completely analogous to the one introduced by Feshbach~\cite{Feshbach:58} to develop the optical potential for elastic nucleon scattering (see below), hence the $\mathcal{F}$ subscript.
Additional forms of the hole Hamiltonian are discussed in Ref.~\cite{CapMah:00}.

From the hole Hamiltonians considered above, it is possible to calculate the hole spectral function with access to overlap functions and their normalization.
This reflects their analytic properties in the complex-energy plane which exhibits poles on the real axis for negative energy and a left-hand cut from $-\infty$ to the threshold energy $T^{(-)}_0$ associated with nucleon emission from the $(A-1)$-system.

Particle one-body Hamiltonians can be introduced in complete analogy with the hole version.
The main difference consists is the presence of elastic overlap states as eigenstates
\be
\chi^{0+}_E(\bm{r})= \langle \bm{r} | \chi^{0+}_E \rangle = \bra{\Psi^A_0} a_{\bm{r}} \ket{\Psi^{0 A+1}_{E}} ,
\label{eq:elwf}
\ee
where $E$ is the asymptotic kinetic energy of the incident nucleon
$E= E^{A+1}- E^A_0$.
The particle Hamiltonian therefore has eigenstates corresponding to discrete and continuum overlap states
\bea
h^{(p)}(E^{+}_n) \ket{\chi^{n+}} &=& E^{+}_n\  \ket{\chi^{n+}}  \label{eq:pd} \\
h^{(p)}(E) \ket{\chi^{0+}_E} &=& E\  \ket{\chi^{0+}_E} .
\label{eq:pdc}
\eea
For  complex $z$, the particle Green's function is defined as
\be
G^{(p)}(\bm{r}, \bm{r}'; z) =\bra{\Psi_0^A} a_{\bm{r}} \frac{1}{z-(\hat{H} - E^A_0 )} a^\dagger_{\bm{r}'} \ket{\Psi^A_0} ,
\label{eq:pgf}
\ee
and can be written employing complete sets of $A+1$-system states as
\begin{equation}
G^{(p)}(\bm{r} ,\bm{r}' ; z)  =  
\sum_n  \frac{ \chi^{n+}(\bm{r}') \left[\chi^{n+}(\bm{r})\right]^*
}{ z - E^{+}_n}  
 + 
\sum_c \int_{T^{(+)}_c}^{\infty} dE'\  \frac{\chi^{c+}_{E'}(\bm{r}') \left[\chi^{c+}_{E'}(\bm{r})\right]^* }{
z - E'} .
 \label{eq:propp} 
\end{equation}
The particle Green's function may therefore have discrete poles at negative energy on the real axis and a cut chosen to be just below the real-energy axis for $E > 0$.

The particle Hamiltonian, as well as the hole one in Eq.~(\ref{eq:hse}), can be decomposed as follows
\be
h^{(p)}(E) = T + \Sigma^{(p)}(E),
\label{eq:pse}
\ee
with $T$ the kinetic-energy operator and $\Sigma^{(p)}(E)$ the particle self-energy.
For scattering eigenstates of the particle Hamiltonian,
one has
\be
h^{(p)}(E_k) \ket{\phi_{\bm{k}}} = E_k \ket{\phi_{\bm{k}}} ,
\label{eq:pevc}
\ee
with the index $\bm{k}$ referring to the asymptotic boundary condition.
The relevant Lippmann-Schwinger equation is given by
\bea
\ket{\phi_{\bm{k}}} & = & \ket{\bm{k}} +\frac{1}{E_k -h^{(p)}(E_k) + i\eta} \Sigma^{(p)}(E_k) \ket{\bm{k}}
\nonumber \\
& = & i\eta G^{(p)}(E_k+i\eta ) \ket{\bm{k}} .
\label{eq:LSeq}
\eea
with $\ket{\bm{k}}$ the usual plane-wave eigenstates of the kinetic energy (with suitable modifications for protons).

For the $(A+1)$-body problem, consider the elastic-scattering eigenstate $\ket{\Psi^{0+}_{\bm{k}}}$ in which the target is in the ground state
\be
\hat{H} \ket{\Psi^{0+}_{\bm{k}}} = \mathcal{E}^0_k \ket{\Psi^{0+}_{\bm{k}}} ,
\label{eq:pscat}
\ee
with 
\be
\mathcal{E}^0_k = E^A_0 +E_k .
\label{eq:escat}
\ee
The corresponding Lippmann-Schwinger equation is given by
\be
\ket{\Psi^{0+}_{\bm{k}}}  =  \frac{i\eta}{E_k -(\hat{H}-E^A_0) + i\eta} a^\dagger_{\bm{k}}
\ket{\Psi^A_0} .
\label{eq:LSeqp}
\ee
The elastic overlap in coordinate space is given by
\be
\chi^{0+}_{\bm{k}}(\bm{r}) = \bra{\Psi^A_0} a_{\bm{r}} \ket{\Psi^{0+}_{\bm{k}}} .
\label{eq:elov}
\ee
It is now possible to demonstrate~\cite{CapMah:00} that
\be
\ket{\chi^{0+}_{\bm{k}}} = \ket{\phi_{\bm{k}}} ,
\label{eq:equivp}
\ee
thereby clarifying the importance of the particle self-energy as the scattering potential that generates the elastic scattering wave.

Following similar steps as for the hole Green's function, a Feshbach Hamiltonian for the particle Green's function can be generated employing the relevant projection operators.
As this particle Hamiltonian also has a divergence at large $E$, one may proceed as before to eliminate it, although in the present case, the relevant auxiliary operator is given by
\be
\mathcal{N} = 1- \rho .
\label{eq:NpF}
\ee
The corresponding Hamiltonian then reads
\bea
h^{(p)}_{\mathcal{F}}(E) & = & (1-\rho) h^{(p)}(E) +E \rho  \nonumber \\ 
& = & \left[ R^{(p)}  + D^{(p)}(E) \right] \frac{1}{1-\rho} \nonumber \\
 & = & T + \left( \mathcal{V}^{(p)} +D^{(p)} \right) \frac{1}{1-\rho} .
\label{eq:hphamF}
\eea
The equivalence of this derivation to the original Feshbach optical potential~\cite{Feshbach:58}
is demonstrated in Ref.~\cite{CapMah:00}.

The Green's-function theory of the optical potential introduced in Ref.~\cite{Bell:59} identifies the optical potential with the self-energy $\Sigma(E)$ associated with the time-ordered Green's function.
Unlike the separate hole and particle self-energies, this self-energy contains information on sp properties of both the ($A-1$)- and ($A+1$)-system, simultaneously. 
Indeed, the one-body Hamiltonian
\be
h(E) = T + \Sigma(E) 
\label{eq:gfh}
\ee
has $\ket{\chi^{n+}} $, $\ket{\chi^{0+}_E}$, and $\ket{\chi^{n-}}$ as eigenstates.

It is convenient also for later practical applications employing this framework to introduce the average Fermi energy
\be
\varepsilon_F = {\scriptstyle{\frac{1}{2}} } \left[\varepsilon_F^+  - \varepsilon_F^- \right] ={\scriptstyle{\frac{1}{2}} }\left[ \left(E^{A+1}_0 - E^A_0 \right) - \left(E^A_0 - E^{A-1}_0\right) \right] .
\label{eq:Fenergy}
\ee
The spectral function is the sum of the hole and particle spectral functions
\bea
S(E) & = & S^{(h)}(E) + S^{(p)}(E)
\label{eq:GFsfs} \\
S(E) &=& S^{(h)}(E) \hspace{4.0cm} E < \varepsilon_F \\
S(E) &=& S^{(p)}(E) \hspace{4.0cm} E > \varepsilon_F
\eea
which can be utilized to write the time-ordered Green's function according to
\begin{equation}
G( z)  =  
\int_{-\infty}^{\infty} dE'\  \frac{S( E') }{z - E'} 
 \label{eq:propLR} 
\end{equation}
which is analytic for all complex values of $z$.
It exhibits singularities on the real axis with a left-hand cut along the real axis from $-\infty$ to $T^-_0$ and right-hand cut from 0 to $\infty$.
It is therefore clear that
\be
G(E) = G^{(h)}(E) + G^{(p)}(E) .
\label{eq:Gph}
\ee
For complex $z$, one may write
\bea
h(z) =  z - \left[ G(z) \right]^{-1} \nonumber \\
\left[z - h(z) \right] G(z) = 1 .
\label{eq:resG}
\eea
With similar limits as $G$ for real $E$, one can write the one-body Hamiltonian as
\be
h(E) = T + \Sigma(E) .
\label{eq:HGF}
\ee
The self-energy or mass operator is in general complex and its perturbation expansion can be analyzed in the standard manner as shown for example in Ref.~\cite{Dickhoff:08}.

\subsection{Alternative derivation linking the nucleon self-energy to the optical potential}
\label{sec:viilars}
While it is intuitively clear that the sp propagator contains information 
related to the elastic scattering of particles from a target ground state, it
is appropriate to provide an additional derivation for this observation.
Historically, the first attempt for such a connection was presented 
in Ref.~\cite{Bell:59}. We will follow the presentation of~\cite{vill67} 
and~\cite{blri86} because it has the merit that it can be extended to composite projectiles and ejectiles.
We will not deal with the complications associated with treating the recoil of 
the nucleus~\cite{revi70}.
Neither will we attempt to discuss this problem in terms of the more 
correct wave-packet formulation.
The initial and final state of the target nucleus are denoted by 
$\ket{\Psi^A_0}$, representing an even-even nucleus with total angular momentum
zero.
As before, $\hat{H} \ket{\Psi^A_0} = E^A_0 \ket{\Psi^A_0}$.
The states describing the projectile before and after the reaction will
have the same magnitude of momentum and correspond to plane waves with momentum
$\bm{p}_i$ and $\bm{p}_f$, respectively. 
An extension to charged-particle scattering is possible by replacing the momentum states by Coulomb ones.
We suppress spin (isospin) for convenience and represent the initial state of
the combined projectile-target system by
\begin{equation}
a^\dagger_{\bm{p}_i} \ket{\Psi^A_0} = \ket{\phi_i}  ,
\label{eq:23.22}
\end{equation}
with energy
\begin{equation}
E_i = E^A_0 + \frac{\bm{p}^2_i}{2m} .
\label{eq:23.23}
\end{equation}
For the final state we write similarly
\begin{equation}
a^\dagger_{\bm{p}_f} \ket{\Psi^A_0} = \ket{\phi_f} ,
\label{eq:23.24}
\end{equation}
with
\begin{equation}
E_f = E^A_0 + \frac{\bm{p}^2_f}{2m} = E_i .
\label{eq:23.25}
\end{equation}
Note that these states are not eigenstates of the Hamiltonian in the 
wave-packet sense, except in the 
distant past or future, respectively.
This becomes clear by evaluating 
\begin{equation}
[\hat{H},a^\dagger_{\bm{p}_i}] =\frac{\bm{p}^2_i}{2m} a^\dagger_{\bm{p}_i}
+ J^\dagger_i = \varepsilon_i a^\dagger_{\bm{p}_i}
+ J^\dagger_i
\label{eq:23.26}
\end{equation}
and
\begin{equation}
[\hat{H},a_{\bm{p}_f}] =-\frac{\bm{p}^2_f}{2m} a_{\bm{p}_f}
- J_f = -\varepsilon_f a_{\bm{p}_f} - J_f .
\label{eq:23.27}
\end{equation}
The first term on the right in Eqs.~(\ref{eq:23.26}) and (\ref{eq:23.27})
comes from the commutator with the kinetic-energy operator $\hat{T}$ and 
the other terms
can be obtained from the interaction, involving two addition and
one removal operator for $J^\dagger_i$ and two removal and one addition 
operator for $J_f$, if one is restricted to two-body interactions.
From Eq.~(\ref{eq:23.26}) we infer
\begin{eqnarray}
\hat{H} \ket{\phi_i} & = & [\hat{H},a^\dagger_{\bm{p}_i}]\ket{\Psi^A_0} +
E^A_0 a^\dagger_{\bm{p}_i}\ket{\Psi^A_0} \nonumber \\
& = & E_i \ket{\phi_i} + J^\dagger_i \ket{\Psi^A_0} ,
\label{eq:23.28}
\end{eqnarray}
and similarly from Eq.~(\ref{eq:23.27})
\begin{equation}
\bra{\phi_f} \hat{H} = E_f \bra{\phi_f} + \bra{\Psi^A_0} J_f  .
\label{eq:23.29}
\end{equation}
The stationary scattering states
are eigenstates of $\hat{H}$ and obey
\begin{equation}
(\hat{H}-E_i) |\Psi^{(+)}_i\rangle = 0 .
\label{eq:23.30}
\end{equation}
Inverting Eq.~(\ref{eq:23.28}), while adding a solution of the homogeneous
equation (\ref{eq:23.30}), yields
\begin{equation}
|\Psi^{(+)}_i\rangle = \ket{\phi_i} +\frac{1}{E_i-\hat{H}+i\eta} J^\dagger_i
\ket{\Psi^A_0} ,
\label{eq:23.31}
\end{equation}
where the $i\eta$ signals the outgoing-wave boundary condition and it can be 
shown that $|\Psi^{(+)}_i\rangle$ is properly normalized.
Similarly, one finds
\begin{equation}
|\Psi^{(-)}_f\rangle = \ket{\phi_f} +\frac{1}{E_f-\hat{H}-i\eta} J^\dagger_f
\ket{\Psi^A_0} .
\label{eq:23.32}
\end{equation}
The $S$-matrix element for the transition $i \rightarrow f$ is then 
~\cite{gott94}
\begin{equation}
S_{fi}= \langle \Psi^{(-)}_f | \Psi^{(+)}_i \rangle .
\label{eq:23.33}
\end{equation}
Inserting Eqs.~(\ref{eq:23.31}) and (\ref{eq:23.32}) judiciously in this 
expression, one finds
\begin{equation}
\langle \Psi^{(-)}_f | \Psi^{(+)}_i \rangle  = 
\langle \phi_f | \phi_i \rangle + \bra{\phi_f} 
\frac{1}{E_i -\hat{H} +i \eta} J^\dagger_i \ket{\Psi^A_0} 
 +  \bra{\Psi^A_0} J_f \frac{1}{E_f -\hat{H} +i \eta} |\Psi^{(+)}_i\rangle .
\label{eq:23.34}
\end{equation}
Using Eq.~(\ref{eq:23.27}) one may show
\begin{equation}
a_{\bm{p}_f} \frac{1}{E-\hat{H}\pm i\eta} = 
\frac{1}{E-\varepsilon_f -\hat{H} \pm i\eta} a_{\bm{p}_f}
+ \frac{1}{E-\varepsilon_f -\hat{H} \pm i\eta} J_f 
\frac{1}{E-\hat{H} \pm i \eta} .
\label{eq:23.35}
\end{equation}
With this result, the second term of Eq.~(\ref{eq:23.34}) can be rewritten
as
\begin{eqnarray}
\lefteqn{\bra{\Psi^A_0} a_{\bm{p}_f} \frac{1}{E_i-\hat{H}+i\eta} J^\dagger_i
\ket{\Psi^A_0} = \frac{1}{E_i-E_f+i\eta} \bra{\Psi^A_0} a_{\bm{p}_f}
J^\dagger_i \ket{\Psi^A_0} } \hspace{1.5truecm} \nonumber \\
& & + \frac{1}{E_i -E_f +i \eta} \left\{ \bra{\Psi^A_0} J_f 
|\Psi^{(+)}_i\rangle
- \bra{\Psi^A_0} J_f \ket{\phi_i} \right\} , \hspace{0.5truecm}
\label{eq:23.36}
\end{eqnarray}
where Eq.~(\ref{eq:23.31}) has been employed as well.
Inserting Eq.~(\ref{eq:23.36}) into (\ref{eq:23.34}) yields, after further 
manipulation,
\begin{eqnarray}
\langle \Psi^{(-)}_f | \Psi^{(+)}_i \rangle & = &
\langle \phi_f | \phi_i \rangle + \frac{1}{E_i -E_f +i \eta} 
\bra{\Psi^A_0} \left( a_{\bm{p}_f} 
 J^\dagger_i -J_f a^\dagger_{\bm{p}_i} \right) \ket{\Psi^A_0} \nonumber \\
& - & 2\pi i \delta(E_f - E_i) \bra{\Psi^A_0}J_f |\Psi^{(+)}_i\rangle .
\label{eq:23.37}
\end{eqnarray}
Using Eqs.~(\ref{eq:23.28}) and (\ref{eq:23.29}), the second term on the
right side can be shown to be proportional to $E_i - E_f=0$, and 
therefore vanishes for all $\eta > 0$.
We may therefore identify the
transition matrix from Eq.~(\ref{eq:23.37}) according to
\begin{equation}
\bra{\phi_f}\mathcal{T}\ket{\phi_i} = \bra{\Psi^A_0} J_f 
|\Psi^{(+)}_i\rangle .
\label{eq:23.38}
\end{equation}
A similar derivation yields the equivalent relation
\begin{equation}
\langle \phi_f |\mathcal{T}\ket{\phi_i} = \langle \Psi^{(-)}_f | J^\dagger_i 
\ket{\Psi^A_0} 
\label{eq:23.39}
\end{equation}
noting that the condition $E_f = E_i$ must hold.

It remains to be shown how to relate these expressions to the sp propagator.
For this purpose we use Eq.~(\ref{eq:23.31}) to transform 
Eq.~(\ref{eq:23.38}) into
\begin{eqnarray}
\bra{\phi_f} \mathcal{T} \ket{\phi_i} & = &
\bra{\Psi^A_0} \left\{J_f,a^\dagger_{\bm{p}_i}\right\}\ket{\Psi^A_0} 
\nonumber \\
& - & \bra{\Psi^A_0} a^\dagger_{\bm{p}_i} J_f \ket{\Psi^A_0}
+ \bra{\Psi^A_0}J_f \frac{1}{E_i - \hat{H} +i \eta} J^\dagger_i 
\ket{\Psi^A_0} . \hspace{0.5truecm}
\label{eq:23.40}
\end{eqnarray}
The second term in this relation can be rewritten by utilizing again
Eq.~(\ref{eq:23.26}) yielding
\begin{equation}
\bra{\Psi^A_0} a^\dagger_{\bm{p}_i} = \bra{\Psi^A_0} J^\dagger_i
\frac{1}{E^A_0 - \hat{H} - \varepsilon_i} ,
\label{eq:23.41}
\end{equation}
where the denominator never vanishes.
This observation follows by inserting a complete set of $A-1$ states on the 
right and noting that the ground-state energy $E^A_0$ can never be equal
to the sum of the energies of the two fragments.
We can then write
\begin{eqnarray}
\bra{\phi_f} \mathcal{T} \ket{\phi_i} & = &
\bra{\Psi^A_0} \left\{J_f,a^\dagger_{\bm{p}_i}\right\}\ket{\Psi^A_0} 
 -  \bra{\Psi^A_0} J^\dagger_i \frac{1}{E^A_0 - \hat{H} -
\varepsilon_i} J_f \ket{\Psi^A_0} \nonumber \\
& + & \bra{\Psi^A_0}J_f \frac{1}{E_i - \hat{H} +i \eta} J^\dagger_i 
\ket{\Psi^A_0} . \hspace{0.5truecm}
\label{eq:23.42}
\end{eqnarray}
It is now straightforward but somewhat tedious to show that
\begin{equation}
\bra{\phi_f} \mathcal{T} \ket{\phi_i} =
\lim_{E \rightarrow \varepsilon_i} (\varepsilon_i - E)^2 
G(\bm{p}_f,\bm{p}_i;E) ,
\label{eq:23.43}
\end{equation}
and furthermore, using the Dyson equation in terms of the reducible 
self-energy (see Sec.~\ref{sec:Mah}), that
\begin{equation}
\bra{\phi_f} \mathcal{T} \ket{\phi_i} = 
\bra{\bm{p}_f} \Sigma_{red}(E=\varepsilon_i) \ket{\bm{p}_i} ,
\label{eq:23.44}
\end{equation}
which completes the proof of the relation between the elastic-scattering 
amplitude and the sp propagator.
From now on we will use both the self-energy notation $\Sigma_{red}$ as well as $\mathcal{T}$ to denote the same quantity in the context of elastic nucleon-nucleus scattering.
We will also employ the $\mathcal{T}$ symbol for the two-body quantity in free space which should be clear from the context of the discussion.

\subsection{Multiple scattering}
\label{sec:MSc}
The formal approach sketched in Sec.~\ref{sec:Feshbach} demonstrates that it is possible to formulate the effective one-body problem to describe elastic scattering using a projection operator $P$ onto the elastic channel and the complementary one $Q$ that projects on all others.
However, using the NN interaction to generate the energy-independent term of Eq.~(\ref{eq:hphamF}) is recognized as an inappropriate starting point as realistic NN interactions contain strong repulsion that is only tamed by an all-order summation that is contained in the energy-dependent term.
With this recognition, it was suggested one could  include important multiple-scattering events from the very start either in terms of the NN $\mathcal{T}$-matrix~\cite{Chew:51,Watson:53} or the $\mathcal{G}$-matrix that only allows intermediate states in accord with the Pauli principle. 
A comprehensive overview of this approach is given in Ref.~\cite{Amos:00} together with an extensive bibliography.
Only a brief discussion is therefore provided here in order to make the transition to schemes that allow for practical calculations.

It is assumed that two-particle interactions between projectile and target nucleons dominate.
With the projectile tagged 0 and the target particles $i$, one can write
\be
V = \sum_{i=1}^A V(0,i) = A V(0,i) .
\label{eq:Vpt}
\ee
The NN $\mathcal{T}$-matrix associated with $V(0,i)$ can be obtained from the Lippmann-Schwinger equation as
\be
\mathcal{T}(0,i) = V(0,i) + V(0,i) G^+_0(E) \mathcal{T}(0,i) ,
\label{eq:Tmat}
\ee
with $G^+_0$ representing propagation in free space while generating the correct boundary condition.
Replacing the NN interaction in the optical potential by the $\mathcal{T}$-matrix yields a workable approach to the optical-model potential.
While this formulation is expected to work at higher energies, it is possible to make corrections for the Pauli principle and mean-field propagation in $G^+_0$ which extend the approach to lower energies.
The most utilized approximation is the $\mathcal{G}$-matrix discussed in Sec.\ref{sec:JLM} for nuclear matter.

A full-folding approach is an important implementation of the spectator expansion of multiple-scattering theory.
The $(A+1)$-particle Hamiltonian for this problem can be written as
\be
\mathcal{H} = H_0 + \mathcal{H}_A + \sum_{i=1}^A V(0,i) = \mathcal{H}_0 + \sum_{i=1}^A V(0,i) .
\label{eq:Ham+}
\ee
Using the Green's function
\be
\mathcal{G}_0(E) = \frac{1}{E-\mathcal{H}_0 + i \eta} ,
\label{eq:GF+}
\ee
the full $\mathcal{T}$-matrix
for the elastic scattering from the target's ground state of $A$ nucleons can be written as
\be
\mathcal{T} = V + V G_0(E) \mathcal{T} ,
\label{eq:Tmat+}
\ee
where $V$ includes all pairwise interactions of the projectile with all target nucleons.
Assuming the elastic-scattering channel to be described in terms of an optical potential $U$, one can write
\be
\mathcal{T} = U + U G_0(E) P \mathcal{T} ,
\label{eq:Tmat+U}
\ee
with
\be
U = V + V G_0(E) Q \mathcal{T} ,
\label{eq:Upot+}
\ee
employing the relevant projection operators, \textit{e.g.} $P= \ket{\Psi^A_0}\bra{\Psi^A_0}$.
The full elastic-scattering $\mathcal{T}$-matrix can then be written as
\be
\mathcal{T}_{el} = PUP + PUP G_0(E) P \mathcal{T}_{el} .
\label{eq:Tmat+el}
\ee
A spectator expansion then involves an ordering of the scattering process in a sequence of active projectile-target interactions.
In first order, the interaction between projectile and target is written in standard notation as
\be
U = \sum_{i=1}^A \tau(0,i),
\label{eq:Utau}
\ee
where
\be
\tau(0,i)= V(0,i) + V(0,i) \mathcal{G}_0(E)Q\tau(0,i) =\hat{\tau}(0,i)  - \hat{\tau}(0,i) \mathcal{G}_0(E) P \tau(0,i) ,
\label{eq:tau}
\ee
with reduced amplitudes
\be
\hat{\tau}(0,i)= V(0,i) + V(0,i) \mathcal{G}_0(E) \hat{\tau}(0,i) .
\label{eq:taur}
\ee
For elastic scattering, only $\bra{\Psi^A_0} \tau(0,i) \ket{\Psi^A_0}$ contributes to the optical potential.
The task is then to find suitable approximations and solutions of Eqs.~(\ref{eq:tau}) and (\ref{eq:taur}).
The effective interaction can be obtained from the free $\mathcal{T}$-matrix by correcting the propagator for the binding associated with the nucleon that the projectile interacts with as well as allowing for its possible mean-field corrections.
Introducing initial and final momentum $\bm{k}$ and $\bm{k}'$, as well as
\bea
\bm{q} & = & \bm{k}' - \bm{k} \nonumber \\
\bm{K} & = & \frac{1}{2} (\bm{k} + \bm{k}')
\label{eq:moms} \\
\bm{P} & = & \frac{1}{2} (\bm{p} + \bm{p}') \nonumber
\eea
the latter is associated with the integration over the internal momenta of the target nucleons.
Convolution of the effective interaction with the target's ground state then yields a nonlocal optical potential which can be written in momentum space as~\cite{Amos:00}
\be
\label{eq:spectr}
U(\bm{q},\bm{K};E) = \sum_\alpha \int d\bm{P}\ \eta(\bm{P},\bm{q},\bm{K}) \hat{\tau}_\alpha(\bm{k}',\bm{k};\omega)\ \rho_\alpha\left(\bm{P}-(A-1)\bm{q}/(2A), \bm{P}+(A-1)\bm{q}/(2A)\right) ,
\ee
where $\hat{\tau}_\alpha$ is evaluated in the NN rest frame, $\eta$ imposes Lorentz invariant flux in transforming from the NN to the nucleon-$A$ system, and $\rho$ represents the one-body density matrix.
In the spectator model, the free $\mathcal{T}$-matrix is used at an energy associated with the beam energy and the binding of the struck particle.
When $\mathcal{G}$-matrix interactions are employed, the coupling with the integration variable $\bm{P}$ can be included~\cite{Amos:00}. 

As noted above, an extensive review was published about this approach in Ref.~\cite{Amos:00} with many citations to original work.
We will only discuss some recent applications of this approach in Sec.~\ref{sec:Trho}.

\subsection{Nucleon self-energy approach: dispersive optical model}
\label{sec:Mah}

We introduce here some relevant results from Green's-function theory as it pertains to the application of the dispersive optical model, promoted successfully by Mahaux~\cite{Mahaux:1991}.
Some of the material below is a condensed version of results recently reviewed in Ref.~\cite{Dickhoff:17}.
The aim of this section is to introduce relevant quantities related to experiment that can be obtained from the single-particle propagator.
In addition, this section presents the pertinent one-body equations that need to be solved including their interpretation.
The link between the particle and hole domain plays an important role in this formulation.

The nucleon propagator with respect to the $A$-body ground state in the sp basis with good radial position, orbital angular momentum (parity) and total angular momentum while suppressing the projection of the total angular momentum and the isospin quantum numbers can be obtained from Eq.~(\ref{eq:Gph}) and written as
\begin{eqnarray}
G _{\ell j}(r ,r' ; E) 
& = &  \sum_m \frac{\bra{\Psi^A_0} a_{r\ell j}
\ket{\Psi^{A+1}_m} \bra{\Psi^{A+1}_m} a^\dagger_{r' \ell j} \ket{\Psi^A_0}
}{ E - (E^{A+1}_m - E^A_0 ) +i\eta }  
\nonumber \\
& + &  \sum_n \frac{\bra{\Psi^A_0} a^\dagger_{r' \ell j} \ket{\Psi^{A-1}_n}
\bra{\Psi^{A-1}_n} a_{r \ell j} \ket{\Psi^A_0} }{
E - (E^A_0 - E^{A-1}_n) -i\eta} , 
\label{eq:prop}
\end{eqnarray}
where the continuum solutions in the $A\pm1$ systems are also implied in the completeness relations.
The numerators of the particle and hole components of the propagator represent the products of overlap functions associated with adding or removing a nucleon from the $A$-body ground state.
The Dyson equation can be obtained by analyzing the perturbation expansion or the equation of motion for the propagator~\cite{Dickhoff:08}.
It has the following form
\begin{equation}
\label{eq:dyson}
G_{\ell j}(r,r';E) = G^{(0)}_{\ell j}(r,r';E)  
+ \int \!\! d\tilde{r}\ \tilde{r}^2 \!\! \int \!\! d\tilde{r}'\ \tilde{r}'^2 G^{(0)}_{\ell j}(r,\tilde{r};E)
\Sigma_{\ell j}(\tilde{r},\tilde{r}';E) G_{\ell j}(\tilde{r}',r';E) ,
\end{equation}
with $G^{(0)}$ representing the noninteracting propagator containing only kinetic-energy contributions.
The nucleon self-energy $\Sigma$ contains all linked-diagram contributions that are irreducible with respect to propagation represented by $G^{(0)}$.

As discussed formally in Sec.~\ref{sec:Feshbach},
the solution of the Dyson equation generates all discrete poles corresponding to bound $A\pm1$ states explicitly given by Eq.~(\ref{eq:prop}) that can be reached by adding or removing a particle with quantum numbers $r \ell j$.
The hole spectral function is obtained from
\begin{equation}
S_{\ell j}(r;E) = \frac{1}{\pi}  \textrm{Im}\ G_{\ell j}(r,r;E)  
\label{eq:holes}
\end{equation}
for energies in the $A-1$-continuum.
For discrete energies as well as all continuum ones, overlap functions for the addition or removal of a particle are generated as well.

For discrete states in the $A-1$ system, one can show that the overlap function obeys a Schr{\"o}dinger-like equation~\cite{Dickhoff:08}.
Introducing the notation
\begin{equation}
\psi^n_{\ell j}(r) = \bra{\Psi^{A-1}_n}a_{r \ell j} \ket{\Psi^A_0} ,
\label{eq:overlap}
\end{equation}
for the overlap function for the removal of a nucleon at $r$ with discrete quantum numbers $\ell$ and $j$, one finds
\begin{eqnarray}
\left[ \frac{ p_r^2}{2m} +
 \frac{\hbar^2 \ell (\ell +1)}{2mr^2}\right]  & \psi^{n}_{\ell j}(r) 
+   \int \!\! dr'\ r'^2 
\Sigma_{\ell j}(r,r';\varepsilon^-_n)  &\psi^{n}_{\ell j}(r') =
\varepsilon^-_n \psi^{n}_{\ell j}(r) ,
\label{eq:DSeq}
\end{eqnarray}
where
\begin{equation}
\varepsilon^-_n=E^A_0 -E^{A-1}_n 
\label{eq:eig}
\end{equation}
and in coordinate space the radial momentum operator is given by $p_r = -i\hbar(\frac{\partial}{\partial r} + \frac{1}{r})$.
Discrete solutions to Eq.~(\ref{eq:DSeq}) exist in the domain where the self-energy has no imaginary part and these are normalized by utilizing the inhomogeneous term in the Dyson equation.
For an eigenstate of the Schr{\"o}dinger-like equation [Eq.~(\ref{eq:DSeq})], the so-called quasihole state labeled by $\alpha_{qh}$, the corresponding normalization or spectroscopic factor is given  by~\cite{Dickhoff:08}
\begin{equation}
S^n_{\ell j} = \bigg( {1 - 
\frac{\partial \Sigma_{\ell j}(\alpha_{qh},
\alpha_{qh}; E)}{\partial E} \bigg|_{\varepsilon^-_n}} 
\bigg)^{-1} .
\label{eq:sfac}
\end{equation}
Discrete solutions in the domain where the self-energy has no imaginary part can therefore be obtained by expressing Eq.~(\ref{eq:DSeq}) on a grid in coordinate space and performing the corresponding matrix diagonalization. 
Likewise, the solution of the Dyson equation [Eq.~(\ref{eq:dyson})] for continuum energies in the domain below the Fermi energy, can be formulated as a complex matrix inversion in coordinate space.
This is advantageous in the case of a non-local self-energy representative of all microscopic approximations that include at least the HF approximation.
Below the Fermi energy for the removal of a particle $\varepsilon_F^-$,
the corresponding discretization is limited by the size of the nucleus as can be inferred from the removal amplitude given in Eq.~(\ref{eq:overlap}), which demonstrates that only coordinates inside the nucleus need to be considered.
Such a finite interval therefore presents no numerical difficulty.

The particle spectral function for a finite system can be generated by the calculation of the reducible self-energy $\mathcal{T}$.
In some applications relevant for elucidating correlation effects, a momentum-space scattering code~\cite{PhysRevC.84.044319} to calculate $\mathcal{T}$ can be employed.
In an angular-momentum basis, iterating the irreducible self-energy $\Sigma$ to all orders, yields
\begin{equation}\label{eq:redSigma1}
\mathcal{T}_{\ell j}(k,k^\prime ;E) = \Sigma_{\ell j}(k,k^\prime ;E) 
  +  \!\!       \int \!\! dq\ q^2\ \Sigma_{\ell j}(k,q;E)\ G^{(0)}(q;E )\ \mathcal{T}_{\ell j}(q,k^\prime ;E) ,
\end{equation}
where $G^{(0)}(q; E ) = (E - \hbar^2q^2/2m + i\eta)^{-1}$ is the free propagator.
The propagator is then obtained from an alternative form of the Dyson equation in the following form~\cite{Dickhoff:08}
\begin{equation}
G_{\ell j}(k, k^{\prime}; E) = \frac{\delta( k - k^{\prime})}{k^2}G^{(0)}(k; E)  
 \label{eq:gdys1}  
		+		      G^{(0)}(k; E)\mathcal{T}_{\ell j}(k, k^{\prime}; E)G^{(0)}(k'; E) .	
\end{equation}
The on-shell matrix elements of the reducible self-energy in Eq.~(\ref{eq:redSigma1}) are sufficient to describe all aspects of elastic scattering like differential, reaction, and total cross sections as well as polarization data~\cite{PhysRevC.84.044319}.
The connection between the nucleon propagator and elastic-scattering data can therefore be made explicit by identifying the nucleon elastic-scattering $\mathcal{T}$-matrix with the reducible self-energy obtained by iterating the irreducible one to all orders with $G^{(0)}$~\cite{Bell:59,Dickhoff:08,vill67,blri86}.

The spectral representation of
the particle part of the propagator referring to the $A+1$ system, appropriate for a treatment of the continuum and possible open channels, is given by Eq.~(\ref{eq:propp}) in the present basis as
\begin{equation}
G_{\ell j}^{p}(k ,k' ; E)  =  
\sum_n  \frac{ \chi^{n+}_{\ell j}(k) \left[\chi^{n+}_{\ell j}(k')\right]^*
}{ E - E^{*A+1}_n +i\eta }   \label{eq:proppl} 
 + 
\sum_c \int_{T_c}^{\infty} dE'\  \frac{\chi^{cE'}_{\ell j}(k) \left[\chi^{cE'}_{\ell j}(k')\right]^* }{
E - E' +i\eta} ,
\end{equation}
which is a generalizing of the discrete version of Eq.~(\ref{eq:prop}) using a momentum-space formulation instead of one in coordinate space.
Overlap functions for bound $A+1$ states are given by $ \chi^{n+}_{\ell j}(k)=\bra{\Psi^A_0} a_{k\ell j}
\ket{\Psi^{A+1}_n}$, whereas those in the continuum are given by $ \chi^{cE}_{\ell j}(k)=\bra{\Psi^A_0} a_{k\ell j} \ket{\Psi^{A+1}_{cE}}$ indicating the relevant channel by $c$ and the energy by $E$.
Excitation energies in the $A+1$ system are with respect to the $A$-body ground-state $E^{*A+1}_n = E^{A+1}_n -E^A_0$.
Each channel $c$ has an appropriate threshold indicated by $T_c$ which is the experimental threshold with respect to the ground-state energy of the $A$-body system.
The overlap function for the elastic channel can be explicitly calculated by solving the Dyson equation while it is also possible to obtain the complete spectral density for $E>0$ 
\begin{eqnarray}
S_{\ell j}^{p}(k ,k' ; E) 
=
\sum_c \chi^{cE}_{\ell j}(k) \left[ \chi^{cE}_{\ell j}(k') \right]^* .
\label{eq:specp}
\end{eqnarray}
In practice, this requires solving the scattering problem twice at each energy so that one may employ
\begin{eqnarray}
\!\! S_{\ell j}^{p}(k ,k' ; E) 
= \frac{i}{2\pi} \left[ G_{\ell j}^{p}(k ,k' ; E^+) - G_{\ell j}^{p}(k ,k' ; E^-) \right]
\label{eq:specpp}
\end{eqnarray}
with $E^\pm =E\pm i\eta$, and only the elastic-channel contribution to Eq.~(\ref{eq:specp}) is explicitly known.
Equivalent expressions pertain to the hole part of the propagator $G_{\ell j}^{h}$~\cite{Mahaux:1991}.

The calculations are performed in momentum space according to Eq.~(\ref{eq:redSigma1}) to generate the off-shell reducible self-energy and thus the spectral density by employing Eqs.~(\ref{eq:gdys1}) and (\ref{eq:specpp}).
Because the momentum-space spectral density contains a delta-function associated with the free propagator, it is convenient for visualization purposes to also consider the Fourier transform back to coordinate space 
\begin{eqnarray}
S_{\ell j}^{p}(r ,r' ; E) = \frac{2}{\pi} \label{eq:specpr} 
  \int \!\! dk k^2 \! \int \!\! dk' k'^2 j_\ell(kr) S_{\ell j}^{p}(k ,k' ; E) j_\ell(k'r') ,
\end{eqnarray}
which has the physical interpretation for $r=r'$ as the probability density $S_{\ell j}(r;E)$ for adding a nucleon with energy $E$ at a distance $r$ from the origin for a given $\ell j$ combination.
By employing the asymptotic analysis to the propagator in coordinate space following \textit{e.g.} Ref.~\cite{Dickhoff:08}, one may express the elastic-scattering wave function that contributes to Eq.~(\ref{eq:specpr}) in terms of the half on-shell reducible self-energy obtained according to
\begin{equation}
\chi^{el E}_{\ell j}(r)  = \left[ \frac{2mk_0}{\pi \hbar^2} \right]^{1/2} \bigg\{ j_\ell(k_0r)  \label{eq:elwfl} \\
 +  \left. \int \!\! dk k^2 j_\ell(kr) G^{(0)}(k;E) \mathcal{T}_{\ell j}(k,k_0;E) \right\} ,
\end{equation}
where $k_0=\sqrt{2 m E}/\hbar$ is related to the scattering energy in the usual way.

The presence of strength in the continuum associated with mostly-occupied orbits (or mostly empty but $E<0$  orbits) is obtained by double folding the spectral density in Eq.~(\ref{eq:specpr}) in the following way
\begin{eqnarray}
\!\!\! S_{\ell j}^{n+}(E) 
=  \int \!\! dr r^2 \!\! \int \!\! dr' r'^2 \phi^{n-}_{\ell j}(r) S_{\ell j}^{p}(r ,r' ; E) \phi^{n-}_{\ell j}(r') ,
\label{eq:specfunc}
\end{eqnarray}
using an overlap function 
\begin{equation}
\sqrt{S^n_{\ell j}} \phi^{n-}_{\ell j}(r)=\bra{\Psi^{A-1}_n} a_{r\ell j} \ket{\Psi^{A}_0} , 
\label{eq:overm}
\end{equation}
corresponding to a bound orbit with $S^n_{\ell j}$, the relevant spectroscopic factor, and $\phi^{n-}_{\ell j}(r)$  normalized to 1~\cite{PhysRevC.84.044319}.

In the case of an orbit below the Fermi energy, Eq.~(\ref{eq:specfunc}) identifies where its now-depleted strength has been shifted into the continuum.
The occupation number of this orbit is given by an integral over a corresponding folding of the hole spectral density
\begin{eqnarray}
\!\!\!\!\!\!\! S_{\ell j}^{n-}(E) 
= \!\! \int \!\! dr r^2 \!\! \int \!\! dr' r'^2 \phi^{n-}_{\ell j}(r) S_{\ell j}^{h}(r ,r' ; E) \phi^{n-}_{\ell j}(r') ,
\label{eq:spechr}
\end{eqnarray}
where $S_{\ell j}^{h}(r,r';E)$ provides equivalent information below the Fermi energy as $S_{\ell j}^{p}(r,r';E)$ above.
An important sum rule is valid for the sum of the occupation number $n_{n \ell j}$ for the orbit characterized by 
$n \ell j$
\begin{equation}
n_{n \ell j} = \int_{-\infty}^{\varepsilon_F} \!\!\!\! dE\ S_{\ell j}^{n-}(E)
\label{eq:nocc}
\end{equation}
and its corresponding depletion number $d_{n \ell j}$
\begin{equation}
d_{n \ell j} = \int_{\varepsilon_F}^{\infty} \!\!\!\! dE\ S_{\ell j}^{n+}(E) ,
\label{eq:depl} 
\end{equation}
as discussed in general terms in Ref.~\cite{Dickhoff:17}.
It is simply given by~\cite{Dickhoff:08}
\begin{eqnarray}
\!\! 1 =  n_{n \ell j} + d_{n \ell j} \!\! =\bra{\Psi^A_0} a^\dagger_{n \ell j} a_{n \ell j} +a_{n \ell j}a^\dagger_{n \ell j}  \ket{\Psi^A_0} ,
\label{eq:sumr} 
\end{eqnarray}
reflecting the properties of the corresponding anticommutator of the operators $a^\dagger_{n \ell j}$ and $a_{n \ell j}$.

Strength above $\varepsilon_F$, as expressed by Eq.~(\ref{eq:specfunc}), reflects the presence of the imaginary self-energy at positive energies.
Without it, the only contribution to the spectral function comes from the elastic channel.
The folding in Eq.~(\ref{eq:specfunc}) then involves integrals of orthogonal wave functions and yields zero.
Because it is essential to describe elastic scattering with an imaginary potential, 
it automatically ensures that the elastic channel does not exhaust the spectral density and therefore some spectral strength associated with bound orbits in the independent-particle model also occurs in the continuum.

The (irreducible) nucleon self-energy in general obeys a dispersion relation between its real and imaginary parts given by~\cite{Dickhoff:08}
\begin{equation} 
\!\!\!\!\!\!\!\! \mbox{Re}\ \Sigma_{\ell j}(r,r';E)\! = \! \Sigma^s_{\ell j} (r,r')\! \label{eq:disprel} 
- \! {\cal P} \!\!
\int_{\varepsilon_T^+}^{\infty} \!\! \frac{dE'}{\pi} \frac{\mbox{Im}\ \Sigma_{\ell j}(r,r';E')}{E-E'}  
+{\cal P} \!\!
\int_{-\infty}^{\varepsilon_T^-} \!\! \frac{dE'}{\pi} \frac{\mbox{Im}\ \Sigma_{\ell j}(r,r';E')}{E-E'} , 
\end{equation}
where $\mathcal{P}$ represents the principal value.
The static contribution arises from the correlated HF term involving the exact one-body density matrix and the dynamic parts start and end at corresponding thresholds in the $A\pm1$ systems that have a larger separation than the corresponding difference between the Fermi energies for addition $\varepsilon_F^+$ and removal $\varepsilon_F^-$ of a particle.
The latter feature is particular to a finite system and generates possibly several discrete quasiparticle and hole-like solutions of the Dyson equation in Eq.~(\ref{eq:DSeq}) in the domain where the imaginary part of the self-energy vanishes.

The standard definition of the self-energy requires that its imaginary part is negative, at least on the diagonal, in the domain that represents the coupling to excitations in the $A+1$ system, while it is positive for the coupling to $A-1$ excitations.
This translates into an absorptive potential for elastic scattering at positive energy, where the imaginary part is responsible for the loss of flux in the elastic channel.
Subtracting Eq.~(\ref{eq:disprel}) calculated at the average Fermi energy [see Eq.~(\ref{eq:Fenergy})], from Eq.~(\ref{eq:disprel}) generates the so-called subtracted dispersion relation 
\begin{eqnarray} 
\!\!\!\!\! \mbox{Re}\ \Sigma_{\ell j}(r,r';E)\! &=& \!  \mbox{Re}\ \Sigma_{\ell j} (r,r';\varepsilon_F) \hspace{2.0cm}  \label{eq:sdisprel} \\
&-& \! {\cal P} \!\!
\int_{\varepsilon_T^+}^{\infty} \!\! \frac{dE'}{\pi} \mbox{Im}\ \Sigma_{\ell j}(r,r';E') \left[ \frac{1}{E-E'}  - \frac{1}{\varepsilon_F -E'} \right]  \nonumber  \\
&+& {\cal P} \!\!
\int_{-\infty}^{\varepsilon_T^-} \!\! \frac{dE'}{\pi} \mbox{Im}\ \Sigma_{\ell j}(r,r';E') \left[ \frac{1}{E-E'}
-\frac{1}{\varepsilon_F -E'} \right]  .
\nonumber
\end{eqnarray}
The beauty of this representation was recognized by Mahaux and Sartor~\cite{Mahaux:1986,Mahaux:1991} since it allows for a link with empirical information both at the level of the real part of the non-local self-energy at the Fermi energy (probed by a multitude of HF calculations) and also through empirical knowledge of the imaginary part of the optical potential (constrained by experimental data) that consequently yields a dynamic contribution to the real part by means of Eq.~(\ref{eq:sdisprel}).
In addition, the subtracted form of the dispersion relation emphasizes contributions to the integrals from the energy domain nearest to the Fermi energy on account of the $E'$-dependence of the integrands of Eq.~(\ref{eq:sdisprel}). 
Recent DOM applications reviewed briefly in this paper include experimental data up to 200 MeV of scattering energy and are therefore capable of determining the nucleon propagator in a wide energy domain as all negative energies are included as well.

\section{Empirical potentials for nucleons} 
\label{sec:empirical}
There is a long history of fitting elastic-scattering angular distributions with empirical local complex potentials;
\begin{equation}
U(r) = V(r) - i W(r) .
\end{equation}
 At the very least one needs to include a real potential well, typically Wood-Saxon in shape, and the Coulomb component if the reactants are both charged. Nucleons unlike clusters, can penetrate into and through the target nucleus without losing energy, especially at high bombarding energies. Therefore they can probe differential absorption between the central and peripheral regions of the target nucleus. This effect can be parametrized as a combination of a volume imaginary potential, with typically a Wood-Saxon form factor, and a surface dominated imaginary part. The latter is typically parametrized as a derivative of a Wood-Saxon or a Gaussian shape. For clusters, one typically considered either just a surface or a volume imaginary potential, there is no need to include both, as absorption takes place strongly on the surface of the target nucleus.

If analyzing powers are also to be described, then a real-spin orbit force should be included. Based on the sp potentials for bound levels, this is usually parametrized as surfaced peaked. At low-energies, this potential is found to be largely real, but a significant imaginary component appears to be needed at high energies~\cite{Nadasen:1981,Schwandt:1982}.

With these ingredients, experimental angular distributions and analyzing powers can be fit for a single bombarding energy. The fitted parameters are the strength, radii, and diffusenesses of the various components of the complex potential. It is well known that such fits suffer from ambiguities, i.e, more than one set of fit parameters can equally describe the experimental data. However it has been found that integrated potentials are better defined~\cite{Greenlees:1968,Hodgson:1971,Mahaux:1991}, \textit{i.e.}, 
\begin{eqnarray}
J_{V} &=& \frac{1}{A} \int V(\bm{r}) d\bm{r}, \label{eq:volV} \\
J_{W} &=& \frac{1}{A} \int W(\bm{r}) d\bm{r}.
\label{eq:volumeInteg}
\end{eqnarray}
The integrated spin-orbit potential has also  been considered in some studies.
Mahaux and Sartor \cite{Mahaux:1986} also consider moments of the optical potentials, \textit{i.e.},
\begin{eqnarray}
[r^q]_V &=& \frac{4\pi}{A} \int V(r) r^q dr, \label{eq:momentReal}\\ 
{[r^q]_W} & =& \frac{4\pi}{A} \int W(r) r^q dr, \label{eq:momentImag} 
\end{eqnarray}
where $q=2$ corresponds to the integrated potentials. They suggest that these are fairly well defined for $q$ from 0.8 to 2.

It is clear that more complicated decompositions of the imaginary potential beyond  surface and volume components is not necessary and cannot be discerned 
from the analysis of elastic-scattering data. In analogy to the imaginary potential, there is the question of whether the real potential also has a surface and volume contribution.  While generally a surface component is not included, at high energy  a ``wine-bottle'' shaped real potential is required to fit scattering data~\cite{Elton:1966,Satchler:1983}.

\subsection{Global potentials}
\label{sec:global}
For simplicity, many applications of optical potentials prefer to have a single global parametrization of the potential as a function of energy, $A$, and $N-Z$ asymmetry. However, this might represent a compromise of simplicity over accuracy. Certainly Hodgson~\cite{Hodgson:1984} thought ``the influence of nuclear structure on neutron scattering, particularly those associated with shell closure and deformation, are sufficiently large for some nuclei to ensure that there is no simple and accurate global potential.''  Indeed typical global potentials ignore these effects and fit all available data with a spherically-symmetric potential and these global fits are of poorer quality  compared to fits for a single target nucleus. Hodgson advocated for more regional fits to particular types of nuclei and regions of the periodical table.

Global parametrizations must specify the energy, $A$, and $N-Z$ dependencies of the magnitudes, radius, and diffuseness of all the different contributions to the total potential. Generally the energy and $N-Z$ dependencies are confined to the strengths of the potentials with their geometry being only dependent on $A$. 
The depths of extracted real potentials have strong energy dependencies which is known to be due to using a local potential instead of a more correct nonlocal version. Perey and Buck showed that
an energy-dependent local potential could provide equivalent results to  
an energy-independent nonlocal potential~\cite{Perey:1961}.  The depth of the equivalent local potential decreases with bombarding energy and eventually changes sign at a nucleon energy of around 200 MeV. This energy dependence has been represented as either an exponential, linear, or higher-order polynomial decrease in global fits.  

At low energies, optical-model fits indicate that the absorption is predominately of the surface type. The strength of this surface absorption peaks at roughly around 20~MeV, and above $\sim$50~MeV, volume absorption becomes dominant. These observations must be incorporated into the energy dependencies of the surface and volume absorption.

The $N-Z$ dependence is generally taken from the Lane potential~\cite{Lane:1962} which was formulated in regard to the description of quasi-elastic ($p$,$n$) reactions. The Lane potential~\cite{ Lane:1962} has the form
\begin{equation}
V = U_{0} + \frac{1}{A} (\bm{t} \cdot \bm{T}) U_{1}
\label{eq:Lane}
\end{equation}   
where $\bm{t}$ and $\bm{T}$ are the isospins of the incident nucleon and target nucleus, respectively and $U_0$ and $U_1$ depend on position but not $N$ or $Z$.   For elastic scattering where the isospin of the nucleus does not change, this reduces to
\begin{equation}
V = U_{0}    \pm U_{1} (N-Z)/4
\label{eq:Lane2}
\end{equation}
where the upper plus sign is for neutrons and the lower minus sign for protons. The $U_1$ coefficient can be related to the potential contribution to the nuclear symmetry energy. 

While Lane only considered the $N-Z$ asymmetry dependence of the real potential, Eq.~(\ref{eq:Lane2})  has also been applied to the imaginary potentials. 
Satchler~\cite{Satchler:1969} gives a microscopic justification of this standard
asymmetry dependence based on the average
interaction of a projectile nucleon with the individual target nucleons.
Taking into account the difference between the mean in-medium neutron-proton 
$\left\langle \sigma _{np}\right\rangle $ and proton-proton $\left\langle
\sigma _{pp}\right\rangle $ scattering cross sections, the imaginary
potential is%
\begin{eqnarray}
W^{p} &=&\frac{\hbar v\rho _{N}}{2}\left( \frac{N}{A}\left\langle \sigma
_{np}\right\rangle +\frac{Z}{A}\left\langle \sigma _{pp}\right\rangle \right) \nonumber
\\
&=&\frac{\hbar v\rho _{N}}{4}\left( \left\langle \sigma _{np}\right\rangle
+\left\langle \sigma _{pp}\right\rangle \right)   \nonumber \\
&&+\frac{N-Z}{A}\frac{\hbar v\rho _{N}}{4}\left( \left\langle \sigma
_{np}\right\rangle -\left\langle \sigma _{pp}\right\rangle \right) \nonumber \\
&=&W_{0}+\frac{N-Z}{A}W_{1}
\end{eqnarray}%
where $v$ is the incident nucleon's velocity inside the target nucleus which
has nucleon density of $\rho _{N}$. Similarly for an incident neutron%
\begin{eqnarray}
W^{n} &=&\frac{\hbar v\rho _{N}}{2}\left( \frac{N}{A}\left\langle \sigma
_{nn}\right\rangle +\frac{Z}{A}\left\langle \sigma _{np}\right\rangle \right) \nonumber
\\
&=&\frac{\hbar v\rho _{N}}{4}\left( \left\langle \sigma _{np}\right\rangle
+\left\langle \sigma _{nn}\right\rangle \right)   \nonumber \\
&&-\frac{N-Z}{A}\frac{\hbar v\rho _{N}}{4}\left( \left\langle \sigma
_{np}\right\rangle -\left\langle \sigma _{nn}\right\rangle \right)
\end{eqnarray}%
as $\left\langle \sigma _{nn}\right\rangle \simeq \left\langle \sigma
_{pp}\right\rangle $, then we can obtain the general form
\begin{equation}
W=W_{0}\pm \frac{N-Z}{A}W_{1}.
\end{equation}

\begin{figure}[tpb]
\begin{center}
\includegraphics[scale=.6]{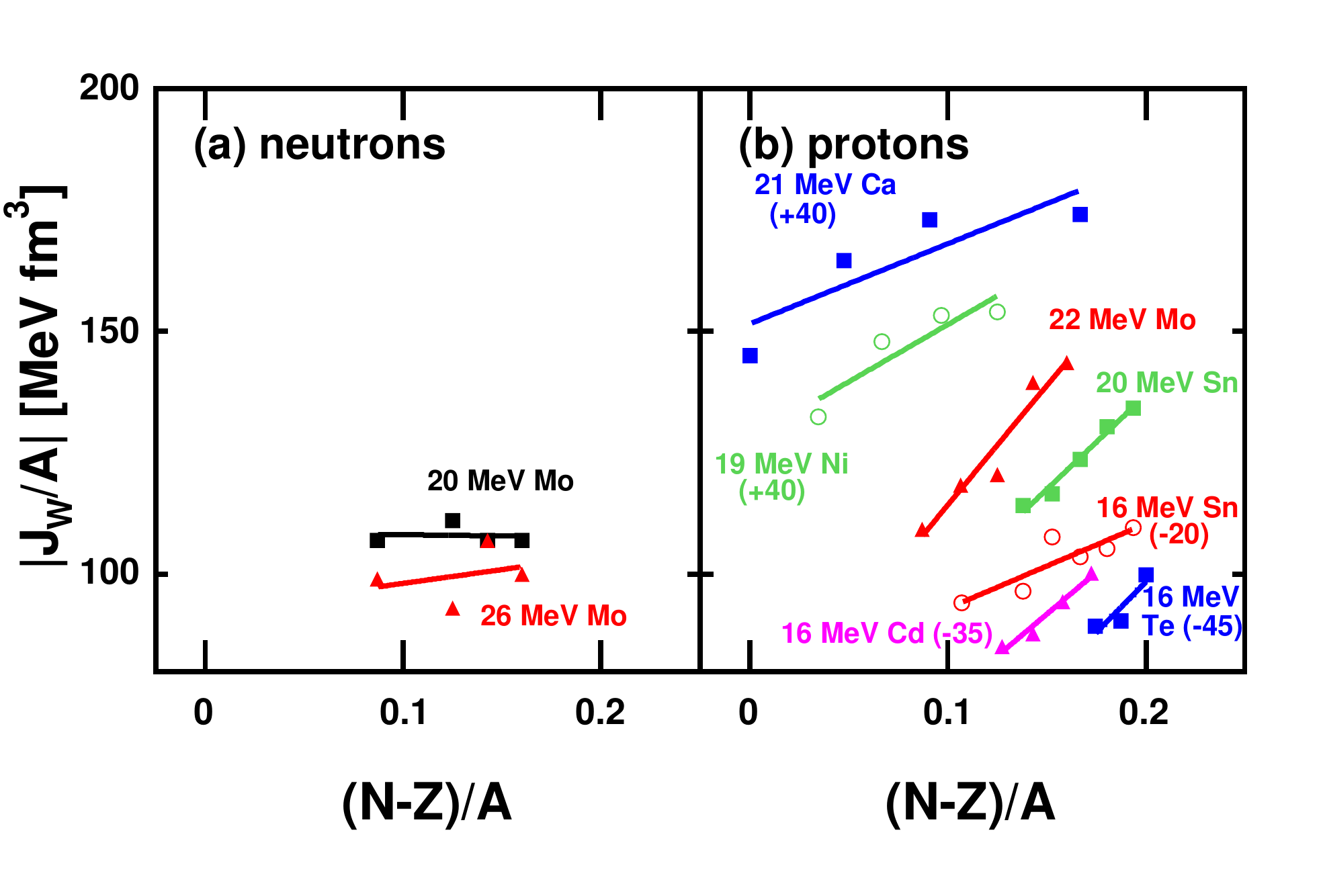}
\caption{Integrated imaginary potentials for (a) neutrons and (b) protons as a function a neutron-proton asymmetry. Data points were obtained from individual fits for elastic-scattering data. Lines are linear fits to results from the same element and bombarding energy. Reprinted figure with permission from \cite{Charity:2007} \textcopyright2007 by the American Physical Society.}
\label{fig:JwMoAsy}
\end{center}
\end{figure}

The correlations implicit is this derivation are short range and thus are
applicable to the volume imaginary potential. However most global potentials do not include an asymmetry dependence for this potential, rather the surface imaginary potential is fitted with a strong $N-Z$ dependence. While, a linear $N-Z$ dependence of the surface is well known for protons, data for neutrons are more sparse. In fact what data are available, plotted in Fig.~\ref{fig:JwMoAsy}, suggest that the dependence is not of the opposite sign to protons as given by Eq.~(\ref{eq:Lane2}), but rather flat \cite{Charity:2007}.
We note that in Fig.~\ref{fig:JwMoAsy} the convention is used to define volume integrals without dividing by $A$ as in Eqs.~(\ref{eq:volV}) and (\ref{eq:volumeInteg}).

In the 1960's a number of attempts at obtaining separate proton and neutron global optical-model potentials were made \cite{Perey:1963,Wilmore:1964, Menet:1971,Becchetti:1969}. Of these, the most influential by far (based on the Scopus citation index) are the proton and neutron potentials from Becchetti and Greenlees \cite{Becchetti:1969}. For protons, they fit elastic differential cross sections, polarization data and reactions cross sections for $A>$40 and $E<$50~MeV. For neutrons, available data restricted the fits to $E<$ 24~MeV. The neutron data included elastic-scattering angular distributions and polarization information plus total  cross sections. Some of these data were from natural targets and, in these cases, the fits considered the weighted contributions from individual isotopes. Only a few of the parameters were determined from these neutron data, most were fixed to their value obtained in fitting the proton data.
For both nucleon types, the real potential contained a linear dependence on energy. The surface and volume imaginary potentials  contained a linear decrease and increase with energy unless they went negative and then they were fixed to zero.
Radii were simply parametrized by a radius parameter times $A^{1/3}$.
Only  the real and surface imaginary potentials were parametrized with $N-Z$ dependencies.

\begin{figure}[tpb]
\begin{center}
\includegraphics[scale=.6]{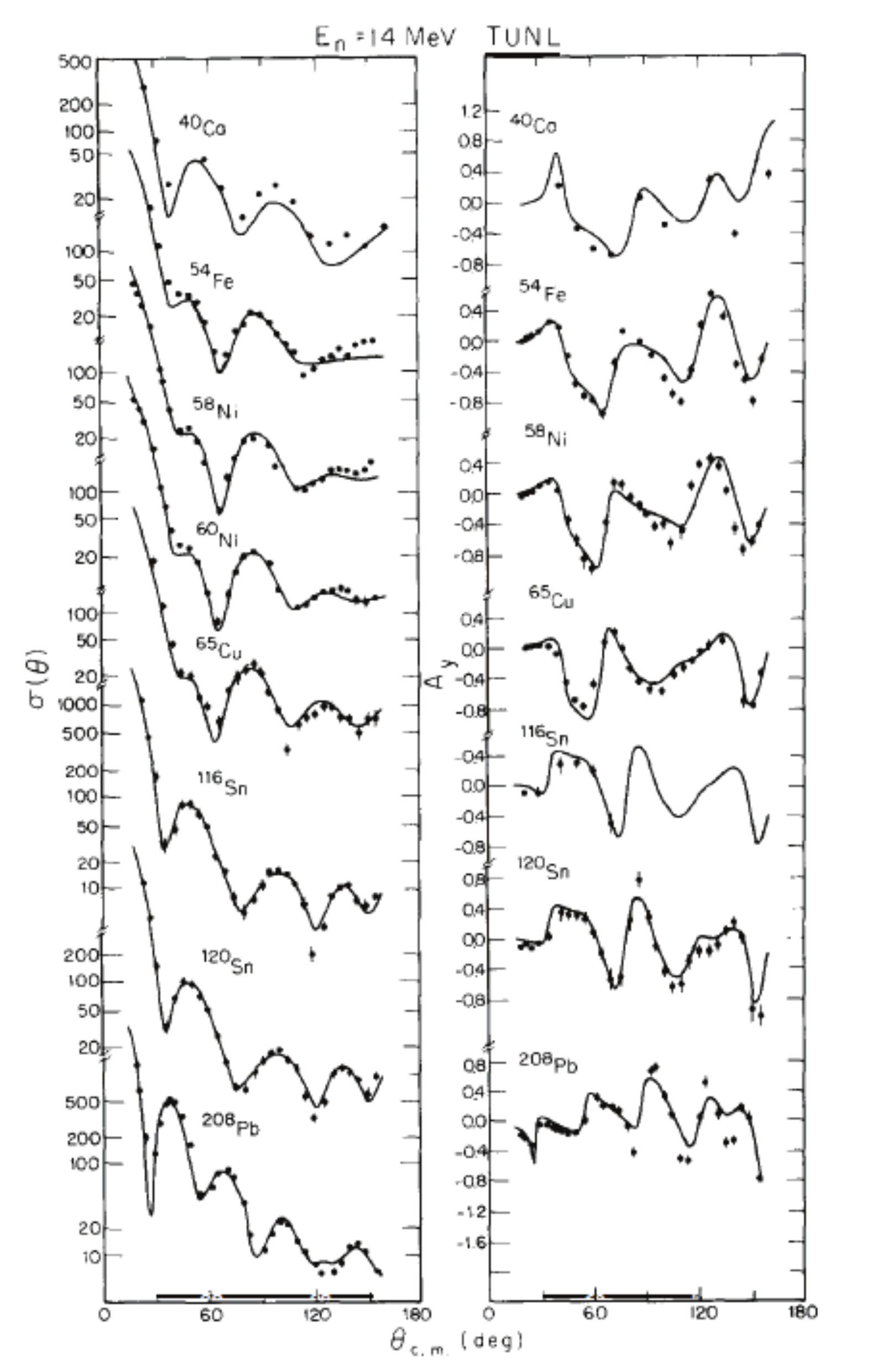}
\caption{Fits to elastic-scattering angular distributions and analyzing powers for polarized 14-MeV neutrons obtained with the CH89 global potential.  Reprinted figure with permission from \cite{Varner:1991} \textcopyright1991 by the Elsevier.}
\label{fig:varner}
\end{center}
\end{figure}

With the increase in available data with time, especially neutron measurements on separated isotopes, there was the possibility of a better global fit. 
This was realized in 1989 with the Chapel-Hill or CH89 potential  produced by Varner \textit{et al.}~\cite{Varner:1991}. It was fit for 40$\leq A\leq$209 with proton energies from 16 to 65~MeV and neutron energies from 10 to 26~MeV. These fitted regions of $A$ and $E$ are very similar to those of  Becchetti and Greenlees.  Again only the real and surface imaginary components have $N-Z$ dependencies and the energy dependence of the real potential in still linear. The surface imaginary part is parametrized to decrease in energy with a Fermi-function form, while the volume imaginary increases with a inverse-Fermi-function form.  Radii were parametrized as $r_{0}A^{1/3} + r_{1}$ with now two fit parameters, $r_0$ and $r_1$. Figure~\ref{fig:varner} shows an example of the quality of the fit to elastic-scattering angular distributions and analyzing powers for polarized 14-MeV neutrons.

\begin{figure}[tpb]
\begin{center}
\includegraphics[scale=.6]{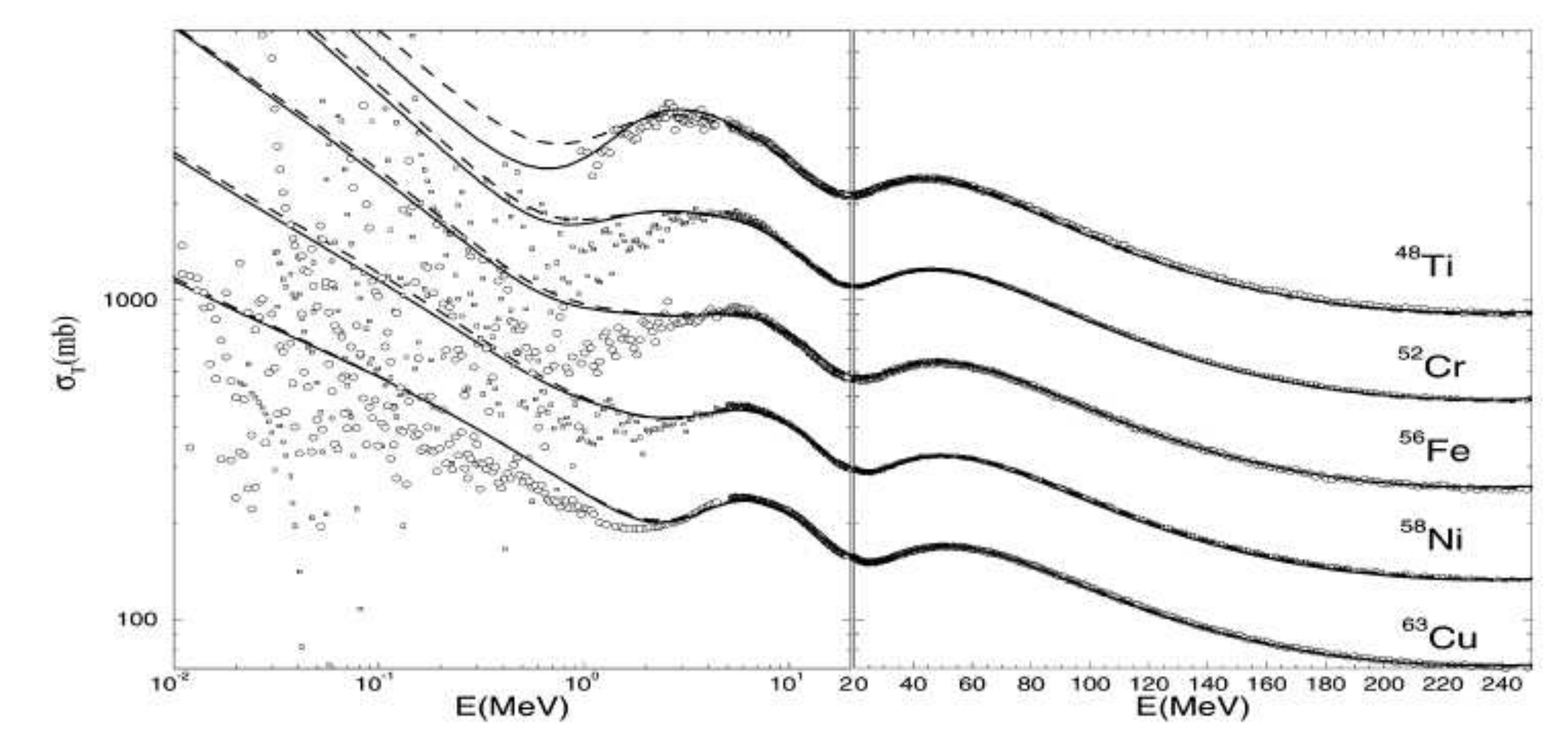}
\caption{Comparison of experimental and fitted total neutron cross sections for nuclei in the Ti-Cu region obtained by Koning \& Delaroche in their global optical-model parametrization. Reprinted figure with permission from \cite{Koning:2003} \textcopyright2003  from Elsevier}
\label{fig:Koning}
\end{center}
\end{figure}

Due to application needs, Koning \& Delaroche \cite{Koning:2003} developed their own global nucleon optical potential in 2003 to cover an enlarged range of  bombarding energies; 1~keV to 200~MeV covering both lower and higher values than the Chapel-Hill fit. The low-energy region $E<$ 5 MeV is largely constrained by neutron total cross sections and the fits only reproduce the average behavior here as they cannot produce the resonance peaks which are significant in this energy regime. An example of such a fit is shown for the Ti-Cu region in Fig.~\ref{fig:Koning}. The neutron total cross sections were also important to extent the fit to higher energy as there is a scarcity of neutron elastic-scattering angular distribution measurements.  This fit is for 24$\leq A\leq$209, but only for near spherical target nuclei, the deformed rare-earth and actinide regions were excluded reflecting the original concerns of Hodgson. Due to the larger range of energies considered, the linear dependence of the real potential was inadequate and replaced by a cubic dependence. An imaginary spin-orbit component is added and the energy dependencies of surface and volume are parametrized with different functions. Again only the real and surface imaginary components have $N-Z$ dependencies. Although this parameterization should not be used for deformed systems in standard optical-model codes, it was found that in coupled-channels calculations the  Koning \& Delaroche potential is useful. 
Nobre \textit{et al.} found that by staticly deforming this potential and including couplings to members of the ground-state rotation band, a good reproduction of elastic and inelastic neutron scattering in the rare-earth region can be obtained \cite{Nobre:2015}.

\subsection{Other projectiles}
Apart from nucleons, there is also a need for optical-model potentials for light clusters including $d$, $t$, $^3$He, and $\alpha$ particles. These have application in transfer reaction for example ($d$,$p$), ($d$,$t$), ($^{3}$He,$d$), and ($\alpha$,$t$) where in the distorted-wave Born approximation (DWBA) calculations, distorted waves are required for these fragments (Sec.~\ref{sec:transfer}). In addition, in statistical-model codes considering the decay of an excited compound nucleus, the transmission coefficients for the evaporation of nucleons and light fragments are usually taken from global-optical-model fits.

The earliest global deuteron potential is from 1963 by Perey and Perey~\cite{Perey:1963a}. It was fit for deuteron energies from 12 to 27~MeV. It has an energy-dependent real and surface imaginary potential, but no spin-orbit. 
In 1974, a complementary parametrization for 8$<E<$13~MeV was produced by Lohr and Haeberli~\cite{Lohr:1974} with energy-independent real and surface imaginary components.  With the inclusion of vector polarization data,  Burgl \textit{et al.}~\cite{Burgl:1980} and Daehnick \textit{et al.} constructed global fits including a spin-orbit term. Again these were limited to low energies with the former from 12-19~MeV and the latter from 9-15~MeV and in this case restricted to medium mass targets (46$<A<$90).

Global parameters for higher-energy deuterons can be obtained from two more recent works; An and Cai~\cite{An:2006} and Han, Shi, and Shen~\cite{Han:2006}. They were fit with data for energies up to 183 and 200~MeV, respectively. Lastly in 2016, a more restrictive parametrization was published for light $p$-shell nuclei  by Zang, Pang, and Lou~\cite{Zhang:2016} covering deuteron energies from 5.25 to 170~MeV.  Neither of these last three parametrizations were  fit to any polarization data.

Global parameters for $t$ and $^3$He fragments have been obtained by fitting elastic-scattering angular distributions and reaction cross sections only. Perey and Perey~\cite{Perey:1976} list separate parameters obtained by Becchetti and Greenlees for both fragments. The former was fit with data at 15 and 20~MeV, while the latter with $E<$40~MeV, otherwise no details are available. Li, Liang, and Cai \cite{Li:2007,Liang:2009} provide separate fits to both fragments with tritons restricted below 40~MeV and $^3$He fragments below 270~MeV. Pang \textit{et al.}~\cite{Pang:2009,Pang:2010} produced a single set of parameters (GDP08)  for both fragments with 30$<E<$217~MeV. Finally we note a more limited set from Urone \textit{et al.}~\cite{Urone:1972} for $E\sim$37~MeV and 40$\leq A\leq$91.






In the 1960's, energy-independent global optical-model parameters for $\alpha$-particle scattering were provided by McFadden and Satchler~\cite{Mcfadden:1966} and Huizenga and Igo~\cite{Huizenga:1962} and were used extensively to calculate transmission coefficients in statistical-model codes. These parametrization are only appropriate for energies of tens of MeV. Indeed the former set was fit to elastic-scattering data measured at 24.7 MeV for  19 targets from Oxygen to Uranium. In 1987, Nolte, Machner, and Bojowald~\cite{Nolte:1987}  obtained a parametrization for $E>$80~MeV including linear-energy dependencies of the real and a volume imaginary potential. More recently Kumar \textit{et al.}~\cite{Kumar:2006} has produced a parametrization applicable  from the Coulomb barrier to 140 MeV and Su and Han~\cite{su:2015} obtained a potential for $E<$286~MeV.

 $\alpha$-particle induced reactions below the barrier are important for nucleosynthesis in astrophysics~\cite{Demetriou:2002}. These include ($\alpha$,$\gamma$) and $\alpha$ transfer reactions and Hauser-Feshbach calculations require knowledge of the $\alpha$-particle optical potential at very low energies. A number of studies  have been made to obtain global parametrization for these applications~\cite{Avrigeanu:1994,Demetriou:2002,Avrigeanu:2010,Avrigeanu:2014} including a dispersive optical-model parametrization~\cite{Demetriou:2002}.   

\section{Microscopic calculations}
\label{sec:calc}

In this section, we discuss some of the most recent developments related to calculations in which an underlying NN interaction is at the basis of the results.
Typically, these NN interactions are called realistic if they fit NN data up to the pion-production threshold.
In some cases additional fitting to elastic-nucleon-nucleus scattering data is attempted as discussed in Sec.~\ref{sec:JLM}.

\subsection{Nuclear matter approach by Jeukenne, Lejeune, and Mahaux}
\label{sec:JLM}
In this section we outline the procedure pioneered in Ref.~\cite{PhysRevC.16.80} which will be referred to by JLM.
Dealing with the strong repulsive core of the NN interaction was the paramount difficulty in obtaining sensible treatments of nuclei and nuclear matter.
A brief sketch of the ingredients of the nuclear-matter considerations that pertain to the development of optical potentials obtained from suitable local-density approximations is given below.
Many-body calculations were initiated by Keith Brueckner who realized that a treatment of the interaction in the medium is required that closely follows how one deals with two free nucleons.
It should therefore be equivalent to solving a Schr\"{o}dinger-like two-particle equation which, in free space, adequately treats the repulsive core (even a hard core).
Such a solution must therefore contain the sum of ladder diagrams, as for
the scattering of nucleons in free space~\cite{PhysRev.97.1344}.
To account for the Pauli principle in symmetric nuclear matter, only 
propagation of particles above $k_F$ is included.
The method is referred to as the 
Brueckner--Hartree--Fock (BHF) approach.
The in-medium scattering equation is solved according to
\begin{eqnarray}
\bra{\bm{k} m_\alpha m_{\alpha'} } G (\bm{K},E)
\ket{\bm{k}' m_\beta m_{\beta'}  } \!\!\!\! & = & \!\!\!\!
\bra{\bm{k} m_\alpha m_{\alpha'} } V
\ket{\bm{k}' m_\beta m_{\beta'} }   + \frac{1}{2} \sum_{m_\gamma m_{\gamma'}}
\int \!\! \frac{d^3q}{(2\pi)^3}\
\bra{\bm{k} m_\alpha m_{\alpha'} } V
\ket{\bm{q} m_\gamma m_{\gamma'} }
\label{eq:16.31} \\
 & \times & \!\!
\frac{\theta(|\bm{q}+\bm{K}/2|-k_F)\
\theta(|\bm{K}/2-\bm{q}|-k_F)}
{E-\varepsilon(\bm{q}+\bm{K}/2) - 
\varepsilon(\bm{K}/2-\bm{q}) + i\eta} 
\bra{\bm{q} m_\gamma m_{\gamma'} } G (\bm{K},E)
\ket{\bm{k}' m_\beta m_{\beta'} } ,
\nonumber
\end{eqnarray}
where momentum space is employed. This is relevant for calculations employing interactions that can be Fourier transformed.
Explicit spin/isospin degrees of freedom are indicated by $m$ quantum numbers and momentum conservation is identified by the total momentum $\bm{K}$. The other momentum variables representing relative momenta of the initial, intermediate, and final two-particle state in the medium characterized by a Fermi momentum $k_F$ and its associated density.
The $G$ notation for the effective interaction was introduced by Bethe~\cite{PhysRev.103.1353}
and has since then, been referred to as the $G$-matrix.
Equation~(\ref{eq:16.31}) is also known as the Bethe--Goldstone 
equation which depends on the total momentum, the energy of propagation $E$, and the density.
The original derivation of the corresponding linked contributions to
the Brueckner theory was developed by Goldstone~\cite{Goldstone267}.
The $G$-matrix interaction obeys a dispersion relation
between its real and imaginary parts which can be employed to calculate the corresponding self-energy contribution by closing the diagrams with a hole propagator~\cite{Dickhoff:08}.
Note that the imaginary part of the $G$-matrix only exists 
above $2\varepsilon_F$.
Taking a mean-field sp propagator in combination with the Hartree-Fock contribution, then gives the BHF self-energy 
\begin{equation}
\Sigma_{BHF}(k;E)) \!\! = \!\! \int \frac{d^3k'}{(2\pi)^3} \frac{1}{\nu}
\sum_{m_\alpha m_{\alpha'}} \theta(k_F -k')  \label{eq:16.33} 
 \bra{ {\scriptstyle{\frac{1}{2}}}(\bm{k}-\bm{k}') 
m_\alpha m_{\alpha'} } 
G(\bm{k}+\bm{k}'; E+\varepsilon(\bm{k}') 
\ket{ {\scriptstyle{\frac{1}{2}}}(\bm{k}-\bm{k}')~ m_\alpha m_{\alpha'} }  .
\end{equation}
In a Green's-function approach such a self-energy, when used in the Dyson equation for $k < k_F$, only
produces solutions at the energies given by
\begin{equation}
\varepsilon_{BHF}(k)=\frac{\hbar^2 k^2}{2m} + 
\Sigma_{BHF}(k;\varepsilon_{BHF}(k)) ,
\label{eq:eBHF}
\end{equation}   
since this self-energy is real for energies less than $\varepsilon_F$.
One may determine the real sp energy  self-consistently, HF-like, hence one may speak of BHF.

A critical point in the BHF approach is encountered when the choice of
the auxiliary potential above the Fermi energy is contemplated.
Such a choice is necessary and relevant, since the final outcome will depend
on this selection.
The so-called continuous choice of the potential is relevant for optical-model considerations as the spectrum is continuous when crossing the Fermi momentum,
\begin{equation}
M^{(0)}_\rho(k)= \Sigma_{BHF}(k;\varepsilon_{BHF}(k))
\label{eq:UCBHF}
\end{equation}  
for all values of $k$ and making explicit the dependence on the density $\rho =2 k_F^3 /(3\pi^2)$.
This requires a more involved iteration scheme since the Bethe--Goldstone
equation must be recalculated, as knowledge of sp energies for
wave vectors above $k_F$ is required. 
It should be clear that due to energy conserving intermediate states in the Bethe-Goldstone equation, the on-shell self-energy in Eq.~(\ref{eq:UCBHF}) becomes complex and therefore a suitable ingredient for optical-potential considerations.

The seminal work of Ref.~\cite{PhysRevC.16.80} then developed the first attempt at generating an optical-model potential from nuclear-matter calculations starting from realistic NN interactions.
More recent work employing this approach is presented in Ref.~\cite{PhysRevC.58.1118}.
In these works, the self-energy is represented by the on-shell BHF term with an isovector correction associated with
the nucleon asymmetry
\be
M_\rho^{(1)} = M_\rho^{(0)} + \delta M^{(1)}_\rho ,
\label{eq:JLM}
\ee
where $\delta M^{(1)}_\rho$ is the first-order term in the difference $\Omega = {\scriptstyle{\frac{1}{2}}} (k_{F_n} -k_{F_p})$ around $M^{(0)}_\rho$~\cite{PhysRevC.16.80}, with $k_{F_n}$ and $k_{F_p}$, the Fermi momenta of neutron and proton distributions, respectively.
The isovector term $M^{(1)}_\rho$ is also evaluated on the energy shell, after which the optical potential is written as~\cite{PhysRevC.16.80}
\be
U_{NM}(\rho, E) = V_0(\rho , E) + \alpha \tau V_1(\rho , E) + i \left[ W_0(\rho , E) + \alpha \tau W_1(\rho , E) \right] ,
\label{eq:JLMop}
\ee
where $\alpha = (\rho_n - \rho_p)(\rho_n + \rho_p)$ is the nucleon asymmetry, and $\tau$ is +1 for proton and -1 for neutron projectiles.
The potential components are defined by
\bea
V_0(\rho , E) & = & \textrm{Re} M^{(0)}_\rho (k(E),E) \label{eq:resy} \\
V_1(\rho , E) & = & \frac{\tilde{m}}{m} \textrm{Re} N^{(0)}_\rho (k(E),E) \label{eq:reasy} \\
W_0(\rho , E) & = & \frac{m}{\bar{m}}\textrm{Im} M^{(0)}_\rho (k(E),E) \label{eq:imsy} \\
W_1(\rho , E) & = & \frac{m}{\bar{m}}\textrm{Im} N^{(0)}_\rho (k(E),E) \label{eq:imasy} ,
\eea
respectively, with
\be
N_\rho(k,E) = \frac{1}{\alpha} \delta M^{(1)}_\rho .
\label{eq:JLMN}
\ee
The effective masses used in Eqs.~(\ref{eq:resy})-(\ref{eq:imasy}) include the $k$-mass and $E$-mass representing the true nonlocality and energy dependence of the optical-model potential~\cite{JEUKENNE197683} which are given by
\bea
\frac{m}{\tilde{m}} &=& 1 + \frac{m}{k} \frac{\partial}{\partial k} \left. \left( \textrm{Re}\ M^{(0)}_\rho \right) \right|_{k=k(E)} \label{eq:kmass} \\
\frac{\bar{m}}{m} &=& 1 - \frac{\partial}{\partial E} \left. \left( \textrm{Re}\ M^{(0)}_\rho \right) \right|_{E} . \label{eq:emass}
\eea
To generate the correct mean free path, it was shown in Refs.~\cite{PhysRevLett.47.71,FANTONI198189} that an additional effective-mass correction $\tilde{m}/m$ must be applied to the imaginary part.
Note that for protons, the optical potential should be evaluated at $E-V_c$~\cite{PhysRevC.15.10}.

\begin{figure}[t]
   \begin{minipage}{\columnwidth}
      \makebox[\columnwidth]{
         \includegraphics[scale=0.4]{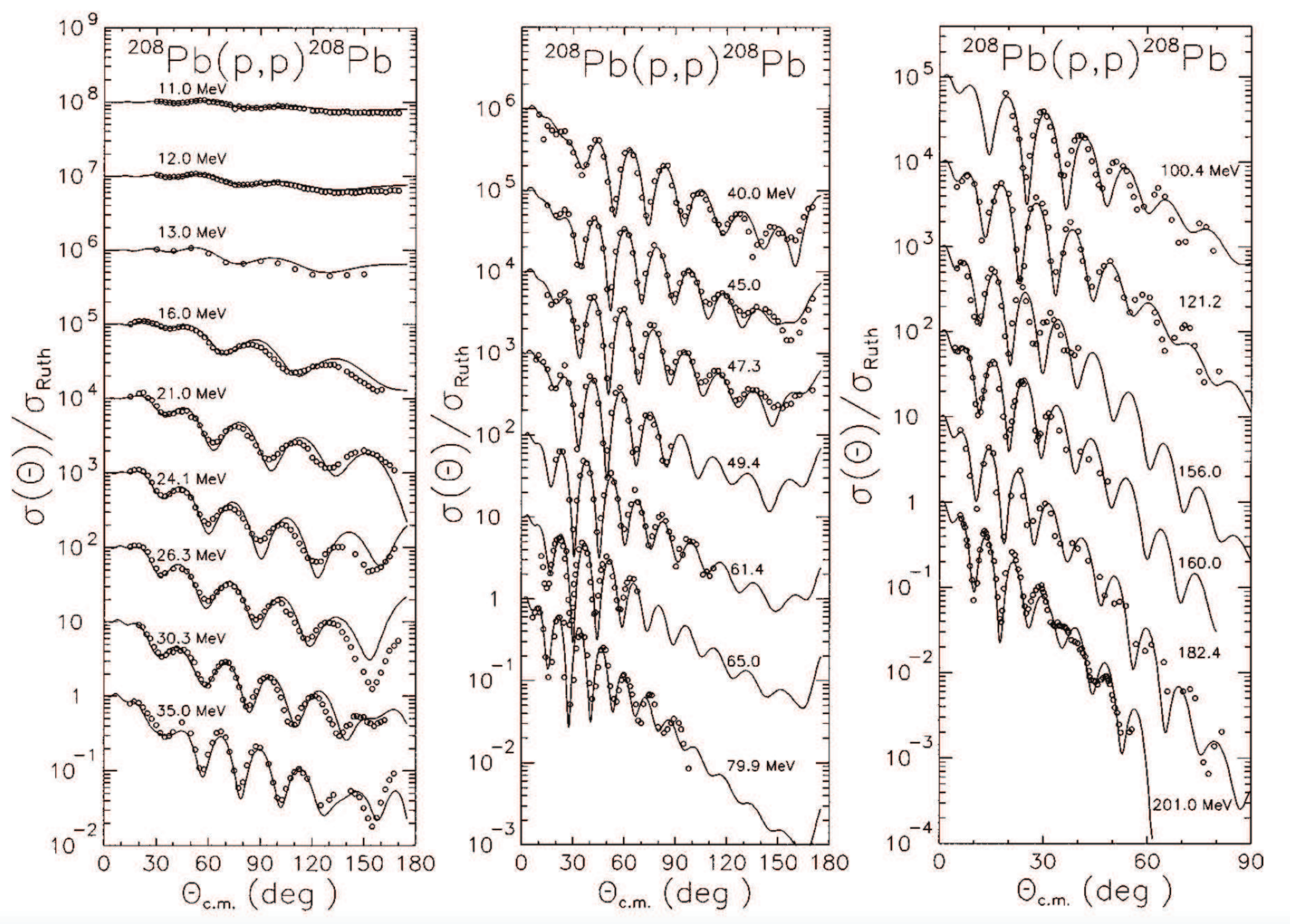}
      }
   \end{minipage}
   \caption{
   Differential cross sections using the extended procedure of Ref.~\cite{PhysRevC.58.1118} compared with experimental data (open circles) for protons incident on ${}^{208}$Pb between 10 MeV and 201 MeV. Note that elastic cross sections are presented as ratios to the Rutherford values. These results are offset by factors of 10. Reprinted figure with permission from Ref.~\cite{PhysRevC.58.1118} \textcopyright1998 by the American Physical Society.
 }
   \label{fig:JLM}
\end{figure} 
The results were parametrized in Ref.~\cite{PhysRevC.16.80} in powers of the density multiplied by powers of the projectile energy below 160 MeV.
An extension to 200 MeV was provided in Ref.~\cite{PhysRevC.58.1118} together with some improvements to avoid the wrong sign of the imaginary part at larger radii in heavy nuclei.
In order to provide the potential for a finite nucleus, a local density approximation (LDA) is required using appropriately calculated densities.
The standard version of the such a LDA is given by
\be
U_{LDA}(r,E) = U_{NM}(\rho(r),E) ,
\label{eq:LDA}
\ee
representing the optical-model potential for a finite nucleus. 
With this simple prescription, volume integrals are well reproduced but root-mean-square radii are underestimated.
An improved LDA is obtained by employing a folding of the nuclear-matter potential with a Gaussian reflecting some of the physics associated with the finite range of the NN interaction.

One possible procedure is to extract the proton density from elastic-electron-scattering data and employ a scaling factor of $N/Z$ to approximate the neutron density.
The work of Ref.~\cite{PhysRevC.58.1118} employs Hartree-Fock-Bogoliubov densities (see \textit{e.g.} Ref.~\cite{PhysRevC.21.1568}).
While the procedure outlined above can describe some elastic-scattering data, it has typically been found necessary to employ correction factors for the various ingredients of the optical potential. 
Such corrections can be quite substantial and in some cases depend strongly on the projectile energy~\cite{PhysRevC.58.1118}.
A typical result of this procedure is shown in Fig.~\ref{fig:JLM}.

With the correction factors adjusted to provide the optimal description of the data, there is a good agreement as illustrated in Fig.~\ref{fig:JLM}.
It is therefore possible to use such potentials to provide distorted waves for other reaction calculations.
Nevertheless, it should be noted that substantial phenomenological adjustments have gone into to the construction of such potentials.
Some of the drawbacks of this approach can be identified as follows. 
The nucleon self-energy for a given nucleus should be nonlocal and obey the standard dispersion relation between its real and imaginary part.
The JLM procedure does not do justice to these basic requirements.
Furthermore, a truly \textit{ab initio} approach would calculate the potential directly for the nucleus in question which remains a very elusive goal at present.

\subsection{Chiral interactions and the isospin-asymmetry of optical potentials}
\label{sec:Holt}
An alternative approach based on nuclear-matter calculations employing modern chiral interactions~\cite{RevModPhys.81.1773,MACHLEIDT20111} was recently introduced in Ref.~\cite{PhysRevC.93.064603}.
Relevant related material can be found in Refs.~\cite{LI2015408,LI201829}.
The softness of chiral interactions with typical cut-offs of 500 MeV allow for a perturbative approach of the nucleon self-energy in nuclear matter.
The calculations of Ref.~\cite{PhysRevC.93.064603} are focused on the isospin-asymmetry dependence of optical potentials with emphasis on its relevance for the study of rare isotopes.
The potential is determined by taking up to second order self-energy contributions into account while self-consistently determining the nucleon sp energies at each different value of $\delta_{np}= (N-Z)/A$.
Using the standard decomposition of Eq.~(\ref{eq:Lane}),
the isovector potential is extracted as the linear term in an expansion in $\delta_{np}$
\be
U = U_0 - U_I \tau_3 \delta_{np} + U_{II} \delta_{np}^2 + ...
\label{eq:UI}
\ee
with $\tau_3 = \pm 1$ for protons and neutrons, respectively.
\begin{figure}[bt]
   \begin{minipage}{\columnwidth}
      \makebox[\columnwidth]{
         \includegraphics[scale=0.45]{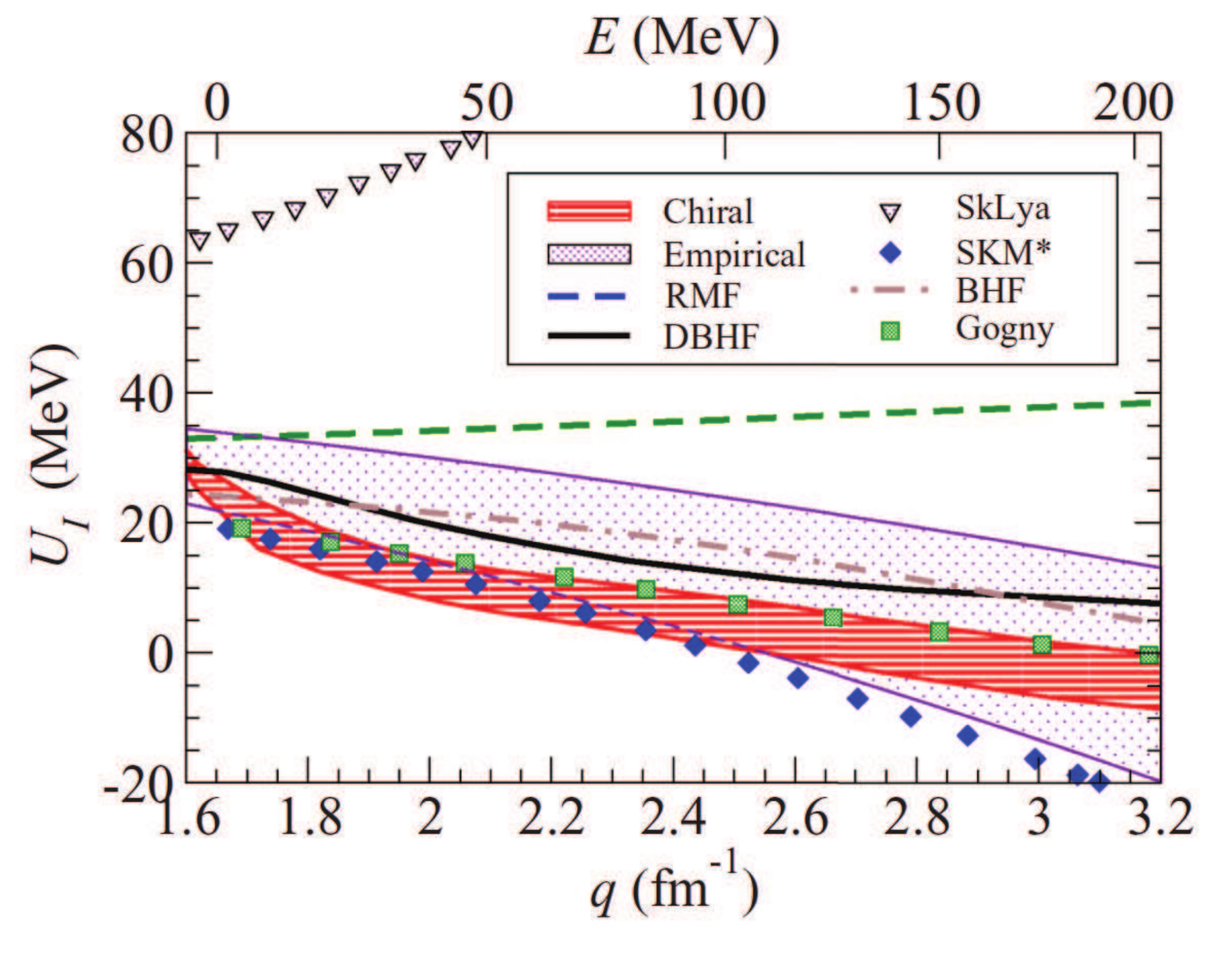}
      }
   \end{minipage}
   \caption{Energy dependence of the isovector real optical-model
potential at saturation density from chiral effective field theory. 
Reprinted figure with permission from Ref.~\cite{PhysRevC.93.064603} \textcopyright2016 by the American Physical Society.
}
   \label{fig:Holt}
\end{figure} 
The energy dependence of the isovector real optical-model
potential at saturation density from such a calculation employing chiral effective field theory is identified by the striped area shown in Fig.~\ref{fig:Holt}.
Shown for comparison are the predictions of other microscopic,
semimicroscopic, and phenomenological models obtained from Ref.~\cite{PhysRevC.72.065803}.
The empirical results are not well constrained and the authors of Ref.~\cite{PhysRevC.93.064603} express the hope that their work may help constrain the use of isovector potentials for exotic nuclei. 

\subsection{Some applications of the multiple-scattering approach}
\label{sec:Trho}
\begin{figure}[t]
   \begin{minipage}{\columnwidth}
      \makebox[\columnwidth]{
         \includegraphics[scale=0.45]{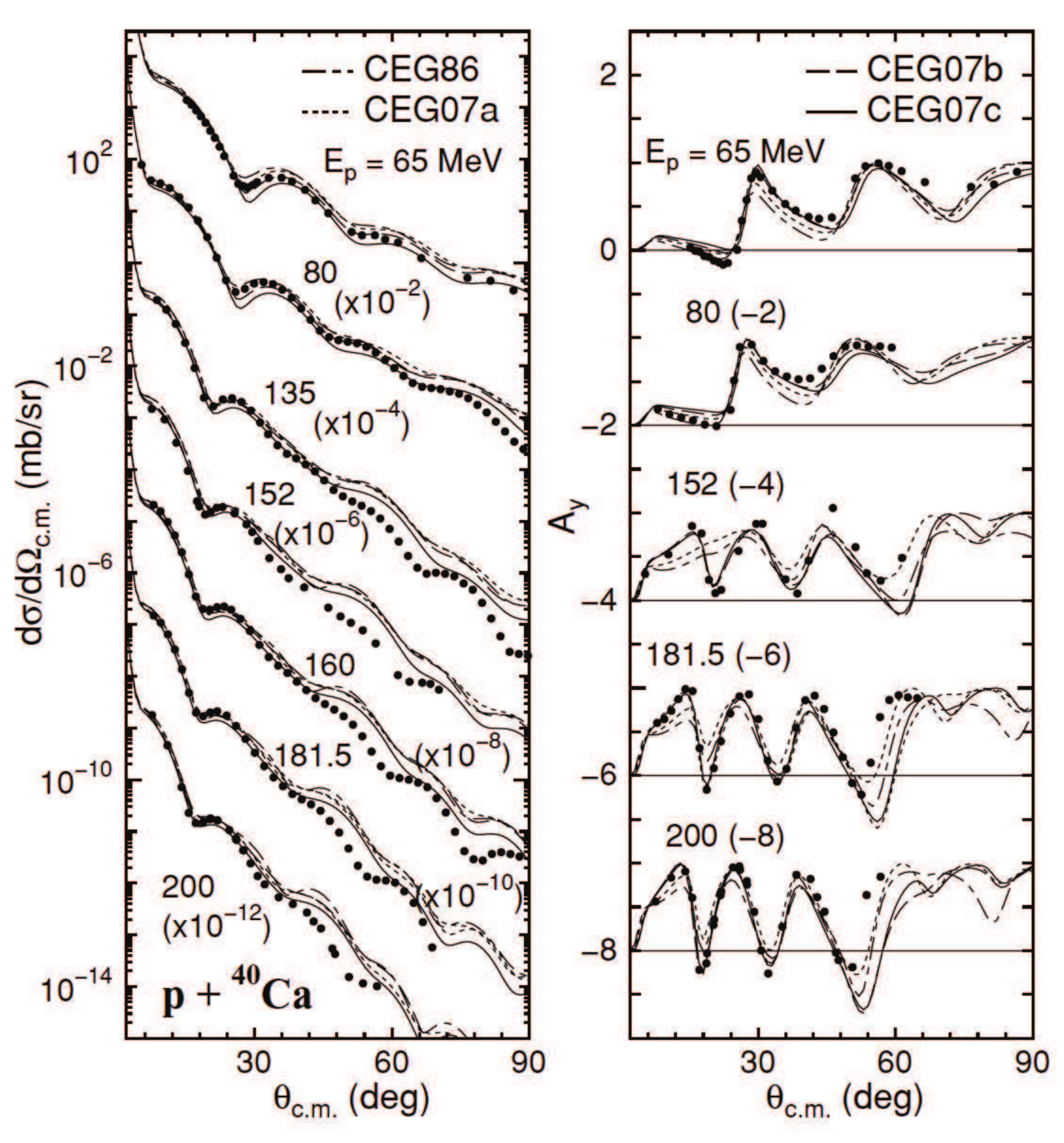}
      }
   \end{minipage}
   \caption{Differential cross sections and analyzing power for elastic proton scattering on a ${}^{40}$Ca target at $E_p$ = 65, 80, 135, 152, 160, 181.5, and 200 MeV.  
   Reprinted figure with permission from Ref.~\cite{PhysRevC.78.044610} \textcopyright2008 by the American Physical Society.
}
   \label{fig:PRC78}
\end{figure}

Applications of the multiple-scattering approach briefly outlined in Sec.~\ref{sec:MSc} mostly employ NN $\mathcal{T}$-matrices or $\mathcal{G}$-matrices that contain medium effects .
The folding of $\mathcal{G}$-matrices with an appropriate density matrix first requires its calculation.
Such calculations are almost exclusively performed in nuclear matter and may follow procedures sketched in Sec.~\ref{sec:JLM}.
A recent calculation along these lines was published in Ref.~\cite{PhysRevC.78.044610}.
A folding procedure involving proton and neutron densities according to Ref.~\cite{PTP70459} was employed.
In Fig.~\ref{fig:PRC78}, calculations for elastic proton scattering from ${}^{40}$Ca are compared with data for the differential cross section and analyzing power.
These calculations employ a soft two-body interaction with an additional three-body force which contains both short-range repulsion and long-range attraction.
Averaging over the third nucleon at a given density, an effective density-dependent interaction is generated that is added to the two-body interaction in generating the nuclear-matter $\mathcal{G}$-matrix.
As usual, there are substantial renormalization factors required to describe reaction cross sections for different nuclei~\cite{PhysRevC.78.044610}.
Results for differential cross sections and analyzing powers for proton elastic scattering from ${}^{40}$Ca are shown in Fig.~\ref{fig:PRC78}.
Different prescriptions for the folding procedures yield similar results and the effect of three-body forces is not significant for differential cross sections.
The authors point out that the description of analyzing powers improves when the effect of three-body interactions is included as can be seen at small angles at higher energies in Fig.~\ref{fig:PRC78}.

The momentum-space approach to the multiple-scattering framework can be implemented when NN $\mathcal{T}$ matrices are used and an appropriate density matrix is employed (see \textit{e.g.} Ref.~\cite{PhysRevC.88.034610} for unstable He isotopes). 
Recent developments focus on generating translation-invariant one-body density matrices obtained from no-core shell-model calculations~\cite{Burrows:18}.
This procedure has the advantage that the one-body density matrix can be generated from the same underlying NN interaction that is employed to calculate the $\mathcal{T}$-matrix. 
This consistency has not been achieved until recently and is currently being implemented.
Another example of this approach has recently been published in Ref.~\cite{Gennari:18}.
The nonlocal densities were obtained from the SRG-evolved NN-N4LO(500)+3Nlnl interaction developed by Refs.~\cite{PhysRevC.91.014002,PhysRevC.96.024004} for NN and for 3N as used in Ref.~\cite{PhysRevLett.120.062503}.
SRG stands for the use of the similarity renormalization group that softens the relevant interactions as reviewed in Ref.~\cite{Bogner:2010}.
\begin{figure}[t]
   \begin{minipage}{\columnwidth}
      \makebox[\columnwidth]{
         \includegraphics[scale=0.45]{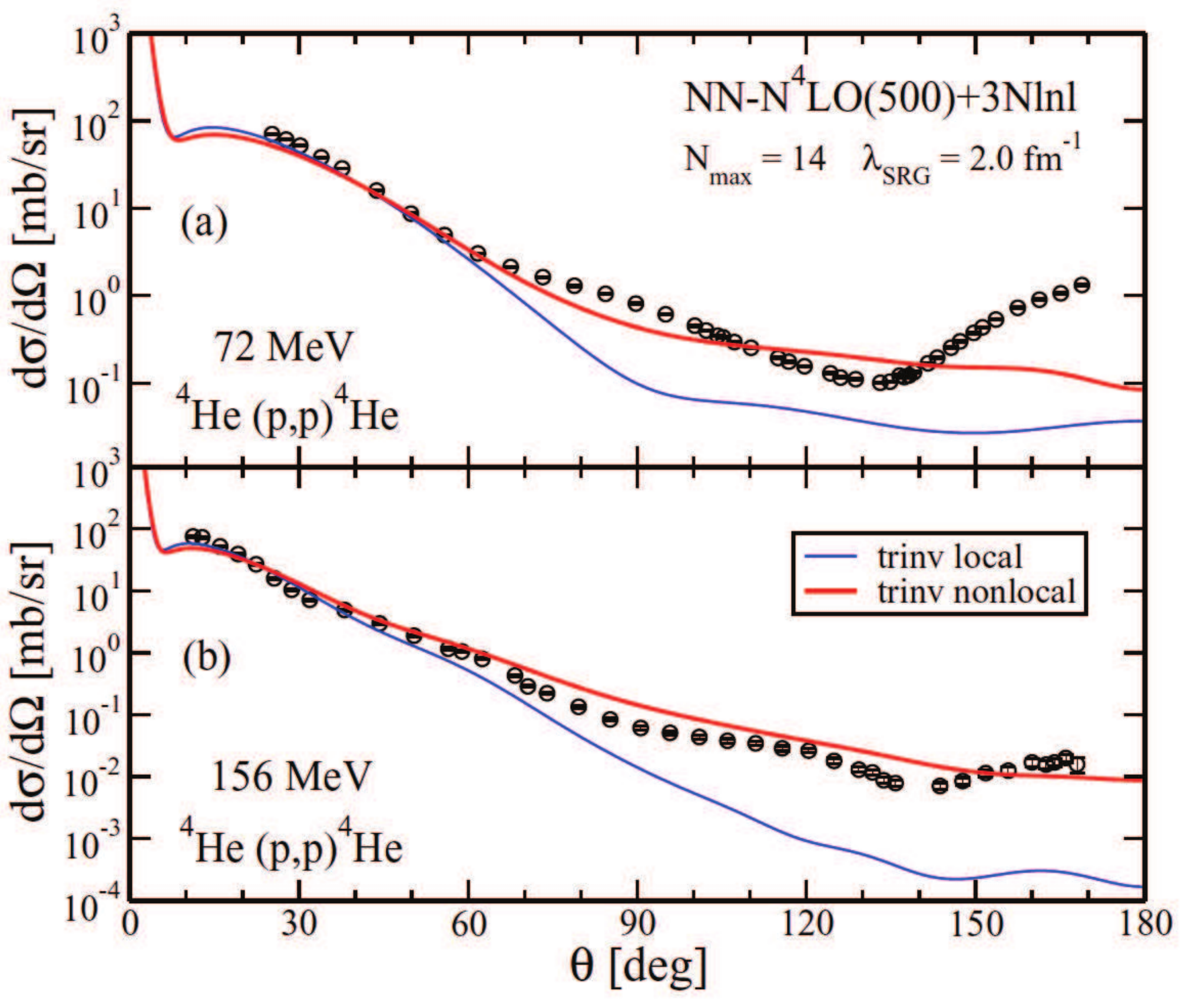}
      }
   \end{minipage}
   \caption{
   Differential cross sections in the $p+A$ center-of-mass
frame computed from local and nonlocal densities for the ${}^4$He$(p,p)^4$He reaction at the incident proton energy in the laboratory
frame of (a) 72 MeV and (b) 156 MeV, respectively. 
Reprinted figure with permission from Ref.~\cite{Gennari:18} \textcopyright2018 by the American Physical Society. }
   \label{fig:TRIUMF-1}
\end{figure} 
Differential cross sections in the $pA$ center-of-mass
frame are shown in Fig.~\ref{fig:TRIUMF-1} based on local and nonlocal (which yield better agreement) densities for the ${}^4$He$(p,p)^4$He reaction at the incident proton energy in the laboratory
frame of (a) 72 MeV and (b) 156 MeV, respectively.
The number of harmonic-oscillator shells corresponds to $N_{max}$ = 14 with an value of $\hbar \omega$  = 20 MeV and SRG cut-off of $\lambda_{SRG}$ = 2.0 fm$^{-1}$. 
The free NN $\mathcal{T}$-matrix was computed with the bare NN-N4LO(500) interaction.
From Fig.~\ref{fig:TRIUMF-1} one may conclude that good results are obtained, especially for nonlocal densities, which leads to a good description of the differential cross section except for the very backward angles at 72 MeV.
Analyzing powers at 72 MeV and 200 MeV are shown in Fig.~\ref{fig:TRIUMF-2} suggesting that it is much more difficult to obtain accurate results for polarization data. 
At 200 MeV, the local-density calculation even gives better results than the more appropriate one containing the nonlocal density. 
\begin{figure}[tb]
   \begin{minipage}{\columnwidth}
      \makebox[\columnwidth]{
         \includegraphics[scale=0.45]{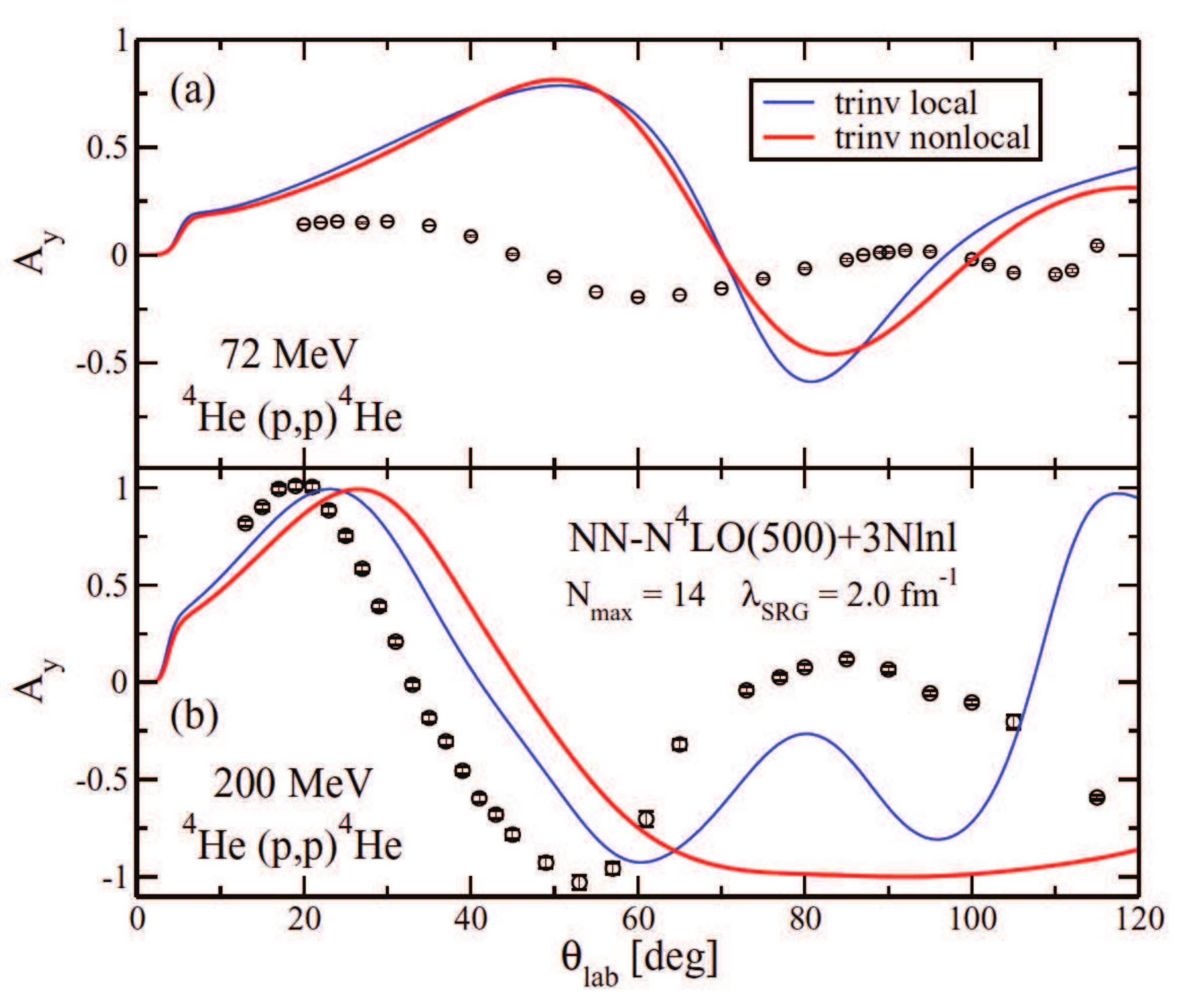}
      }
   \end{minipage}
   \caption{The same as Fig.~\ref{fig:TRIUMF-1} but for the analyzing power for (a) 72 MeV and (b) 200 MeV.
 Reprinted figure with permission from Ref.~\cite{Gennari:18} \textcopyright2018 by the American Physical Society.}
   \label{fig:TRIUMF-2}
\end{figure} 

The Pavia group used different chiral interactions in Ref.~\cite{PhysRevC.93.034619} to construct the optical potential in the multiple-scattering approach along the lines of Ref.~\cite{KERMAN1959551}.
The authors employed interactions from Refs.~\cite{PhysRevC.68.041001} and \cite{EPELBAUM2005362}.
In this work, the densities were obtained from relativistic mean-field calculations based on the work of Ref.~\cite{NIKSIC20141808}.
The authors note that their analysis of proton elastic scattering on ${}^{16}$O yielded the best results for the interaction with the largest cut-off.
As noted above and also in this work, there are difficulties in generating accurate results for polarization data.

All multiple-scattering approaches involving either $\mathcal{T}$- matrices or medium-modified $\mathcal{G}$-matrices suffer from the difficulty of an easy systematic way to improve the calculations.
In the case of $\mathcal{T}\rho$-calculations, one always faces the difficulty that the $\mathcal{T}$-matrix contains the deuteron pole at negative energy which has no relevance for the scattering from a  finite nuclei in which the two-nucleon propagation takes place on top of a correlated finite nucleus.
In a Green's-function formulation this propagation may contain pole structure associated with the $(A+2)$-system.
The corresponding effective interaction should then be folded with a hole propagator to generate the optical potential (see Sec.~\ref{sec:dussanCD}).
In principle, such a folding should sample the hole spectral function instead of the nonlocal density matrix.
In the case of folding calculations employing the $\mathcal{G}$-matrix, there is the inevitable local density approximation.
In addition, nuclear matter $\mathcal{G}$-matrices contain poles associated with Cooper bound states~\cite{VONDERFECHT19911} 
at lower densities that may have very little to do with finite nuclei but influence the results in an uncontrollable way.
Additional need for empirical correction factors are also a drawback that make the predictive power of this approach less impressive.

\subsection{Green's-function based methods}
\label{sec:GF}
In this section several calculations of nucleon self-energies are discussed that are based on the Green's-function method.
These approaches emphasize the influence of short-range and long-range correlations separately but have the advantage that they are implemented for the finite system in question.
While these applications have definite shortcomings, they do provide insights into the functional form of the optical potential and leave no doubt about the relevance of nonlocal contributions.
This particular insight is relevant for the application of optical potentials in the description of other reactions where distorted waves are required.

\subsubsection{Green's-function method for finite nuclei including short-range correlations}
\label{sec:dussanCD}
Recent work on a direct calculation of the nucleon self-energy for ${}^{40}$Ca with an emphasis on SRC was reported in Ref.~\cite{PhysRevC.84.044319}.
Such a microscopic calculation of the nucleon self-energy proceeds in two steps, as employed in Refs.~\cite{PhysRevC.49.R17,PhysRevC.51.3040,POLLS1995117}.
A diagrammatic treatment of SRC always involves the summation of ladder diagrams.
When only particle-particle (pp) intermediate states are included, the resulting effective interaction is the so-called $\mathcal{G}$-matrix as discussed in Sec.~\ref{sec:JLM}.
The corresponding calculation for a finite nucleus (FN) can be represented in operator form by
\begin{equation}
\mathcal{G}_{FN}(E)= V + V G^{pp}_{FN}(E) \mathcal{G}_{FN}(E) ,
\label{eq:gmatfn}
\end{equation}
where the noninteracting propagator $G^{pp}_{FN}(E)$ represents two particles above the Fermi sea of the finite nucleus taking into account the Pauli principle.
The simplest implementation of $G^{pp}_{FN}$ involves plane-wave intermediate states (possibly orthogonalized to the bound states).
Even such a simple assumption leads to a prohibitive calculation to solve Eq.~(\ref{eq:gmatfn}) and subsequently generate the relevant real and imaginary parts of the self-energy over a wide range of energies above and below the Fermi energy.
We are not aware of any attempt at such a direct solution at this time, except for the use of the $\mathcal{G}$-matrix as an effective interaction at negative energy.
Instead, it is possible to employ a strategy developed in Refs.~\cite{BONATSOS198923,BORROMEO1992189} that first calculates a $\mathcal{G}$-matrix in nuclear matter at a fixed density and fixed energy according to
\begin{equation}
\mathcal{G}_{NM}(E_{NM})= V + V G^{pp}_{NM}(E_{NM}) \mathcal{G}_{NM}(E_{NM}) .
\label{eq:gmatnm}
\end{equation}
The energy $E_{NM}$ is chosen below twice the Fermi energy of nuclear matter for a kinetic-energy sp spectrum and the resulting $\mathcal{G}_{NM}$ is therefore real.
Formally solving Eq.~(\ref{eq:gmatfn}) in terms of $\mathcal{G}_{NM}$ can be accomplished by 
\begin{equation}
\mathcal{G}_{FN}(E) = \mathcal{G}_{NM} 
\label{eq:gmatfnp} 
+   \mathcal{G}_{NM} \left\{G^{pp}_{FN}(E) - G^{pp}_{NM} \right\} \mathcal{G}_{FN}(E) ,
\end{equation}
where the explicit reference to $E_{NM}$ is dropped.
The main assumption to make the self-energy calculation manageable is to drop all terms higher than second order in $\mathcal{G}_{NM}$, leading to
\begin{equation}
\mathcal{G}_{FN}(E) = \mathcal{G}_{NM} - \mathcal{G}_{NM}  G^{pp}_{NM}  \mathcal{G}_{NM}
+   \mathcal{G}_{NM} G^{pp}_{FN}(E) \mathcal{G}_{NM} ,
\label{eq:gmatfnq}
\end{equation}
where the first two terms are energy-independent.
Since a nuclear-matter calculation already incorporates all the important effects associated with SRC, it is reasonable to assume that the lowest-order iteration of the difference propagator in Eq.~(\ref{eq:gmatfnq}) represents an accurate approximation to the full result.

The self-energy contribution of the lowest-order term $\mathcal{G}_{NM}$ in Eq.~(\ref{eq:gmatfnq}) is similar to a Brueckner-Hartree-Fock (BHF) self-energy. 
While strictly speaking the genuine 
BHF approach involves self-consistent sp wave functions, as in the HF 
approximation, the
main features associated with using the $\mathcal{G}_{NM}$-matrix 
are approximately the same when employing a summation over the occupied harmonic-oscillator states of ${}^{40}$Ca. 
The correction term involving the second-order in $\mathcal{G}_{NM}$ calculated in nuclear matter is also static and can be obtained from the second term in Eq.~(\ref{eq:16.33}) by replacing the bare interaction by $\mathcal{G}_{NM}$. The corresponding self-energy is also real and generated by summing over the occupied oscillator states in the same way as for the BHF term.

The second-order term containing the correct energy-dependence for $\mathcal{G}_{FN}$ in Eq.~(\ref{eq:gmatfnq}) can now be used to construct the self-energy contribution, representing the coupling to two-particle--one-hole (2p1h) states.
In the calculation, harmonic-oscillator states for the occupied (hole) 
states and plane waves for the intermediate particle-unbound 
states are assumed incorporating the correct energy and density dependence 
characteristic of a finite nucleus $\mathcal{G}_{FN}$-matrix.
In a similar way, one can construct the second-order self-energy contribution which has an imaginary part below the Fermi energy and includes the coupling to one-particle--two-hole (1p2h) states.

Calculations of this kind require several basis transformations, including the one from relative and center-of-mass momenta with corresponding orbital angular momenta to two-particle states with individual momenta and orbital angular momentum. 
Complete details can be found in Refs.~\cite{BONATSOS198923,BORROMEO1992189}.
In practice, the imaginary parts are employed to obtain the corresponding real parts by employing the appropriate dispersion relation.
The resulting (irreducible) self-energy then reads
\begin{eqnarray}
\Sigma^* & = & \Sigma^*_{BHF} + \Delta\Sigma^* \label{eq:defsel}  \\
& = & \Sigma^*_{BHF} + 
\left( \textrm{Re}~\Sigma^*_{2p1h} - \Sigma^*_{c} + \textrm{Re}~\Sigma^*_{1p2h} 
\right) 
+ i \left( \textrm{Im}~\Sigma^*_{2p1h} + \textrm{Im}~\Sigma^*_{1p2h} \right) \; 
 \nonumber
\end{eqnarray}
in obvious notation.

\begin{figure}[bt]
 \begin{minipage}{\columnwidth}
      \makebox[\columnwidth]{
       \includegraphics[totalheight=4.3in,clip=true]{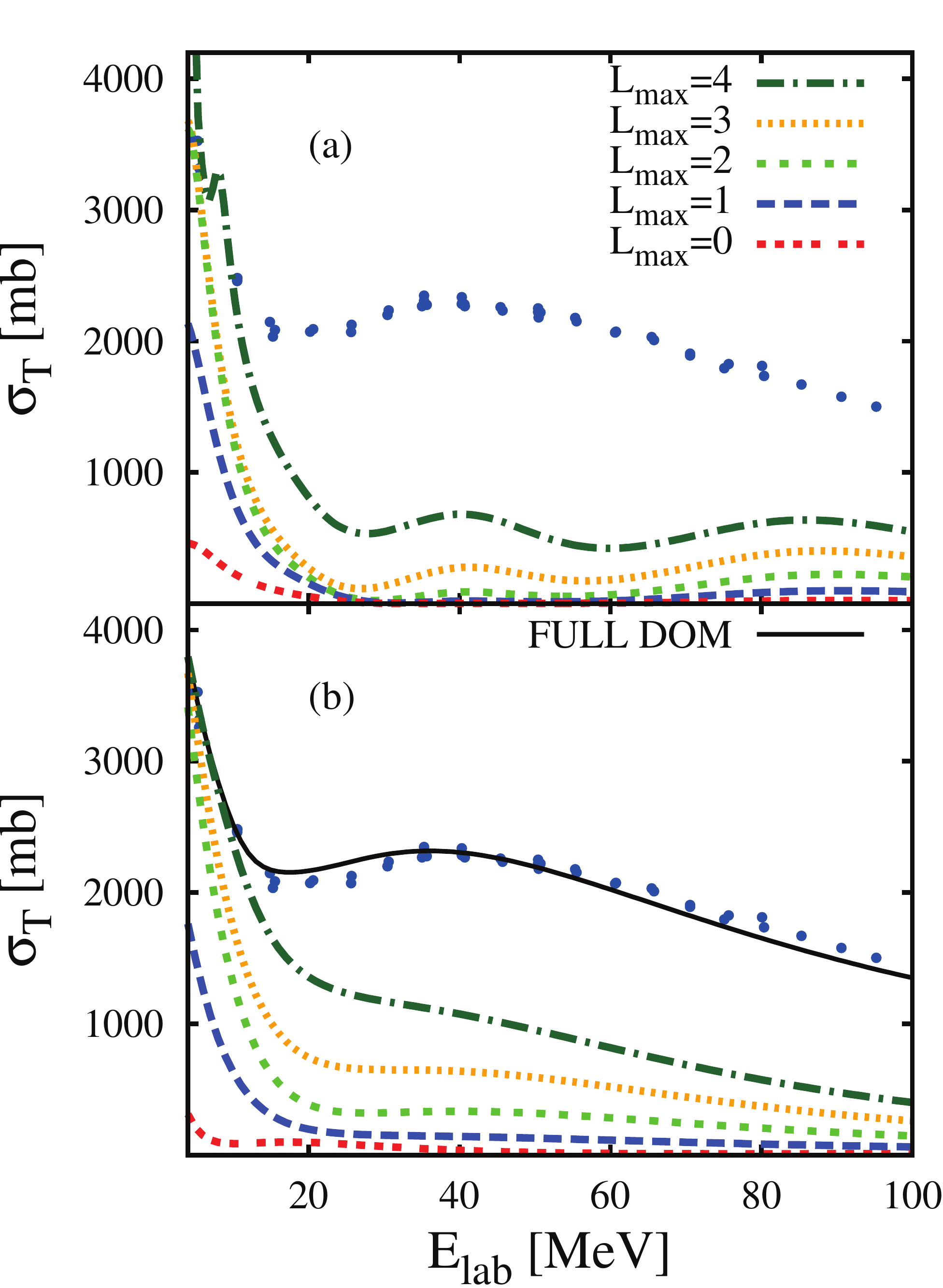}
      }
   \end{minipage}
\caption{Effect of adding subsequent partial waves for the total neutron cross section for ${}^{40}$Ca from the CDBonn self-energy (upper panel) and corresponding DOM result (lower panel). The total DOM cross section with a converged contribution of the partial-wave decomposition is also shown as the solid line in the lower panel.
The experimental data are represented by dots. Reprinted figure with permission from Ref.~\cite{PhysRevC.84.044319} \textcopyright2011 by the American Physical Society.
}
\label{sigmaTot2}
\end{figure}
The work of Ref.~\cite{PhysRevC.84.044319} employed the CDBonn interaction~\cite{PhysRevC.53.R1483,PhysRevC.63.024001}.
The total cross section obtained from the corresponding CDBonn self-energy is comparable to the experimental data for energies between 5 to 10 MeV, while for larger energies it does not 
reproduce the observations. 
Currently such a CDBonn self-energy only includes contributions up to $\ell_{max}=4$ which is a reasonable limit for calculating properties below the Fermi energy.  
At higher positive energies, this represents a serious drawback because a large number of partial waves is then required to generate converged results for differential cross sections.
A reasonable comparison with results from the DOM potential of Ref.~\cite{Mueller:2011} is still possible however. 
The most useful procedure is therefore to compare the DOM and CDBonn calculations with the inclusion of the same number of partial waves.
It is then gratifying to observe that the microscopic calculation generates similar total cross sections as the DOM for energies where surface contributions are expected to be less relevant, \textit{i.e.} above 70 MeV.   
Although calculations for more partial waves are in principle possible, the amount of angular momentum recoupling corresponding to the appropriate basis transformations becomes increasingly cumbersome.
Nevertheless, it is clear that the current limit of $\ell_{max}=4$ is insufficient to describe the total cross section already at relatively low energy irrespective of whether surface effects (LRC) are properly included.

To visualize in more detail the specific contribution from each partial wave,
we display  in Fig.~\ref{sigmaTot2}   cross sections calculated up to a specific $\ell_{max}$ for CDBonn (upper panel) and DOM (lower panel). Despite the limitation of not having CDBonn values for $\ell>4$, both potentials provide a comparable cross sections up to this angular-momentum cut-off in the explored energy range.
In the lower panel, we also include the converged cross section of the DOM potential that provides a good description of the data.

\begin{figure}[bt]
 \begin{minipage}{\columnwidth}
      \makebox[\columnwidth]{
      \includegraphics[width=3.2in]{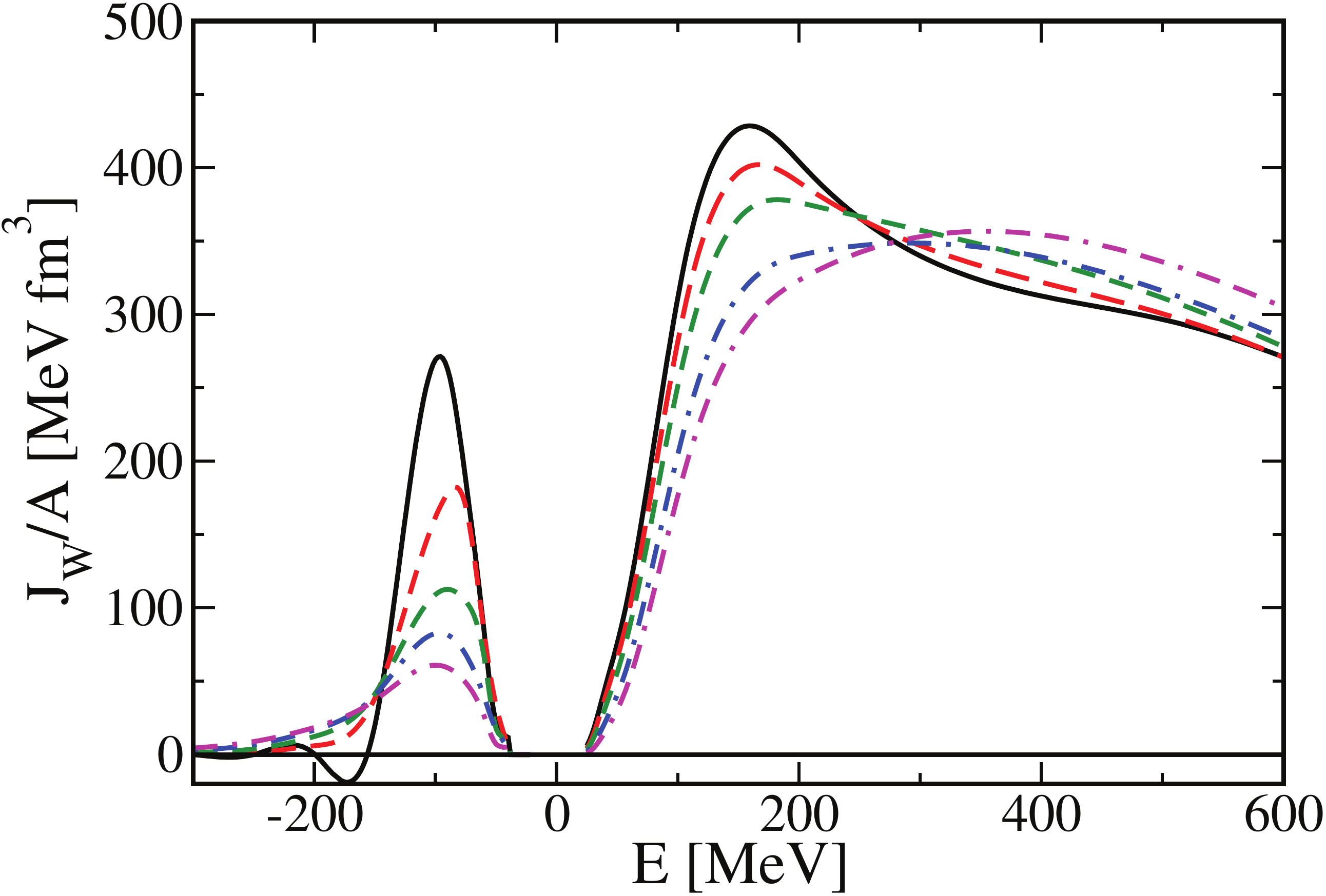}
      }
   \end{minipage}
\caption{Imaginary volume integrals for the CDBonn self-energy as a function of energy for different $\ell$-values: $\ell=0$ (solid), $\ell=1$ (dashed), $\ell=2$ (short-dashed), $\ell=3$ (dash-dot), and $\ell=4$ (dash-dash-dot).\label{fig:Jw}
Reprinted figure with permission from Ref.~\cite{PhysRevC.84.044319} \textcopyright2011 by the American Physical Society.
}
\end{figure}

An important issue concerning the description of the optical potential is the amount of nonlocality that is needed to represent \textit{ab initio} potentials. 
The work of Ref.~\cite{PhysRevC.84.044319} explored  how far the standard Gaussian form of nonlocality~\cite{PEREY1962353} can represent such microscopic potentials.
A convenient tool to identify the quality of such a representation is provided by volume integrals.
Extending the definition of Eq.~(\ref{eq:volumeInteg}) for a given angular momentum $\ell$ and non-local potential according to \cite{PhysRevC.84.034616}
\begin{equation}
J_W^\ell(E) = 4\pi\int{dr\ r^2\int{dr'  r'^{2}\ \textrm{Im } \Sigma_{\ell}^*(r, r' ; E)}} .
\label{eq:intgs_W} 
\end{equation}
Here and in the following we deviate from the convention of Eq.~(\ref{eq:volumeInteg}) by not dividing by $A$ although for plotting purposes these volume integrals are divided by $A$ in the next two figures.
This definition averages over spin-orbit partners for a given value of $\ell$.
For a local potential, it reduces to the standard definition of the volume integral [Eq.~(\ref{eq:volumeInteg})]. 
By fitting the parameters of the Gaussian to $\ell = 0$, values of the integral in Eq.~(\ref{eq:intgs_W}) can be reproduced with reasonable accuracy.
Typical  nonlocality parameters characterizing the width of the Gaussian are close to 1.5 fm with a tendency to decrease with increasing energy suggesting that at higher energy, absorption dictated by SRC becomes more local.
For a local potential there is no $\ell$-dependence of the volume integral, so the behavior of $J_W^\ell$ for different $\ell$-values in a wide-energy domain clarifies these tendencies.
The results of this analysis are shown in Fig.~\ref{fig:Jw}.
The degree of non-locality appears to be largest below the Fermi energy with a substantial separation between the different $\ell$-values. 
The result for $\ell=0$ also demonstrates that it is possible to have the ``wrong'' sign for the volume integral.
This can happen because the microscopic self-energy develops negative lobes off the diagonal and as a result a positive volume integral cannot be guaranteed, as must be the case for a local potential.
Although the imaginary part above the Fermi energy is negative, it is conventional to plot the imaginary volume integral as a positive function of energy~\cite{Mueller:2011}.
At positive energy, the volume integrals for different $\ell$-values at first exhibit a spread although not as large as below the Fermi energy.
However above 300 MeV, the curves apparently become similar suggesting a trend to a more local self-energy.

\subsubsection{Green's-function method for finite nuclei including long-range correlations}
\label{sec:LRC}
Low-lying collective states have considerable impact on the properties of a nucleon in the nucleus.
When sp degrees of freedom are coupled to such excitations, it leads to an important fragmentation of the sp strength distributions~\cite{MAHAUX19851}.
The microscopic calculation of this particle-vibration coupling is captured by the so-called Faddeev random-phase approximation (FRPA).
This is accomplished by using the random-phase approximation (RPA) to calculate phonons of particle-particle (hole-hole) and particle-hole type. 
These are then included to all orders in a Faddeev summation for both two-particle--one-hole (2p1h) and two-hole--one-particle (2h1p) propagation. This approach is referred to as Faddeev random-phase approximation (FRPA)~\cite{PhysRevC.63.034313,PhysRevA.76.052503}.
This method is size extensive and can be successfully applied to finite electron systems, giving results of comparable accuracy to coupled-cluster theory.

The FRPA was originally developed to describe the self-energy of the double-closed-shell nucleus ${}^{16}$O~\cite{PhysRevC.63.034313}.
The method has later been applied to atoms and molecules~\cite{PhysRevA.76.052503,PhysRevA.83.042517} and other nuclei (see \textit{e.g.} Ref.~\cite{PhysRevLett.103.202502}).
An important observation is that the configuration space needed for the incorporation of long-range (surface) correlations, including the coupling to giant resonances, is much larger than the space that can be utilized in large-scale shell-model diagonalizations.
FRPA calculations for ${}^{40}$Ca and ${}^{48}$Ca and ${}^{60}$Ca were performed to shed light on the \textit{ab initio} self-energy properties in medium-mass nuclei including the nucleon asymmetry 
dependence~\cite{PhysRevC.84.034616}.
The main goal of this work was to clarify whether substantial nonlocal contributions should be expected when optical potentials for elastic scattering are considered.
 In particular, one may expect to extract useful information regarding the functional form of the DOM from a study of the self-energy for a sequence of calcium isotopes. 
 The resulting analysis was intended to provide a microscopic underpinning of the qualitative features of empirical optical potentials.
Additional information concerning the degree and form of the non-locality of both the real and imaginary parts of the self-energy were also addressed.

The analysis proceeded from 
the introduction of  some of the basic properties of the self-energy in a finite system.
For a $J = 0$ nucleus, all partial waves $(\ell , j, \tau )$ are decoupled, where $\ell$,$j$ label the orbital and total angular momentum and $\tau$ represents its isospin projection. The irreducible self-energy in coordinate space (for either a proton or a neutron) can be written in terms of the harmonic-oscillator basis used in the FRPA calculation, as follows:
\begin{equation}
\Sigma^\star( \bm{x}, \bm{x^\prime}; E ) = \sum_{\ell j m_j \tau} {\cal I }_{\ell j m_j}( \Omega, \sigma ) 
\left[ \sum_{n_a, n_b} R_{n_a \ell}(r) \Sigma^\star_{ab}(E)R_{n_b \ell}( r^\prime )\right] ( {\cal I }_{\ell j m_j}( \Omega^\prime, \sigma^\prime ) )^* ,
\label{eq:selfr}
\end{equation}
where $\bm{x} \equiv \bm{r}, \sigma, \tau$.  
The spin variable is represented by $\sigma$, $n$ is the principal quantum number of the  harmonic oscillator, and $a\equiv(n_a, \ell , j, \tau)$ (note that for a $J = 0$ nucleus, the self-energy is independent of $m_j$).
The standard radial harmonic-oscillator function is denoted by $R_{n \ell}(r)$, while ${\cal I }_{\ell j m_j}( \Omega, \sigma )$ represents the $j$-coupled angular-spin function. 

The harmonic-oscillator projection of the self-energy can be written as
\begin{eqnarray}
\Sigma^\star_{ab}( E ) &=& \Sigma^{\infty}_{ab}(E) + \tilde{\Sigma}_{ab}(E) \nonumber  \\
&=& \Sigma^{\infty}_{ab}(E) + \sum_{r}{\frac{m_{a}^r( m_{b}^r )^*}{E -\varepsilon_r \pm i\eta}} .
\label{eq:selfHO}
\end{eqnarray}
The term with the tilde is the dynamic part of the self-energy due to long-range correlations calculated in the FRPA, and $\Sigma^{\infty}_{ab}(E)$ is the correlated Hartree-Fock term which acquires an energy dependence through the energy dependence of the $G$-matrix effective interaction (see below). $\Sigma^{\infty}_{ab}(E)$ is the sum of the strict correlated Hartree-Fock diagram (which is energy independent) and the dynamical contributions due to short-range interactions outside the chosen model space.
The self-energy can be further decomposed in a central ($0$) and a spin-orbit ($\ell s$) part according to
\begin{subequations}
\label{eq:ls}
\begin{eqnarray}
\Sigma^{\ell j_>} &=& \Sigma^{\ell}_{0}+\frac{\ell}{2}\Sigma^{\ell}_{\ell s} \label{eq:ls1} \, ,\\
\Sigma^{\ell j_<} &=& \Sigma^{\ell}_{0} - \frac{ \ell + 1}{2}\Sigma^{\ell}_{\ell s}  \, ,
\label{eq:ls2}
\end{eqnarray}
\end{subequations}
with $j_{>,<}\equiv \ell \pm \frac{1}{2}$.
The corresponding static terms are denoted by $\Sigma^{\infty, \ell}_0$ and $\Sigma^{\infty, \ell}_{\ell s}$, and the corresponding dynamic terms are denoted by $\tilde{\Sigma}^{\ell}_0$ and $\tilde{\Sigma}^{\ell}_{\ell s}$.

The FRPA calculation employs a discrete sp basis in a large model space which results in a substantial number of poles in the self-energy~(\ref{eq:selfHO}).
Since the goal is to compare with optical potentials at positive energy, it is appropriate to smooth out these contributions by employing a finite width for these poles.
We note that the optical potential was always intended to represent an average smooth behavior of the nucleon self-energy~\cite{Mahaux:1991}.
In addition, it makes physical sense to at least partly represent the escape width of the continuum states by this procedure.
Finally, further spreading of the intermediate states to more complicated states can also be approximately accounted for by this procedure.
Thus, before comparing to the DOM potentials, the dynamic part of the microscopic self-energy was smoothed out using a finite, energy-dependent width for the poles
\begin{equation}
\tilde{\Sigma}_{n_a, n_b}^{\ell j}(E) = \sum_{r}\frac{m_{n_a}^{r} m_{n_b}^{r}}{ E - \varepsilon_r \pm i\eta } \longrightarrow \sum_{r}\frac{m_{n_a}^{r} m_{n_b}^{r}}{ E - \varepsilon_r \pm i\Gamma(\varepsilon_r) } \, .
\label{eq:width}
\end{equation}
Solving for the real and imaginary parts one obtains
\begin{eqnarray}
\lefteqn{
\tilde{\Sigma}_{n_a, n_b}^{\ell j}(E) = \sum_{r}\frac{(E - \varepsilon_r )}{ ( E - \varepsilon_r )^2 + [\Gamma(\varepsilon_r)]^2 }m_{n_a}^{r} m_{n_b}^{r} 
 } & &	
 \label{eq:smooth} \\
&\qquad +& i\left[\theta( \varepsilon_F - E )\sum_{h}\frac{\Gamma(\varepsilon_h)}{( E - \varepsilon_h )^2 + \Gamma(\varepsilon_h)^2}m_{n_a}^{h}m_{n_b}^h  \right. \nonumber \\
&\qquad -& \left.  \theta( E - \varepsilon_F )\sum_{p}\frac{\Gamma(\varepsilon_p)}{( E - \varepsilon_p )^2 + [\Gamma(\varepsilon_p)]^2}m_{n_a}^{p}m_{n_b}^p\right] ,
\nonumber 
\end{eqnarray}
where $r$ implies a sum over both particle and hole states, $h$ denotes a sum over the hole states only, and $p$ a sum over the particle states only. 
For the width, the following form was used~\cite{Brown:81}:
\begin{equation}
\Gamma( E ) = \frac{1}{\pi}\frac{ a \, (E- \varepsilon_F)^2}{(E - \varepsilon_F)^2 - b^2}
\end{equation}
with $a$=12~MeV and $b$=22.36~MeV.
This generates a narrow width near $\varepsilon_F$ that increases as the energy moves away from the Fermi surface, in accordance with observations. 

In the DOM representation of the optical potential, the self-energy is recast in the form of a  subtracted dispersion relation
\begin{equation}
\Sigma^\star_{ab}( E ) = \Sigma^{\infty}_{ab, \,S} + \tilde{\Sigma}_{ab}(E)_S, 
\label{eq:suba}
\end{equation}
where
\footnote{It is the (real) $\Sigma^\infty_{ab, \, S}$ and the imaginary part of $\tilde{\Sigma}_{ab}(E)_S$ that are 
parametrized in the DOM potential. \hbox{$Re \; \tilde{\Sigma}_{ab}(E)_S$} is then fixed by the subtracted dispersion relation.}
\begin{eqnarray}
       \Sigma^{\infty}_{ab \, S} &= &\Sigma^\star_{ab}(\varepsilon_F) \, ,
\label{eq:SigSubHF} \\
\tilde{\Sigma}_{ab}(E)_S    &= &\Sigma^\star_{ab}(E) - \Sigma^\star_{ab}(\varepsilon_F) \, .
\label{eq:SigSubDyn}
\end{eqnarray}		     
For the imaginary potential, this is the same as the above-defined self-energies (\ref{eq:selfHO}) and it can therefore be directly compared to the DOM potential. For the real parts, we will employ either the normal or the subtracted form in the following as appropriate.

In fitting optical potentials, it is usually found that volume integrals are better constrained by the experimental data (Sec.~\ref{sec:empirical}). For this reason, they have been considered as a reliable measure of the total strength of a potential. For a non-local and $\ell$-dependent potential of the form~(\ref{eq:selfr}), it is convenient to consider separate integrals for each angular-momentum component, $\Sigma^{\ell}_0(r, r^\prime)$ and $\Sigma^{\ell}_{\ell s}(r, r^\prime)$, which correspond to the square brackets in Eq.~(\ref{eq:selfr}) and decomposed according to~(\ref{eq:ls}).
Labeling the central real part of the optical potential with $V$, and the central imaginary part by $W$,  we calculate:
\begin{subequations}
\label{eq:intgs}
\begin{eqnarray}
J_W^\ell(E) = 4\pi\int{drr^2\int{dr^\prime r^{\prime 2} \text{Im } \Sigma^{\ell}_0(r, r^\prime ; E)}}
\label{eq:intgs_Wp} \\
J_V^\ell(E) = 4\pi\int{drr^2\int{dr^\prime r^{\prime 2} \text{Re } \Sigma^{\ell}_0(r, r^\prime ; E)}}
\label{eq:intgs_V}  .
\end{eqnarray}
\end{subequations}

The correspondence between the above definitions and the volume integrals used for local potentials can be obtained by casting a spherical local potential $U(r)$ into a non-local form $U(\bm r, \bm r^\prime ) = U(r) \delta(\bm r - \bm r^\prime)$. Expanding this in spherical harmonics gives
\begin{equation}
 U(\bm r, \bm r^\prime ) = \sum_{\ell m} U^\ell(r, r^\prime)Y^*_{\ell m}(\Omega^\prime)Y_{\ell m}(\Omega) \, ,
\end{equation}
with the $\ell$-projection
\begin{equation}
 U^\ell(r, r^\prime ) = \frac {U(r)}{r^2} \delta(r - r^\prime) \, ,
\end{equation}
which is actually angular-momentum independent.
The definition~(\ref{eq:intgs}) for the volume integrals lead to
\begin{eqnarray}
 J^\ell_U &=& 4\pi  \int{ dr\ r^2\int{dr^\prime r^{\prime 2} U^{\ell}(r, r^\prime)}} 
\\
      &=& 4\pi\int{ U(r) r^2dr} = \int{ U(r)\ d\bm{r}}  \, \hbox{, \hspace{.2in} for any $\ell$}
\nonumber
\end{eqnarray}
and reduces to the usual definition of volume integral for local potentials [Eq.~(\ref{eq:volumeInteg})].
Thus, Eq.~(\ref{eq:intgs}) can be directly compared to the corresponding integrals when nonlocal implementations are employed in the DOM.

\begin{figure}[bt]
\begin{center}
\includegraphics[width=2.5in]{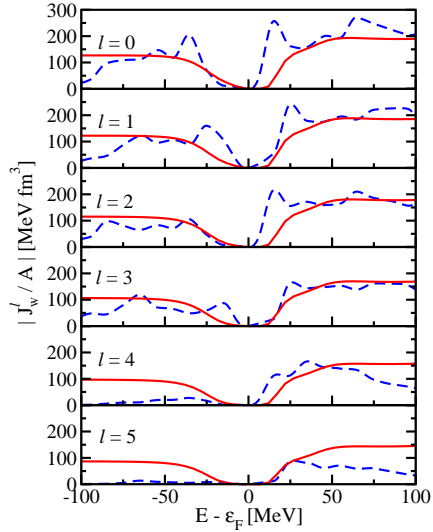}
\caption{ Imaginary volume integrals $J^{\ell}_W$ from Eq.~(\ref{eq:intgs_Wp}) of the ${}^{40}$Ca self-energy for neutrons. The dashed curves represent the FRPA results. The results of the non-local DOM fit of Ref.~\cite{PhysRevLett.112.162503} are shown by the solid lines. Figure adapted from Fig.~11 of Ref.~\cite{Dickhoff:17}.} 
\label{fig:Jw_review}
\end{center}
\end{figure}

Early implementations of the DOM employed local potentials and an energy dependence of its imaginary part symmetrically centered around $\varepsilon_F$~\cite{Mahaux:1991}. Such features are not obtained in the FRPA
as illustrated in Fig.~\ref{fig:Jw_review} where for $\ell$-values up to 5, the volume integrals of the FRPA calculation are displayed by the dashed lines.
The domain of the DOM fit extends beyond those shown in Fig.~\ref{fig:Jw_review}, but is limited to those energies for which a meaningful comparison with FRPA results is possible.
We therefore also include the result of the non-local fit of Ref.~\cite{PhysRevLett.112.162503} shown by the solid lines in Fig.~\ref{fig:Jw_review} which confirms this assessment.
Higher $\ell$-values are less relevant below the Fermi energy and this is clearly illustrated by the FRPA results in the Fig.~\ref{fig:Jw_review}.
Since the absorption above the Fermi energy is strongly constrained by elastic-scattering data, it is encouraging that the $\ell$-dependent FRPA result is reasonably close to the DOM fit in the domain where the FRPA is expected to be relevant on account of the size of the chosen configuration space. 
Note that the calculated $J_W$ values decreases quickly at energies $E-\varepsilon_F >$100~MeV due to the truncation the model space. Instead,  it correctly  remains sizable even at higher energies in  the DOM.
Also at negative energies, the FRPA results do not adequately describe the admixture of high-momentum components that occur at large missing energies.

A further development of these \textit{ab initio} self-consistent Green's function methods is provided by the extension to single-closed shell nuclei with the so-called Gorkov Green's function approach employing anomalous propagators~\cite{Soma13}. 
This approach makes it possible to study sequences of isotopes and provides insights into features of elastic scattering that are unique to those systems.
Predictions of the location of the drip lines will also be of interest.
Some result derived from this method are illustrated in the next section.

\subsubsection{Example of the use optical potentials and structure calculations to extract and describe nuclear radii}
\label{sec:BarLap}
A useful perspective on the use of optical potentials is associated with the determination of matter radii for rare isotopes and in turn for the testing of the application of many-body methods and their use of NN interactions~\cite{PhysRevLett.117.052501}.
As discussed in Ref.~\cite{PhysRevLett.117.052501}, the ground-state energies of the sequence of oxygen isotopes can be well described using different many-body techniques utilizing \textit{e.g.} the Entem-Machleidt version of a realistic chiral NN interaction~\cite{PhysRevC.68.041001}.
A serious problem that remains is the difficulty to simultaneously describe nuclear radii~\cite{PhysRevC.92.014306}.
This was further confirmed by the study of heavier nuclei and the saturation properties of nuclear matter.
Thus motivation was provided to propose an alternate paradigm to study nuclei starting from an NN interaction that is adjusted to describe radii of selected heavier nuclei~\cite{PhysRevC.91.051301}.
A drawback of this approach, which has made this interaction somewhat controversial, is that it only describes NN scattering data up to 35 MeV.
Nevertheless, as long as this feature is understood, there is no doubt that this interaction can provide further insight into the description of nuclear sizes.
It should be noted that the problem with current chiral interactions may well be their softness as cut-offs of the contact interactions are quite a bit lower than those employed in phenomenological interactions like AV18~\cite{PhysRevC.51.38} or an accurate boson-exchange potential like CDBonn~\cite{PhysRevC.63.024001}.  
Employing such harder interactions, it is much easier to keep nucleons apart due to their stronger repulsive nature and, as a result, radii will be more reasonable.
The difficulty of course is that their use in many-body calculations is considerably more difficult, although for light nuclei, AV18 can be implemented on account of its local form with Monte Carlo methods~\cite{RevModPhys.87.1067}.
 
\begin{figure}[bt]
\begin{center}
\includegraphics[width=2.0in]{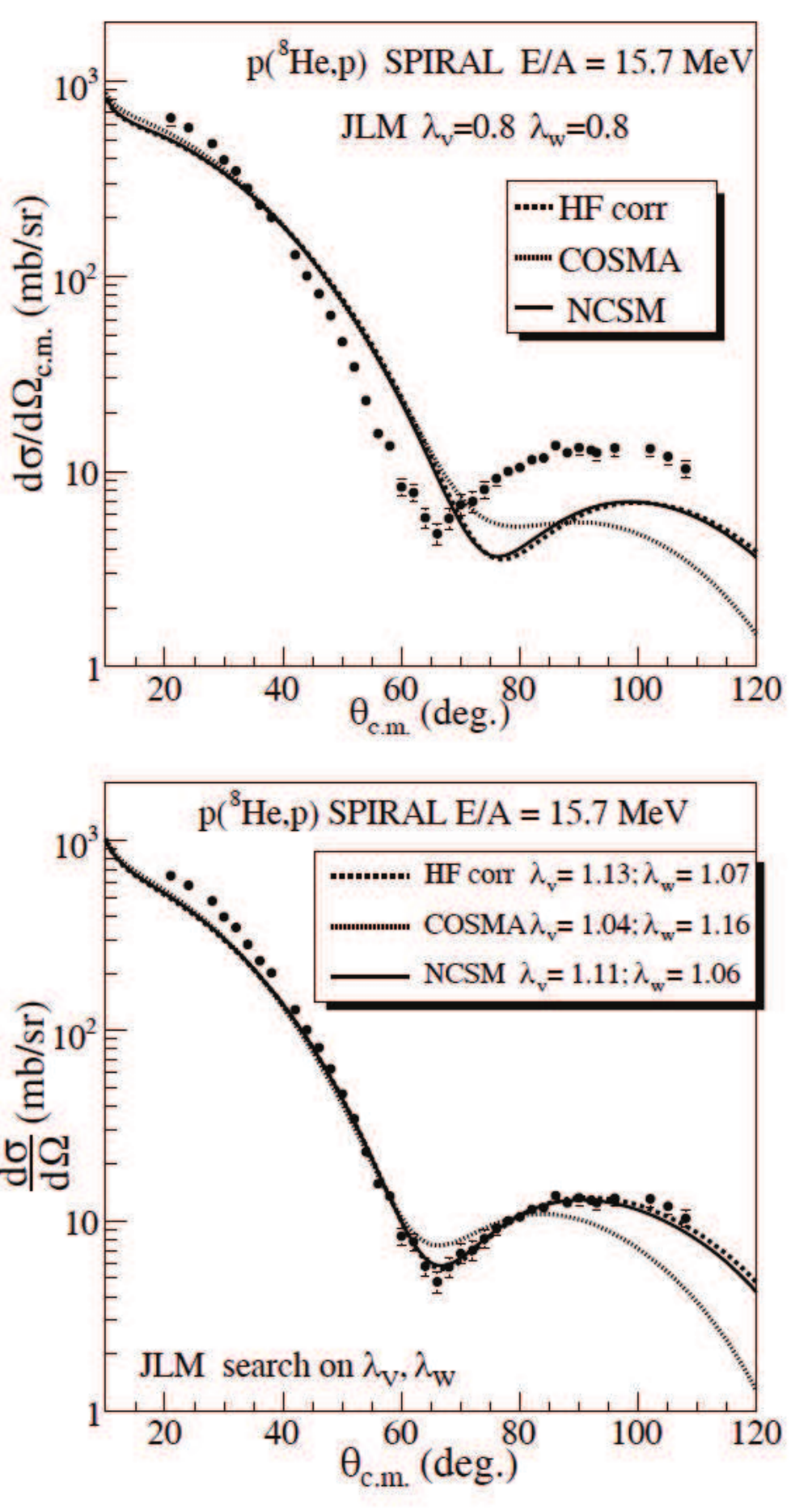}
\caption{Elastic scattering ${}^8$He($p$,$p$) at 15.7 MeV/$A$; the data from Ref.~\cite{SKAZA200582} are compared to the JLM calculations performed using three different densities, and the following normalizations
factors for the real and imaginary parts: (top) renormalization parameters
$\lambda_V$ = 0.8 (for weakly bound nuclei) and standard $\lambda_W$ = 0.8;
(bottom) calculations done with the potentials resulting from
a search on these parameters inorder to reproduce the data.
Reprinted figure with permission from Ref.~\cite{Lapoux2015} \textcopyright2015 by Springer.
} 
\label{fig:Lapoux15}
\end{center}
\end{figure}

Matter radii were determined in Ref.~\cite{PhysRevLett.117.052501} by using proton elastic-scattering data following the strategy of Ref.~\cite{Lapoux2015}.
In this strategy, the JLM potential (Sec.~\ref{sec:JLM}) and appropriately chosen densities are used to  calculate proton elastic scattering  as applied for exotic He isotopes. 
An example is shown in Fig.~\ref{fig:Lapoux15} for ${}^8$He.
While it is possible to generate reasonable agreement with the data thereby validating the radii implied by the densities identified in Ref.~\cite{Lapoux2015}, there are concerns about whether nuclear-matter calculations have relevance for such a light system.
In addition, it is unclear whether the agreement with the data obtained in this figure cannot be the result of cancelling errors in both the optical potential and the density.
Nevertheless, the procedure provides a way to access matter radii for exotic systems as the authors claim uncertainties of only 0.1 fm.
Future electron-ion colliders~\cite{SUDA20171} will provide accurate charge radii in addition to results from laser experiments~\cite{Garcia16}.
Result from parity-violating elastic electron scattering~\cite{PhysRevLett.108.112502} will provide unambiguous information about the weak charge, and therefore the neutron distribution of ${}^{208}$Pb and ${}^{48}$Ca. 

\begin{figure}[bt]
\begin{center}
\includegraphics[width=3.5in]{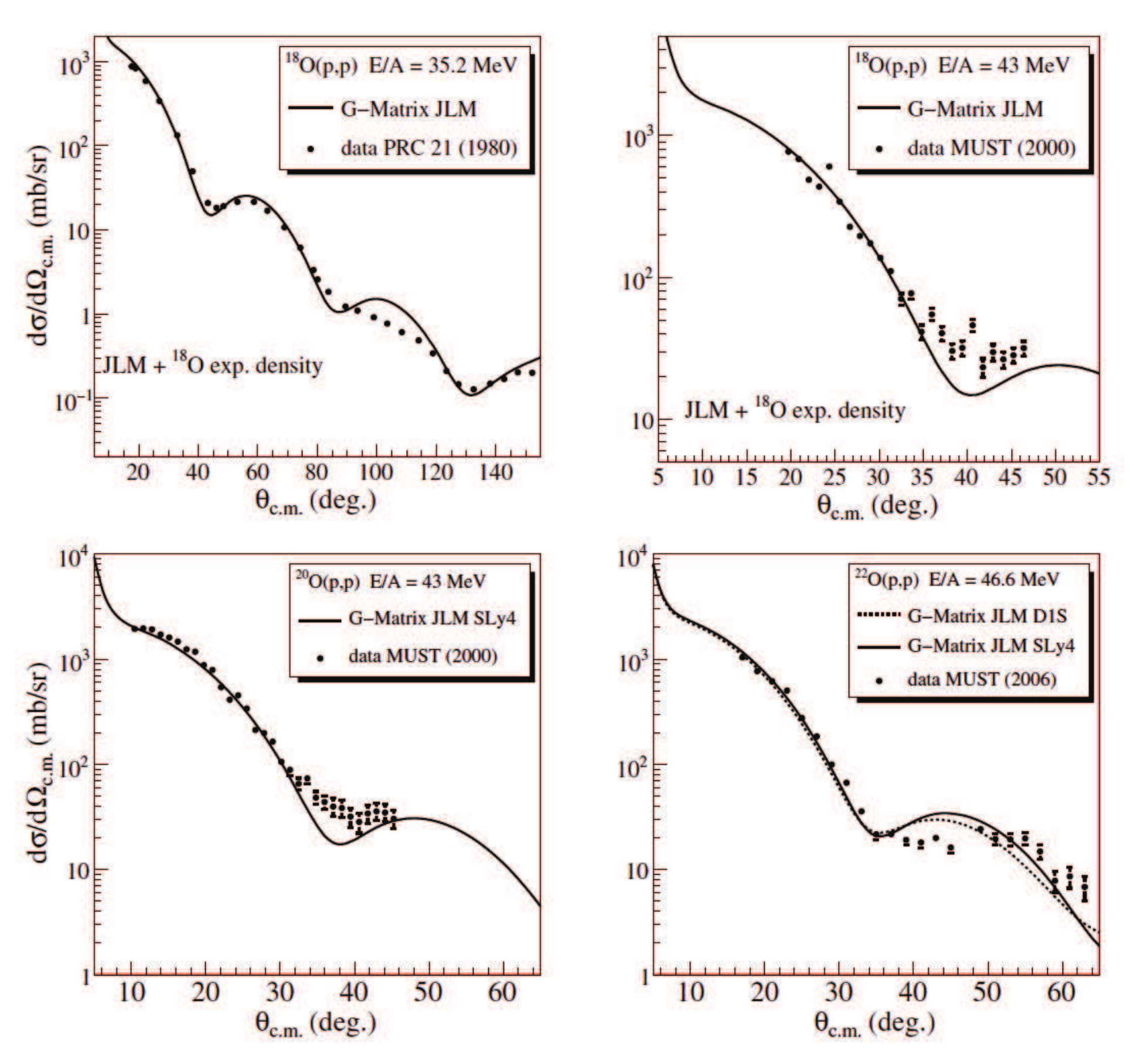}
\caption{Experimental elastic $(p, p)$ distributions compared
to calculations of Ref.~\cite{PhysRevLett.117.052501} including data for the rare isotopes from Refs.~\cite{KHAN200045,PhysRevLett.96.012501}.
Reprinted figure with permission from Ref.~\cite{PhysRevLett.117.052501} \textcopyright2016 by the American Physical Society.
} 
\label{fig:Lapoux17}
\end{center}
\end{figure}

\begin{figure}[bt]
\begin{center}
\includegraphics[width=2.5in]{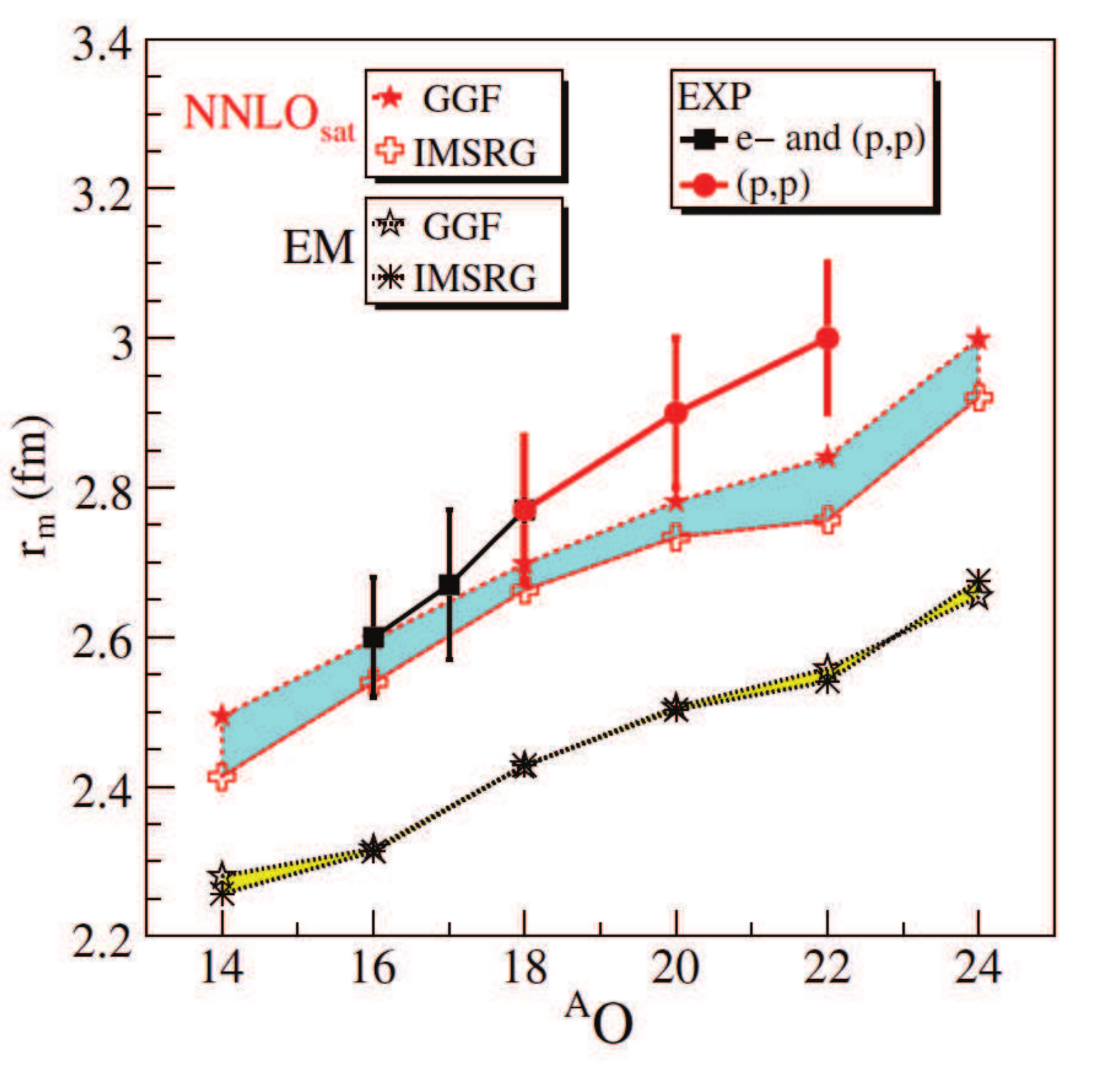}
\caption{Experimental matter radii are compared with results of \textit{ab initio} many-body calculations employing the Entem-Machleidt chiral interaction~\cite{PhysRevC.68.041001} (lower band) and the NNLO$_{sat}$ from Ref.~\cite{PhysRevC.91.051301} (upper band). The band in each case spans the results of different many-body approaches (see Ref.~\cite{PhysRevLett.117.052501}).  
Reprinted figure with permission from Ref.~\cite{PhysRevLett.117.052501} \textcopyright2016 by the American Physical Society.
} 
\label{fig:Lapoux17b}
\end{center}
\end{figure}

The analysis in Ref.~\cite{PhysRevLett.117.052501} anchored their calculation by using proton-elastic-scattering data for the stable ${}^{18}$O isotope made consistent with the known experimental charge and thereby implied matter distributions.
These results are illustrated in Fig.~\ref{fig:Lapoux17}.
Using Hartree-Fock-Bogoliubov densities, it is possible to generate good agreement with the data as shown in Fig.~\ref{fig:Lapoux17}.
In turn, the corresponding radii can be used to test the quality of \textit{ab initio} calculations.
The resulting comparison is made in Fig.~\ref{fig:Lapoux17b}.
The matter radii are compared to \textit{ab initio} calculations with the standard chiral interaction that fits NN data up to 300 MeV (lower band) and the chiral interaction fit to the ${}^{16}$O charge radius (upper band). The bands span results from different many-body techniques as discussed in Ref.~\cite{PhysRevLett.117.052501}.
The authors conclude that although there is considerable improvement employing NNLO$_{sat}$ for the
simultaneous description of the binding energies, radii, and scattering cross sections for stable isotopes, there are still deficiencies for the most
neutron-rich systems. They therefore conclude that crucial challenges related to the development of NN interactions remain. 

A word of caution should be inserted here. As discussed in substantial detail in Ref.~\cite{PIEKAREWICZ200610} and further clarified by the title  ``Insensitivity of the elastic proton-nucleus reaction
to the neutron radius of ${}^{208}$Pb,'' considerable doubt has been expressed whether such experimental data can determine neutron distributions unambiguously.
In Ref.~\cite{PIEKAREWICZ200610}, high-energy data at 500 and 800 MeV on ${}^{208}$Pb were analyzed so that NN $\mathcal{T}$-matrices could be employed with confidence.
The conclusion of this paper states that the diffractive oscillations of the cross sections are controlled by the matter radius of the nucleus, however the large spread in the neutron skin among the various models gets diluted into a mere 1.5\% difference in the matter radius. Quoting these authors directly: ``This renders ineffective the elastic reaction as a precision tool for the measurement of neutron radii.''
This message should be taken seriously and results discussed above and \textit{e.g.} in  Ref.~\cite{PhysRevC.82.044611} should be carefully considered in this context.

We conclude this section by pointing out the approach reported in Ref.~\cite{PhysRevC.91.014612}.
It employs the Gogny interaction to not only describe the traditional structure information but to also  describe reactions by employing this interaction to calculate the imaginary part of the self-energy using the collective states in the random-phase approximations. 
While not purely microscopic, it is an important contribution to the integration of the simultaneous study of structure and reactions.
 Good agreement with experimental data up to 30 MeV is reported in Ref.~\cite{PhysRevC.91.014612}  which corresponds to the domain where long-range correlations are expected to play a dominant role.

\subsubsection{Coupled-cluster method}
\label{sec:CC}
\begin{figure}[bt]
\begin{center}
\includegraphics[width=3.0in]{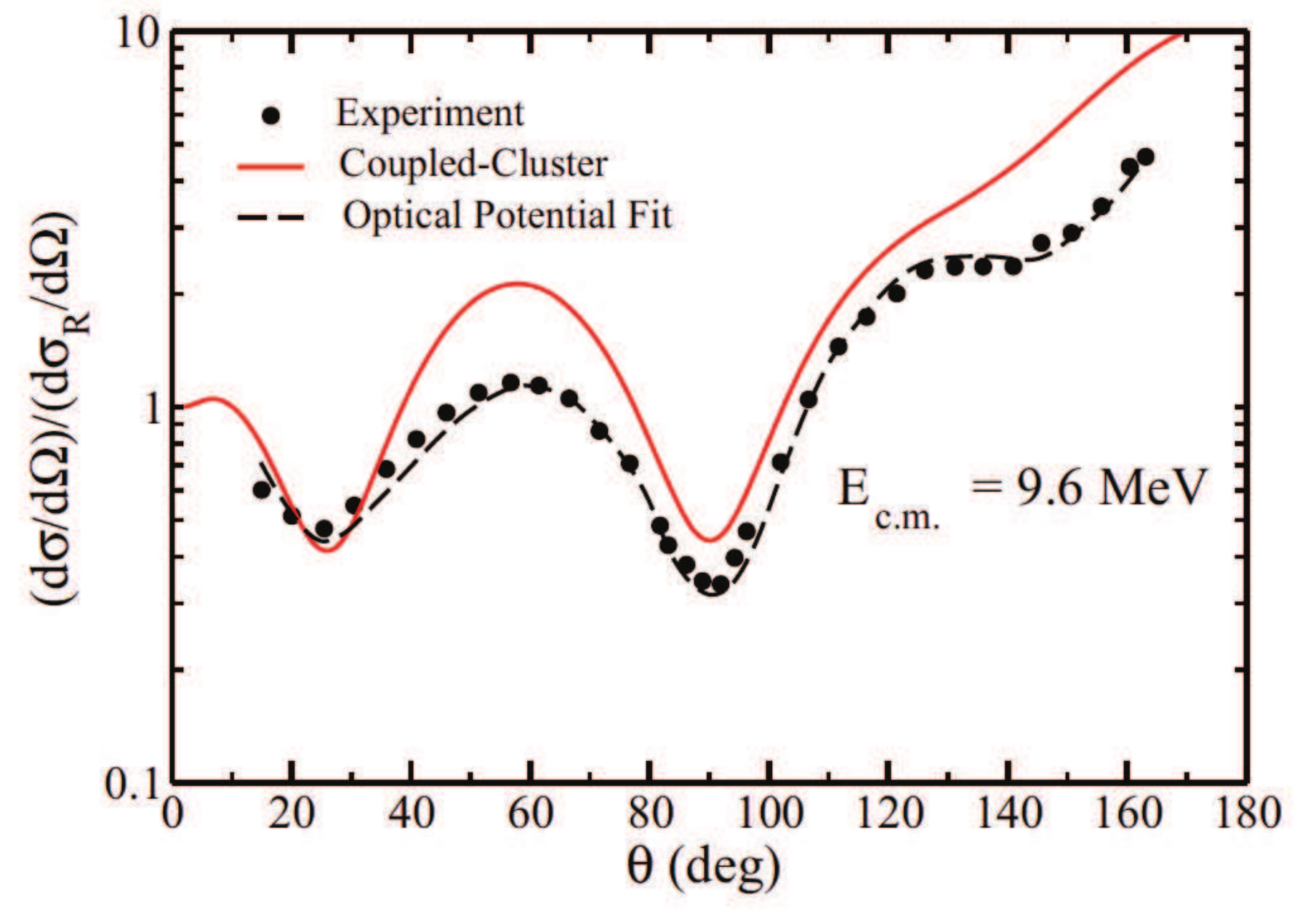}
\caption{Differential cross section from coupled-cluster
calculations divided by the Rutherford cross section for elastic
proton scattering on ${}^{40}$Ca at E$_{\textrm{cm}}$ = 9.6 MeV (solid line), experimental
data (dots), and optical-model potential fits (dashed line), taken from Ref.~\cite{PhysRevC.3.1550}.
Reprinted figure with permission from Ref.~\cite{PhysRevC.86.021602} \textcopyright2012 by the American Physical Society.} 
\label{fig:cc40}
\end{center}
\end{figure}

An important many-body technique is the coupled-cluster method recently reviewed for nuclei in Ref.~\cite{Hagen14}.
The first application to elastic scattering was reported in Ref.~\cite{PhysRevC.86.021602} for low-energy elastic proton scattering from ${}^{40}$Ca.
A typical result is shown in Fig.~\ref{fig:cc40}.
The agreement with experiment is reasonable but there are difficulties at larger scattering angles.
The work includes partial waves with $\ell \le 2$ while higher partial waves do not improve the results.
The authors note that there are some deficiencies in the ingredients of the calculation.
In addition, the coupled-cluster method in realistic applications is limited to the inclusion of two-particle--two-hole excitations (see below).
This implies that the low-energy spectrum of ${}^{40}$Ca is not well described as very few states, mostly of dominant one-particle--one-hole character, are generated at small excitation energies and a lack of absorption is inevitable.
Since few partial waves are needed at low energies to describe the cross section, generating the correct number of bound states for these waves will allow the corresponding phase shifts to start at the right value at $E = 0$ and this may partly explain the reasonable agreement with the data in Fig.~\ref{fig:cc40}.
Another worry is the fact, that at higher energies, a considerably larger number of partial waves are required which  makes a coupled-cluster application considerably more difficult in terms of convergence.

More recently, this method was used in Ref.~\cite{PhysRevC.95.024315} to directly calculate the Green's function in Eq.~(\ref{eq:Gph}) for ${}^{16}$O and subsequently derive the nonlocal (energy-dependent) optical potential by inverting the Dyson equation.
In particular, all contributions obtained within the considered configuration space and approximation scheme to Eq.~(\ref{eq:propp}) were included.
 The optical potential was then constructed by inverting the Dyson equation.
The authors restricted the calculation to the inclusion of two-body interactions.
In the expansion of the correlation operator, the usual restriction to singles and doubles (CCSD) was applied implying that up to two-particle--two-hole amplitudes were included in the adopted basis.
The ground state in the coupled-cluster method is linked to the uncorrelated Slater determinant (usually calculated in the Hartree-Fock approximation) as follows
\be
\ket{\Psi^A_0} = \textrm{e}^T \ket{\Phi^A_0} ,
\label{eq:expT}
\ee
where the cluster operator is given by
\bea
T & = & T_1 + T_2 + ... \nonumber \\
& = & \sum_{i,a} t^a_i a^\dagger_a a_i + \frac{1}{4} \sum_{ijab} t_{ijab} a^\dagger_a a^\dagger_ b a_i a_j + ... 
\label{eq:Teq}
\eea
The operators $T_1$ and $T_2$ therefore introduce the one-particle--one-hole (1p1h) and two-particle--two-hole (2p2h) excitations on top of the Hartree-Fock ground state with the convention that $i,j,..$ refer to hole and $a,b,..$ to particles.
In the CCSD approximation, operators beyond $T_2$ are not taken into account.
The CCSD equations are obtained by projecting the ground state [Eq.~(\ref{eq:expT})] on the reference state and
all 1p1h and 2p2h excitations on top of it.
The ground-state energy and $t$-amplitudes then fulfill the following equations
\bea
E^A_0 & = & \bra{\Phi^A_0} \overline{H} \ket{\Phi^A_0} \nonumber \\
0 & = &  \bra{\Phi^a_i} \overline{H} \ket{\Phi^A_0} \label{eq:CCSD} \\
0 & = &  \bra{\Phi^{ab}_{ij}} \overline{H} \ket{\Phi^A_0} . \nonumber 
\eea
The similarity-transformed Hamiltonian is given by
\be
\overline{H} = \textrm{e}^{-T} H \textrm{e}^{T} = H + [ H,T] + {\scriptstyle{\frac{1}{2}}} [[H,T],T] + ...
\label{eq:stH}
\ee
In the CCSD approximation and with two-body interactions, the series terminates at four-fold nested commutators.
The similarity-transformed Hamiltonian is not Hermitian and therefore has left and right eigenvectors with a correspondingly different resolution of the unit operator given by
\be
\sum_i \ket{\Phi^R_i}\bra{\Phi^L_i} = \hat{1} ,
\label{eq:unitcc}
\ee
where the right ground state is the reference state and the left one also contains particle-hole admixtures.
\be
G^{CC}(\alpha, \beta; E) =\bra{\Phi_0^L} \overline{a_{\alpha}} \frac{1}{E-(\overline{H}-E^A_0) +i\eta} \overline{a^\dagger_{\beta} }\ket{\Phi^R_0} 
+ \bra{\Phi_0^L} \overline{a^\dagger_{\beta}} \frac{1}{E-(E^A_0 - \overline{H})-i\eta} \overline{a_{\alpha}} \ket{\Phi^R_0}  .
\label{eq:ccgf}
\ee
To avoid dealing with many-body states in the continuum when complete sets of states are inserted in Eq.~(\ref{eq:ccgf}), the authors of Ref.~\cite{PhysRevC.95.024315} employ the Lanczos technique to determine Eq.~(\ref{eq:ccgf}) and employed a Berggren basis~\cite{BERGGREN1968265} for some ingredients of the calculation.
Other ingredients involved harmonic-oscillator states for the chosen nucleus ${}^{16}$O for which convergence for $N_{max}$ was checked with a value of $\hbar \omega$ = 20 MeV.
The chosen NN interaction is denoted by NNLO$_{\textrm{opt}}$~\cite{PhysRevLett.110.192502}.

\begin{figure}[bt]
\begin{center}
\includegraphics[width=3.0in]{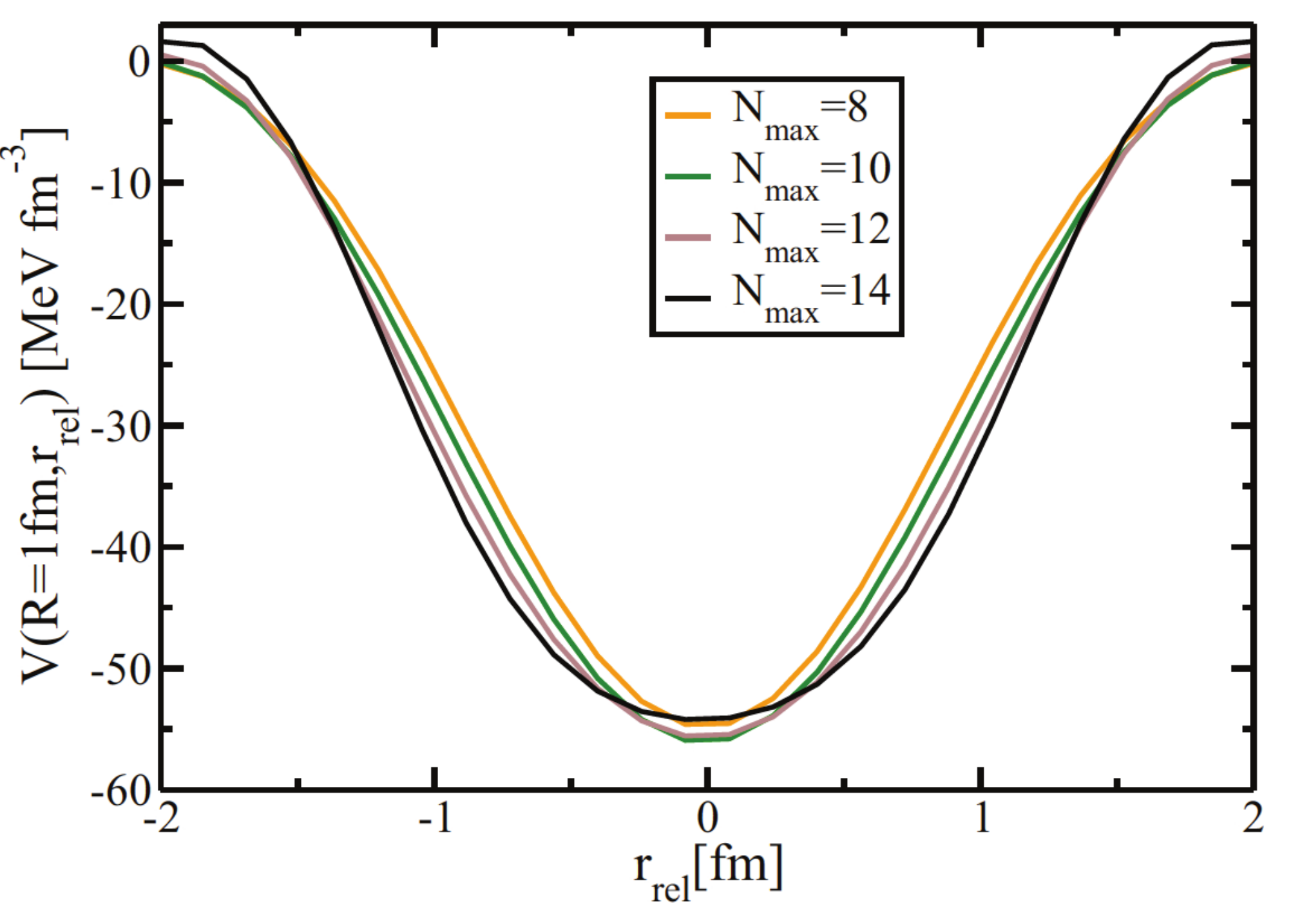}
\caption{Neutron s-wave optical potential at $E$ = 10 MeV plotted
as $V(R+r_{rel}/2,R-r_{rel}/2)$ at fixed $R = 1$ fm. Here $N_{max}$ = 14
and 50 discretized $s$-wave shells are included in the sp
basis. 
The displayed results exhibit a reasonable convergence with $N_{max}$ with a tendency for the potential to deepen slightly with increasing $N_{max}$ except near $r_{rel} = 0$.
Reprinted figure with permission from Ref.~\cite{PhysRevC.95.024315} \textcopyright2017 by the American Physical Society.
} 
\label{fig:16O1}
\end{center}
\end{figure}

The calculated real part of the optical potential for $\ell = 0$ exhibits a substantial nonlocality. It is plotted  in Fig.~\ref{fig:16O1} as a function of the coodinate $r_{rel}= r-r'$ for a fixed value of the center-of-mass coordinate $R = (r+r')/2$ = 1~fm.
The full width at half maximum is more than 2 fm. Clearly, such an \textit{ab initio} potential has no resemblance to standard local phenomenological potentials which concentrate all strength at $r_{rel} = 0$.

\begin{figure}[bt]
\begin{center}
\includegraphics[width=3.0in]{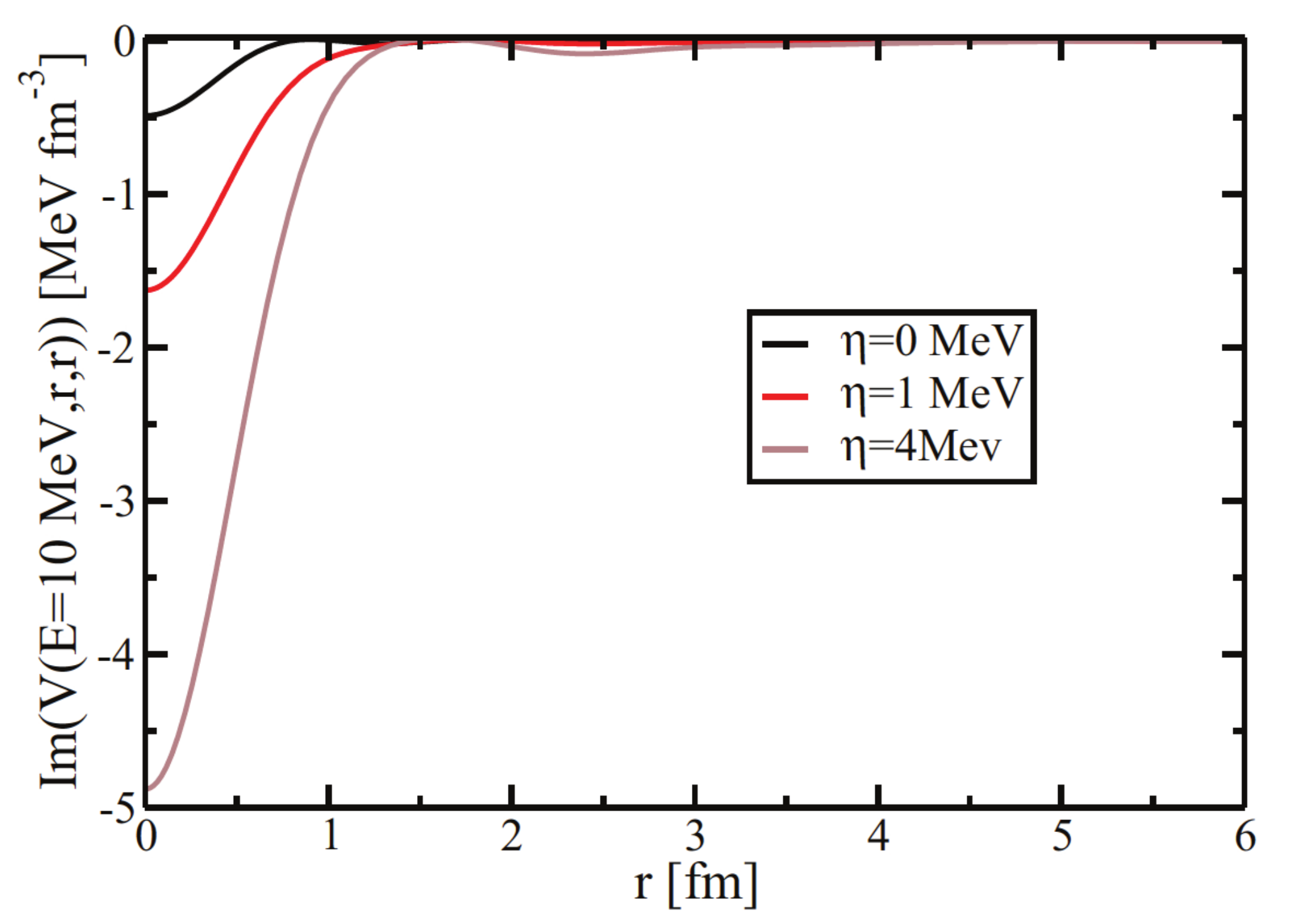}
\caption{Imaginary part of the radial (diagonal) optical potential
for neutron $s$ waves at $E$ = 10 MeV. Same basis ingredients as in Fig.~\ref{fig:16O1} were included.
Results are shown for several values of $\eta$.
With increasing value of $\eta$ a more substantial imaginary part is generated.
Reprinted figure with permission from Ref.~\cite{PhysRevC.95.024315} \textcopyright2017 by the American Physical Society.} 
\label{fig:16O2}
\end{center}
\end{figure}

The imaginary part describes the loss of flux due to inelastic processes.
As discussed above for ${}^{40}$Ca, the coupled-cluster method in CCSD does not describe the excitation spectrum of double-closed shell nuclei very well, in particular at low energy.
This is reflected in the results for the imaginary part of the potential, along the
diagonal $r' = r$ in Fig.~\ref{fig:16O2} for the neutron $s_{1/2}$ wave
at E = $10$ MeV for a model space with N$_{\textrm{max}}$ = 10 and
50 discretized shells for the $s$ wave.
It is possible to simulate some of the missing physics by artificially increasing the value of the $\eta$ infinitesimal in Eq.~(\ref{eq:ccgf}).
In the limit $\eta = 0$, the imaginary part of the potential is
very small, and this is true for the whole range of energies up
to E = 10 MeV. 
Figure~\ref{fig:16O2} demonstrates that, as $\eta$ decreases
to zero, the imaginary part also decreases and becomes very
small for $\eta = 0$.
The same qualitative behavior is obtained  for all other considered partial waves up to $d_{5/2}$, and the result does not change when the model space is increased.
It is not clear at present how to improve the coupled-cluster method to generate better optical potentials without introducing some phenomenological ingredients as the authors intimate.
 
\section{Dispersive optical model}
\label{sec:DOM}
It is no longer practicable and useful to list all optical-model studies that have been performed. Now with increasing use of the DOM, the same could almost be said of it. In the following we describe some large-scale studies performed with the DOM by a number of groups. Some single works with the DOM not mentioned are \cite{Hicks:1988,Delaroche:1989}.  See also review article on the DOM by Hodgson in 1992 \cite{Hodgson:1992}.

\subsection{DOM Mahaux}
\begin{figure}[tpb]
\begin{center}
\includegraphics[scale=.6]{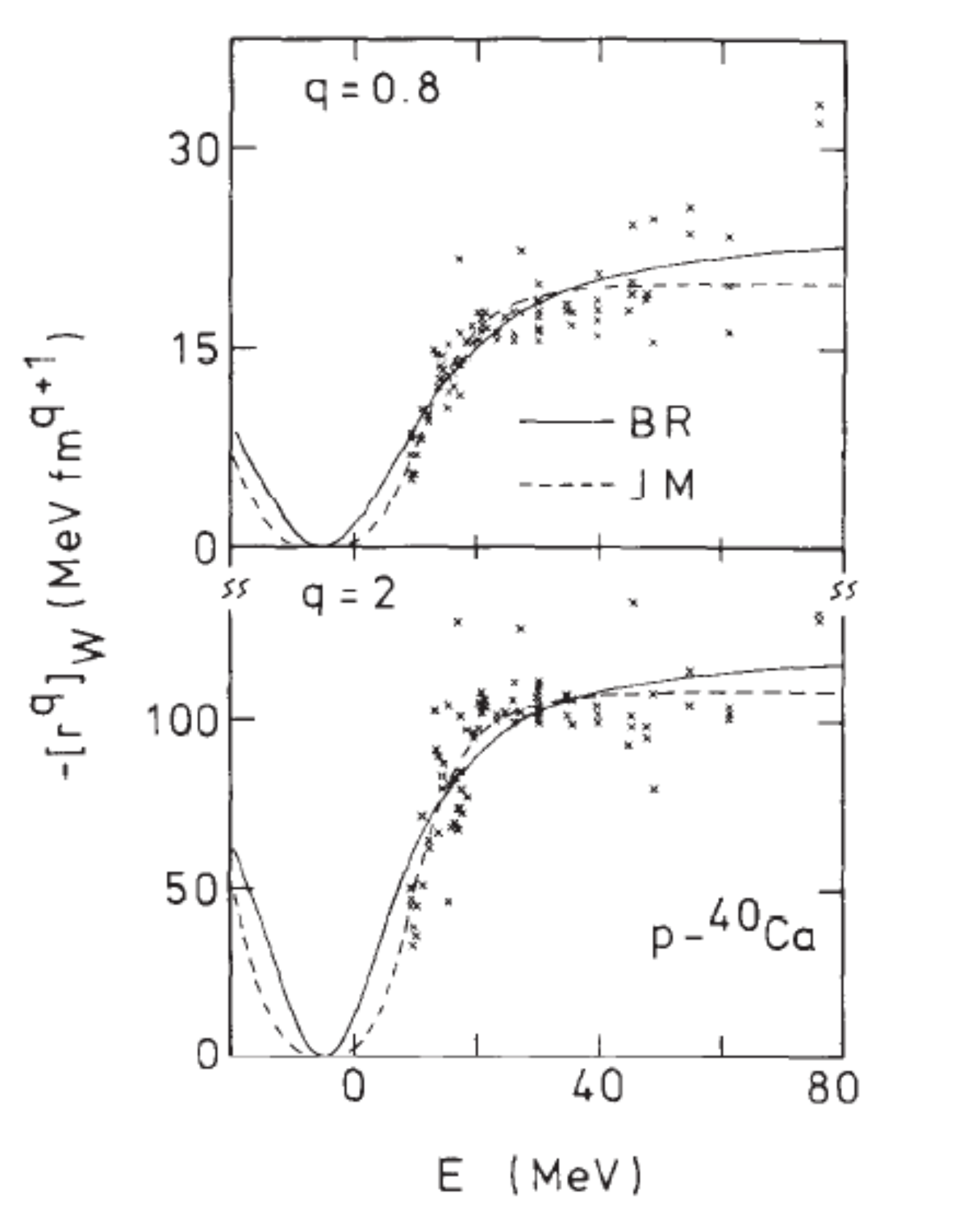}
\caption{Imaginary moments $[r^q]_{W}$ for $q$=0.8 and 2 from an analysis of $n$+$^{40}$Ca elastic scattering by Mahaux and Sartor. Curves show two possible fits.  Reprinted figure with permission from \cite{Mahaux:1988} \textcopyright1988 by Elsevier }
\label{fig:n40CaW}
\end{center}
\end{figure}

\begin{figure}[tpb]
\begin{center}
\includegraphics[scale=.6]{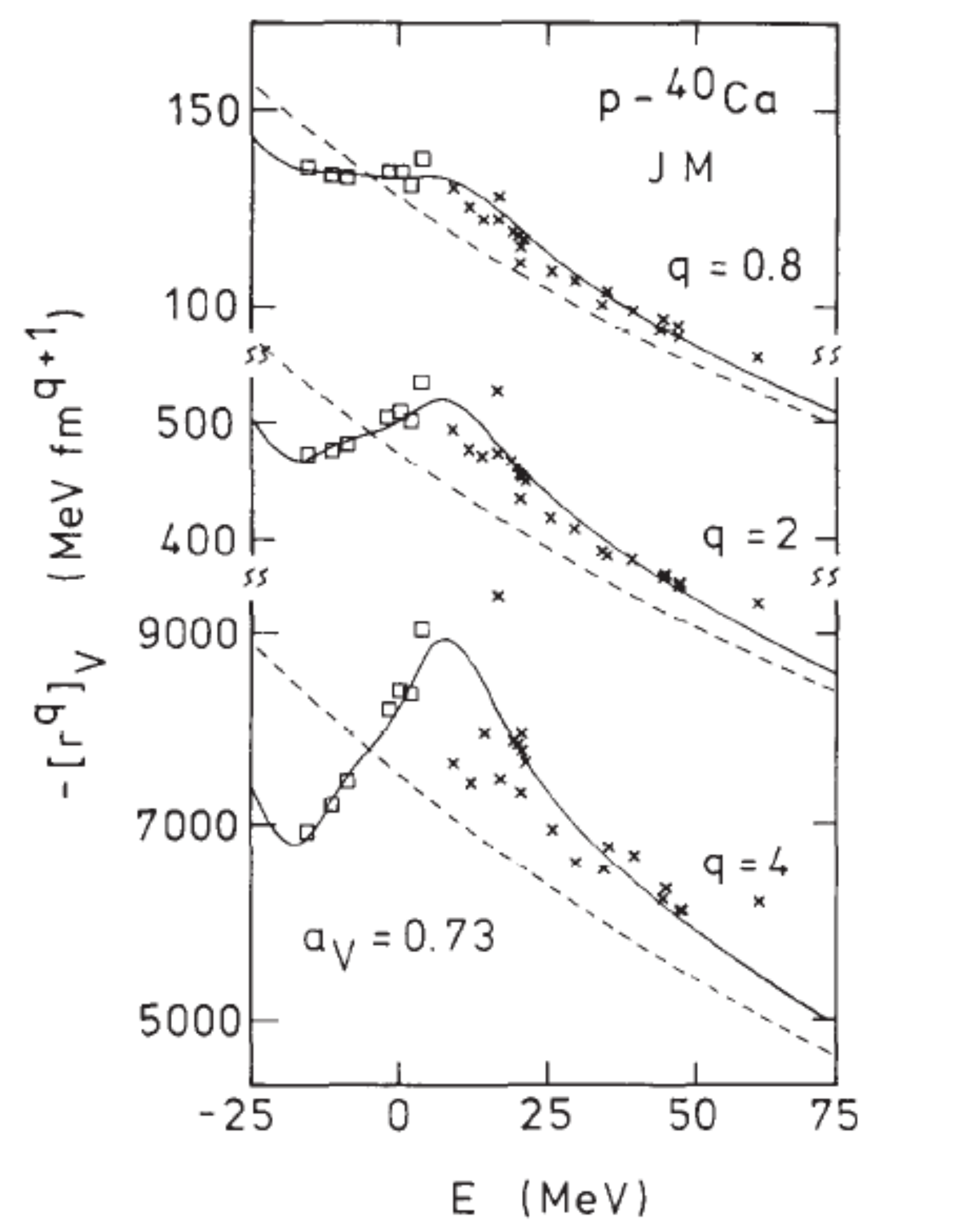}
\caption{Real moments $[r^q]_{V}$ for $q$=0.8, 2, and 4 from an analysis of $n$+$^{40}$Ca elastic scattering by Mahaux and Sartor.  The solid curves are fit with  smooth Hartree-Fock components (dash curves) plus the dispersive corrections. The crossed-shaped data points are from analysis of elastic-scattering data, while the square data points are from the experimental sp energies. Reprinted figure with permission from~\cite{Mahaux:1988} \textcopyright1988 by Elsevier }
\label{fig:n40CaV}
\end{center}
\end{figure}

\begin{figure}[tpb]
\begin{center}
\includegraphics[scale=.6]{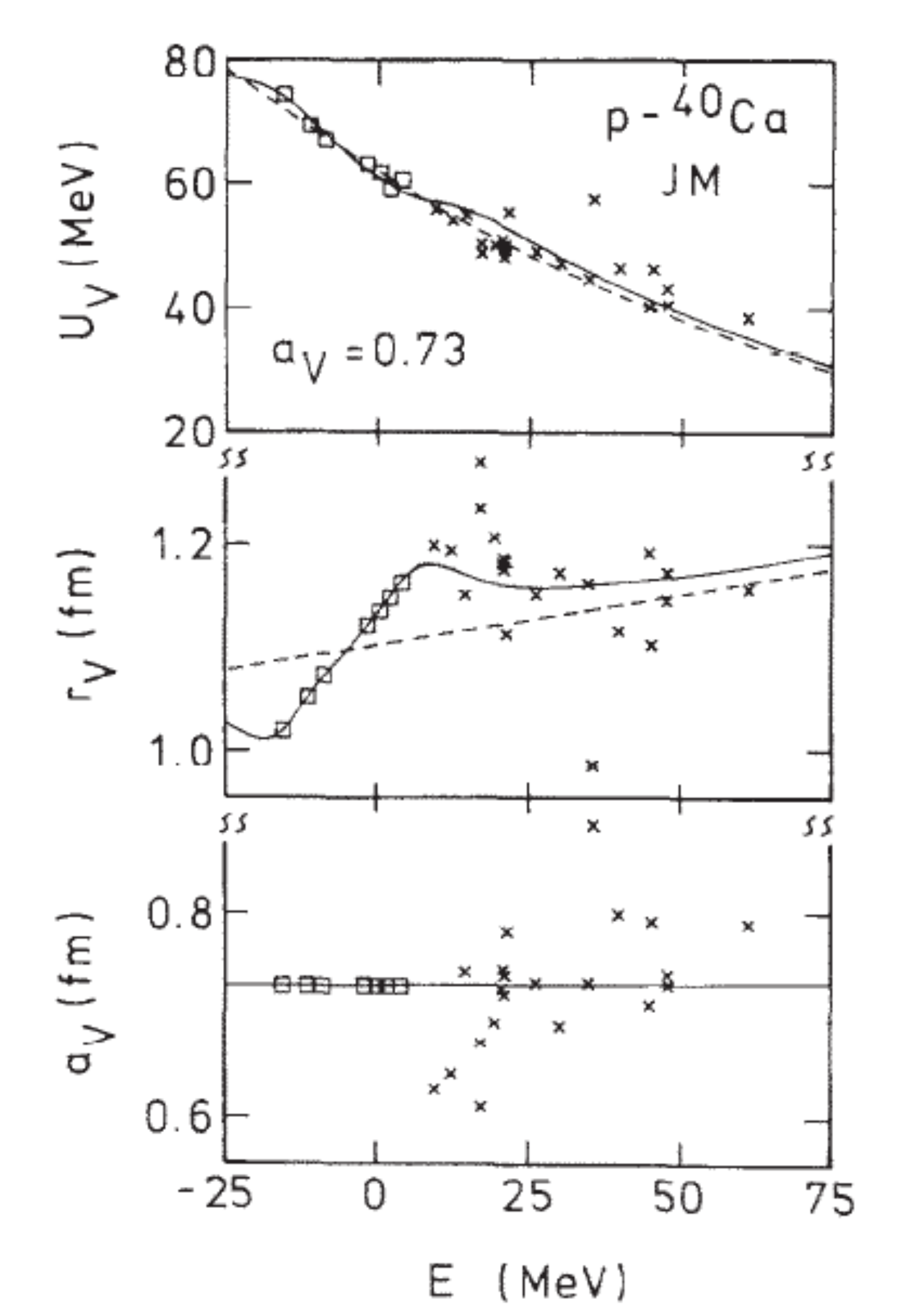}
\caption{Energy dependence of the depth, radius and diffuseness of the real Wood-Saxon potential obtained from the iterative-moment analsis of $n$+$^{40}$Ca by Mahaux and Sartor. The solid curves are the fitted values, while the dashed curves are the smooth dependence from the Hartree-Fock component. See Fig.~\ref{fig:n40CaV} for a description of the data points. The diffuseness $a_{v}$ was assumed energy independent in the analysis. Reprinted figure with permission from \cite{Mahaux:1988} \textcopyright1988 by Elsevier }
\label{fig:n40Cadepth}
\end{center}
\end{figure}

Mahaux and Sartor \cite{Mahaux:1987b,Mahaux:1987,Mahaux:1988,Mahaux:1989a,Mahaux:1989} developed a number of local DOM potentials for the $n$,$p$ + $^{40}$Ca, $^{208}$Pb, and $n$+$^{89}$Y systems in the period from 1987 to 1989.   
Rather than fitting the experimental elastic-scattering angular distributions, and total and absorption cross sections directly, they considered the moments of the real and imaginary potentials [Eqs.~(\ref{eq:momentReal},\ref{eq:momentImag})] obtained by others from fitting such data at a each energy independently.  For example in the iterative-moment approach for $n$+$^{40}$Ca in Ref.~\cite{Mahaux:1988}, the imaginary moments $[r^{q}]_{W}$, $q$=0.8 and 2 are first fit [Fig.~\ref{fig:n40CaW}]. From these, the corresponding dispersion corrections $[r^q]_{\Delta V}$ are calculated. The real moments $[r^{q}]_V$, $q$=0.8, 2 in Fig.~\ref{fig:n40CaV} (crosses) are then fit. These real  moments consist of a component with a smooth energy dependence, the so called Hartree Fock component, and the previously determined dispersion corrections which casues the oscillations around the Fermi energy. In this case, the moments are fit assuming the Hartree-Fock components have exponential-decaying energy dependences (dashed curves).  From these two fitted values of 
$[r^q]_{V}$, the energy-dependent depth and radius of a Wood-Saxon potential can be deduced [Fig.~\ref{fig:n40Cadepth}] with the assumption of a fixed diffuseness [in this case $a$=0.73~fm].

Now additional constrains are applied from the experimental sp energies. With the assumed diffuseness and a spin-orbit potential, the depths of the real Wood-Saxon potential for each sp energy are obtained.  These Wood-Saxon potentials produce the square data points in Fig.~\ref{fig:n40CaV}. Now both these new square and the old crossed-shaped  data points are then refit to give a new Hartree-Fock potential and new depths and radii are found as a function of energy. This procedure is iterated until a stable solution is found. The final depths and radii are shown in Fig.~\ref{fig:n40Cadepth} which illustrates the complex evolution of the real potential with energy due to the dispersive correction. Mahaux and Sartor developed an improved version of this called the variational moment approach~\cite{Mahaux:1989,Mahaux:1989a} where they explicitly decomposed the imaginary potential into surface and volume components. 

\begin{figure}[tpb]
\begin{center}
\includegraphics[scale=.6]{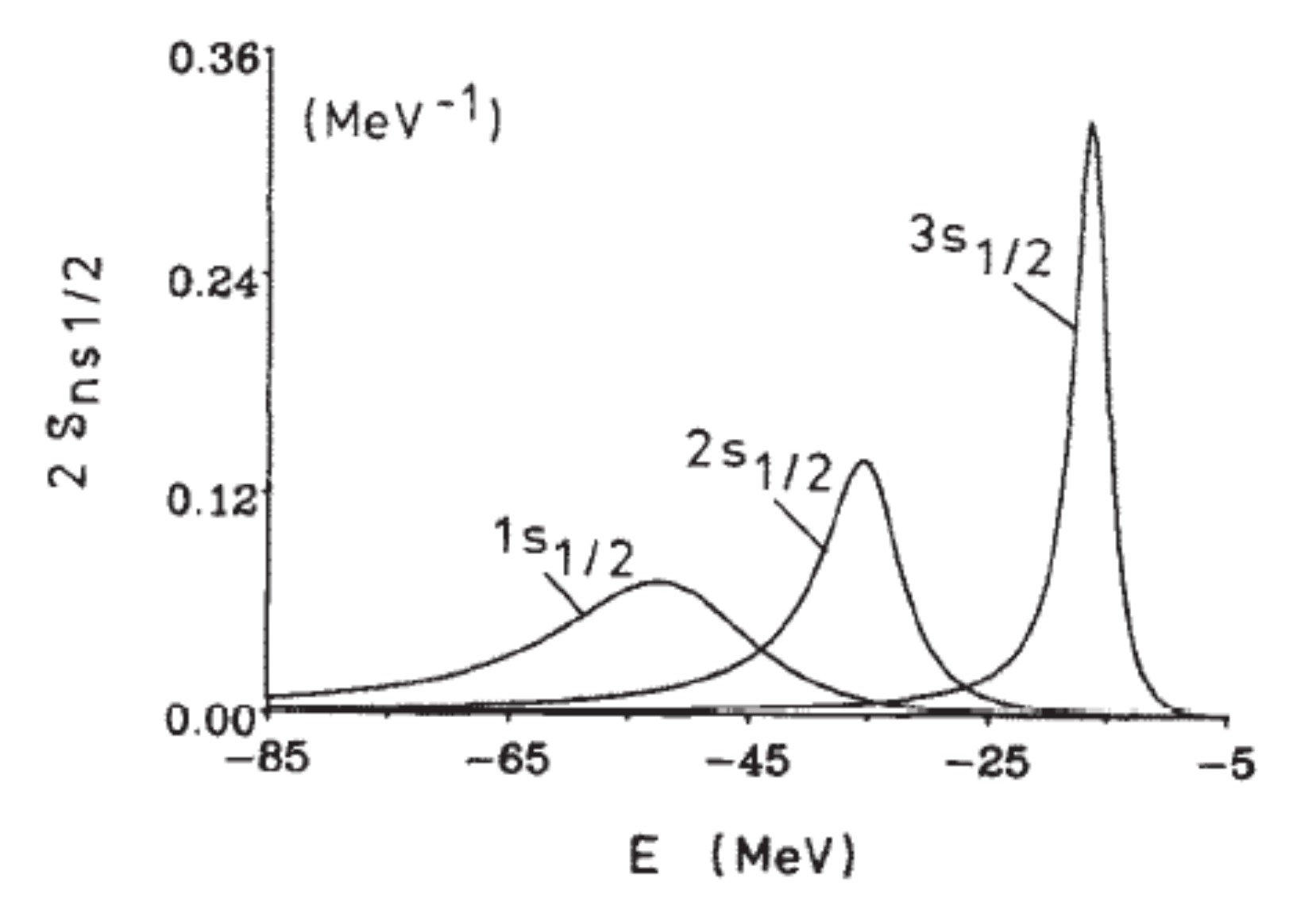}
\caption{(2$j$+1) weighted spectral functions of the 3$s_{1/2}$, 2$s_{1/2}$, and 1$s_{1/2}$ neutron hole states in $^{208}$Pb obtained by Mahaux and Sartor.  Reprinted figure with permission from \cite{Mahaux:1989} \textcopyright1989 by Elsevier }
\label{fig:VMA}
\end{center}
\end{figure}

Mahaux also worked with Johnson and collaborators to constrain the DOM potential in a more conventional manner directly fitting elastic-scattering angular distribution and cross sections~\cite{Johnson:1987,Johnson:1988,Jeukenne:1988}. This approach gave similar optical potentials, but there were some important differences~\cite{Mahaux:1989a}.   The DOM potentials fit by Mahaux and collaborators were used to deduce effective masses, spectral functions, spectroscopic factors, and occupation probabilities of sp orbitals. For example the spectral functions for $s_{1/2}$ neutron hole orbits in $^{208}$Pb deduced with the variational moment analysis are shown in Fig.~\ref{fig:VMA}.
 
\subsection{DOM TUNL}
A number of DOM analyses for neutrons  \cite{Tornow:1990,Walter:1993,Weisel:1996,Chen:2004,AlOhali:2012} were performed  based on angular distributions and analyzing powers measured at Triangle Universities Nuclear Laboratory (TUNL) and additional such data and total sections from other sources. The targets ranged from $^{27}$Al to $^{209}$Bi and the fitted parameters were used to deduce bound-state properties including sp energies, spectroscopic factors following the prescription of Mahaux and Sartor~\cite{Mahaux:1991}.

\subsection{DOM coupled-channels}
 Following Hodgson's distaste for global OM fits and his suggestion that more localized parametrizations  are important (see Sec.\ref{sec:global}), the region of the deformed actinides is a good place for such a more localized study. Indeed this deformed region may not be well described by global parametrizations which are fit to more spherical systems. Note, the Koning and Delaroche Global parametrization deliberately did not fit data from this region. 

OM predictions in this region are important for various reactor systems. In radioactive-waste transmutation by accelerator driven systems, predictions are needed for both protons and neutrons with energies up to several hundred MeV. To this end, Capote \textit {et al.} \cite{Capote:2005,Capote:2008,Soukhovitskii:2016} have fitted a local dispersive OM potential. As low-lying collective states are strongly excited in nucleon elastic scattering in this region, the coupled-channels formalism was used to include coupling to the lower-lying rotational and vibrational levels. They fit to proton and neutron-induced data with bombarding energies from 1~keV to 200 MeV including total cross sections, elastic and inelastic angular distributions, strength functions and scattering radii.  This data base was dominated by reactions with  $^{232}$Th and $^{238}$U targets.
The authors indicate that the dispersive corrective is very important in fitting the data and expect the parametrization will extrapolate well to neighboring nuclei.   This coupled-channel approach has also been used for Fe isotopes, where the potential is made Lane-consistent [Eq.~{\ref{eq:Lane})] and used to also describe ($p$,$n$) quasi-elastic scattering as well the other reactions \cite{Sun:2014}.

\subsection{DOM St. Louis}
A recent, more detailed overview of the work described in this section is summarized in Ref.~\cite{Dickhoff:17}.
The Washington University group in St. Louis  first performed local dispersive optical-model fits to understand the $N-Z$ asymmetry dependence of correlations in nuclei~\cite{Charity:2006,Charity:2007,Mueller:2011}. The forms of the potential were largely consistent with those used by Mahaux and Sartor~\cite{Mahaux:1991}. The St.-Louis group fit neutron and proton elastic-scattering angular distributions, total and absorption cross section, sp energies and spectroscopic factors deduced from ($e,e'p$) reactions. For protons, the long-range correlations associated with the surface imaginary potential showed a strong $N-Z$ asymmetry dependence, with the proton correlations increasing in neutron-rich systems. This was consistent with the assumed asymmetry dependence of  this potential in most global optical-model fits (Sec.~\ref{sec:global}). On the other hand, neutrons seems to have very little, if any, dependence. The OM parametrization were later used to provide distorted waves and overlap functions for the analysis of transfer reactions (Sec.~\ref{sec:DOMknockout}).

The connection of the OM potential and the self-energy in the Green's-function formalism can be made if the energy-dependent real potential in local OM fits is replaced by a energy-independent non-local. Efforts in this regard were made for the case of $^{40}$Ca~\cite{Dickhoff:2010} where the fitted energy-dependent real potential was  replaced with a nonlocal potential. This limited the binding of the lowest $s_{1/2}$ in better agreement with  experiment. However, the charge distribution was not correctly predicted with too much yield at small radii.  This problem was rectified when a fully non-local fit to $n,p$+$^{40}$Ca data was performed with nonlocality in both the real and imaginary potentials~\cite{Mahzoon:2014}. As well as the positive-energy reaction data, this fit also included the experimental sp energies, the experimental charge distribution and the neutron and proton particle numbers. The fit to the charge distribution is shown in Fig.~\ref{fig:Mahzoon4}. 
 \begin{figure}[tpb]
\begin{center}
\includegraphics[scale=.6]{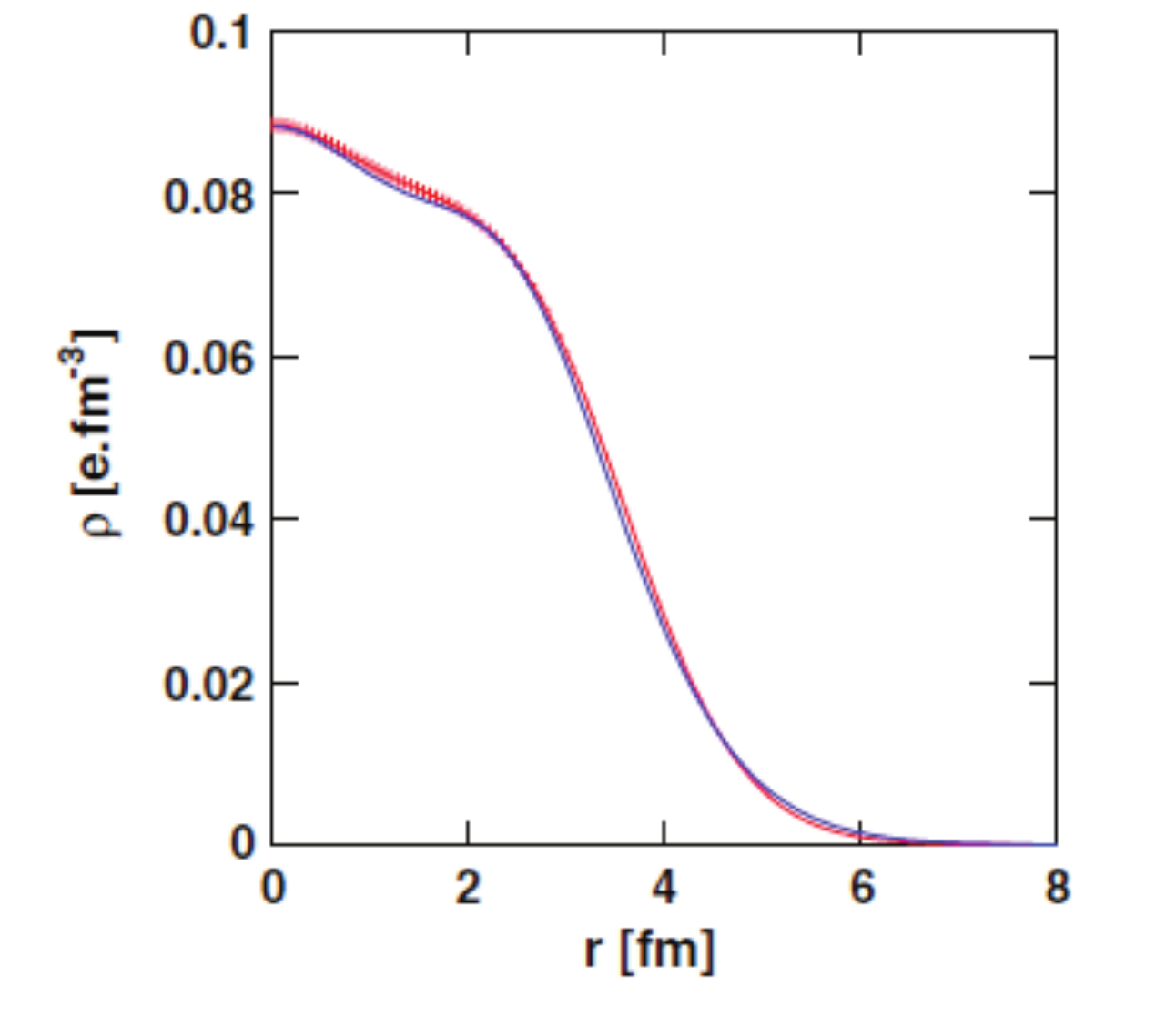}
\caption{Fit (thin curve) to the experimental charge distribution (thick curve) for $^{40}$Ca with the 
nonlocal dispersive optical model by the St.-Louis group.  Reprinted figure with permission from \cite{Mahzoon:2014} \textcopyright2014 by the American Physical Society .}
\label{fig:Mahzoon4}
\end{center}
\end{figure}
The introduction of nonlocality in the imaginary potential changes the absorption profile as a function of angular momentum and was essential in reproducing the particle numbers and the experimental charge distribution. This fit allowed for the reproduction of the yield of high-momentum nucleons determined by data from Jefferson Lab~\cite{Mahzoon:2014}.
We note that these high-momentum components involve the single-nucleon spectral function as extracted from measurements reported in Refs.~\cite{Rohe04,Rohe04A}.
The DOM is therefore not yet capable of addressing the relative momentum distribution as probed by knockout experiments involving two removed nucleons~\cite{Subedi08}.

From this fitted nonlocal dispersive potential, the spectral function for bound orbitals is deduced both at negative energies~\cite{Mahzoon:2014} and at positive energies~\cite{Dussan:2014} as shown in Fig.~\ref{fig:Dussan2}. Distorted waves and overlap function and spectroscopic factors predicted with an updated version of this fitted potential were used in an analysis of $^{40}$Ca($e,e'p$) reaction (Sec.~\ref{sec:DOMknockout}).  
\begin{figure}[tpb]
\begin{center}
\includegraphics[scale=.6]{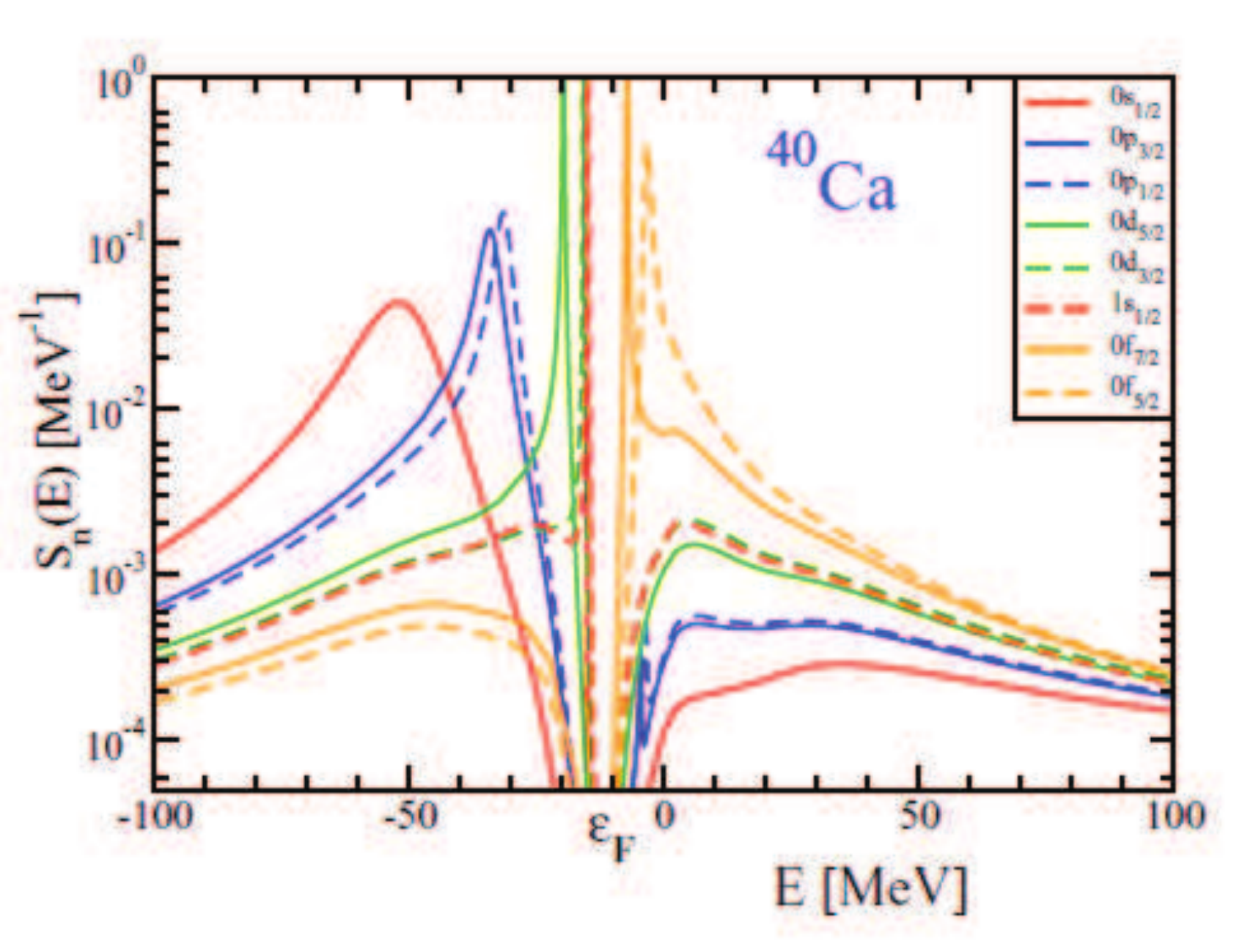}
\caption{Spectral function for neutron sp orbits in $^{40}$Ca obtained by the St.-Louis group.  
The sequence of peaks follows the standard order of the independent particle model. Reprinted figure with permission from~\cite{Dussan:2014} \textcopyright2014 by The American Physical Society.}
\label{fig:Dussan2}
\end{center}
\end{figure}
The nonlocal DOM treatment was extended to the $n,p$+$^{48}$Ca reactions and used to deduce the neutron-matter and weak-charge density distributions (Fig.~\ref{fig:Mahzoon2}). The results imply a neutron skin of 0.249$\pm$0.023~fm \cite{Mahzoon:2017} which is larger than the prediction of 0.12-0.15~fm from the coupled-clusters formalism \cite{Hagen:2016}. Future experiments at Jefferson Lab with parity-violating electron scattering should be able to resolve this disagreement \cite{PhysRevLett.108.112502}.   

\begin{figure}[tpb]
\begin{center}
\includegraphics[scale=.6]{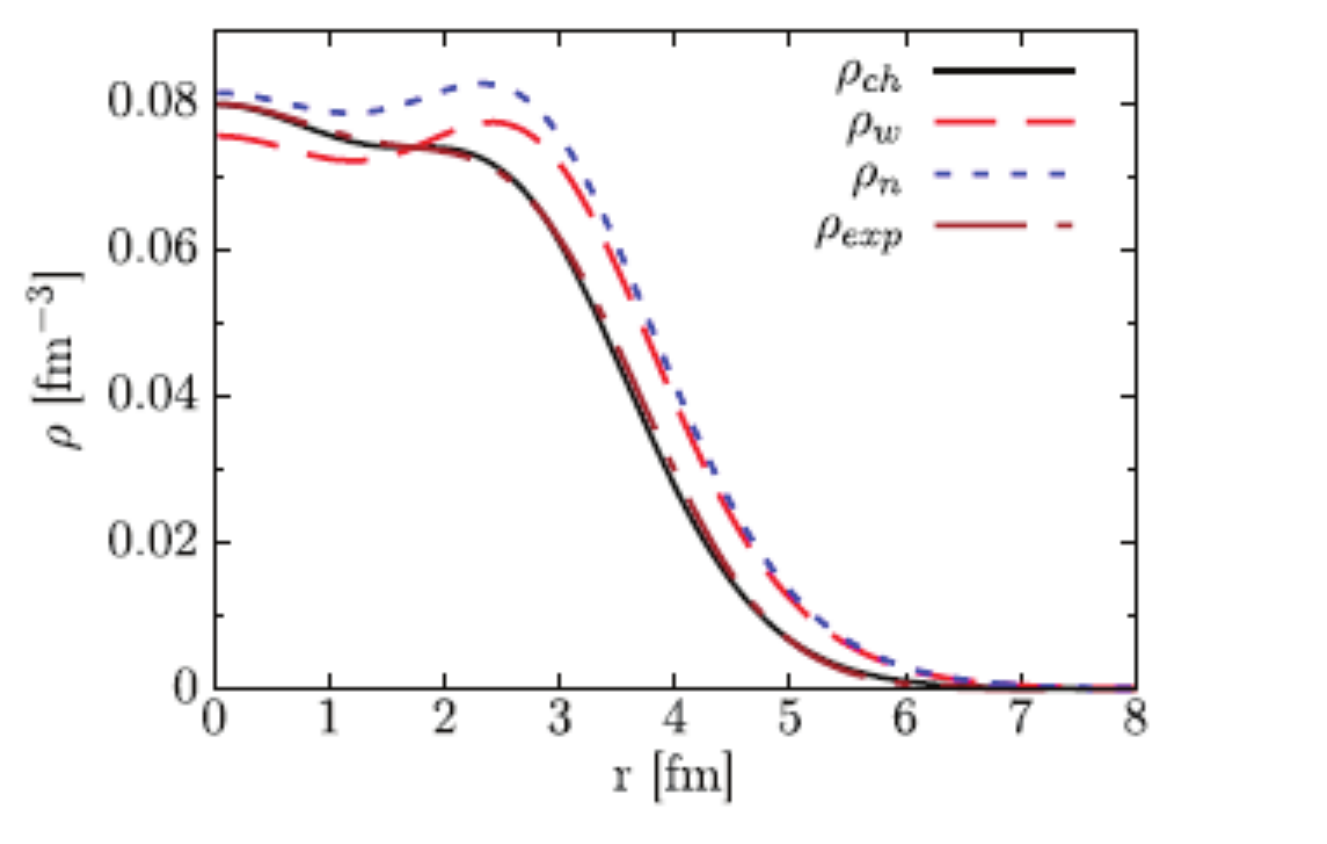}
\caption{Comparison of the density distribution for $^{40}$Ca from the St.-Louis group. These include the experimental  ($\rho_{exp}$) and fitted ($\rho_{ch}$) charge densities, the neutron matter distribution ($\rho_{n}$) and  the weak charge density ($\rho{w}$).  Reprinted figure with permission from \cite{Mahzoon:2017} \textcopyright2017 by the American Physical Society.}
\label{fig:Mahzoon2}
\end{center}
\end{figure}

\section{Practical applications}
\label{sec:applic}
\subsection{Transfer reactions}
\label{sec:transfer}
Single and multi-nucleon transfer direct reactions have and continued to be an important tool to probe nuclear structure. Single-nucleon transfer can be considered as either pickup, where the light projectile gains a nucleon, or stripping, when it loses a nucleons. These are peripheral reactions that specifically probe the sp structure of a nucleus.  Some common transfer reactions are the single-neutron stripping reaction ($d$,$p$), the proton stripping reactions ($d$,$n$) and ($^{3}$He,$d$), neutron pickup reaction ($d$,$t$) , and the proton pickup reaction ($d$,$^{3}$He). 

There is a long history in performing such pickup reactions in normal kinematics with stable targets and bombarding energies just above the barrier. More recently there is renewed interest in these reactions using inverse kinematics with radioactive beams. Some recent reviews can be found in Refs.~\cite{Bardayan:2016,Wimmer:2018}. An important feature of these reactions is that the angular distributions of the outgoing particles reflects the transferred orbital angular momentum $\ell$ with respect to the ``core'' or residual. Also, if polarization measurements are made, the transferred total angular momentum can be extracted. This gives the parity of the final state and information on its spin.
Lastly the spectroscopic factor, \textit{i.e.}, the overlap between the initial and final state of the nucleus, is interpreted to be proportional to the cross section.  

Historically most transfer reactions studies calculate the  angular distribution for transfer using the Distorted Wave Born Approximation (DWBA) using reaction codes such as DWUCK~\cite{DWUCK},  TWOFNR~\cite{TWOFNR}, PTOLEMY~\cite{Macfarlane:1978}, or FRESCO~\cite{Thompson:1988}. The spectroscopic factor $S$  is obtained by scaling this distribution to match the experimental data, \textit{i.e.},
\begin{equation}
\frac{d\sigma}{d\Omega}_{Exp} = S   \frac{d\sigma}{d\Omega}_{DWBA}
\end{equation}

There are different definitions of the spectroscopic factor in use  and thus correspondingly different definitions of $d\sigma/d\Omega_{DWBA}$. Some definitions do, or do not, include the isospin Clebsch-Gordan factor $C$ and hence we often see the notation $C^2 S$ for the spectroscopic factor. In addition, the normalization is typically different for stripping and pickup reactions~\cite{Bertulani:2004}.

 For an $A(a,b)B$ reaction where $a=b+x$ transfers x to the target nucleus $A$,  the differential cross section is 
\begin{equation}
\frac{d\sigma}{d\Omega} = \frac{\mu_{\alpha} \mu_{\beta}}{(2\pi\hbar^2)^2} \frac{k_{\beta}}{k_{\alpha}} \left| \mathcal{T} \right|^2
\end{equation}
where $\mu_{\alpha}$ and $\mu_{\beta}$ are reduced masses in the entrance and exit channels, respectively, $k_{\alpha}$ and $k_{\beta}$ are the corresponding momenta, and $\mathcal{T}$ is transition matrix. The DWBA approximation is valid when the most-likely occurrence to take place when two nuclei collide is elastic scattering. Other reactions can then be considered perturbations corresponding to weak transitions between different elastic-scattering states. The scaling to the experimental data is best made at forward angles where the DWBA hypothesis, that the reaction is a direct one-step process, is most valid.

The DWBA transition amplitude is \cite{Austern:1970,Satchler:1983a,Glendenning:1983,Thompson:2009}
\begin{equation}
\mathcal{T}^{DWBA} = \int\int \left[\chi^{(-)}(\bm{k}_{\beta},\bm{r}_{\beta}) \right]^* 
\braket{\phi_{A:B}|V|\phi_{b:a}} \chi^{(+)}(\bm{k}_{\alpha},\bm{r}_{\alpha}) d\bm{r}_{\alpha} d\bm{r}_{\beta}.
\end{equation}
The distorted waves $\chi^{(\pm)}(\bm{k},\bm{r})$ represent the elastic-scattering wavefunctions and asymptotically are composed of a plane wave of momentum $\bm{k}$ plus an (-) outgoing or (+) incoming spherical scattering wave which, for no Coulomb interaction, has the form
\begin{equation}
\chi^{(\pm)}(\bm{k},\bm{r}) = e^{i\bm{k}\cdot\bm{r}} + f(\theta) \frac{e^{\pm ikr}}{r}.
\end{equation}
The nuclear matrix element is $\braket{\phi_{A:B} | V |\phi_{b:a}}$ 
where $V$ is the residual interaction and $\phi_{A:B}(\bm{r}_{Ax})$ and $\phi_{b:a}(\bm{r}_{bx})$ are the overlap functions for the target and projectile, respectively. 

The standard procedure to obtain spectroscopic factors, is to calculate the distorted waves using optical-model potentials. The overlap functions are typically described by sp states in a Wood-Saxon potential with the depth adjusted to give the correct binding energy. Typically the radius and diffuseness parameters are taken to be the  standard values ($r_{0}\sim$1.2~fm and $a\sim$0.65). The dependence on the choice of the optical-model parameters is important, the extracted spectroscopic factor can vary by factors of 2 or 3. Schiffer \textit{et al.} argued that one should not use a parameter set specifically  fitted to elastic-scattering data at the energy of interest \cite{Schiffer:1967} as there may be fluctuations in these parameters, for instance from resonances. Rather it is better to use parameters obtained from a fit over a range of energies or the global parametrizations (Sec.~\ref{sec:global}). Even so, there are still significant differences with using different sets of global parameters with changes of order 20\% coming from different choices \cite{Liu:2004}. To compare spectroscopic factors, it is best to calculate the DWBA in a consistent manner, \textit{i.e.} with the same global OM parameters, and the overlap state with the same radius and diffuseness parameters.  Lee, Tsang, and Lynch have recently reanalyzed at large body of  ($d$,$p$) and ($p$,$d$) cross section data to this end and found consistency of the spectroscopic factors from these two reactions \cite{Lee:2007} which is demonstrated in Fig.~\ref{fig:Lee}.

\begin{figure}[tpb]
\begin{center}
\includegraphics[scale=.8]{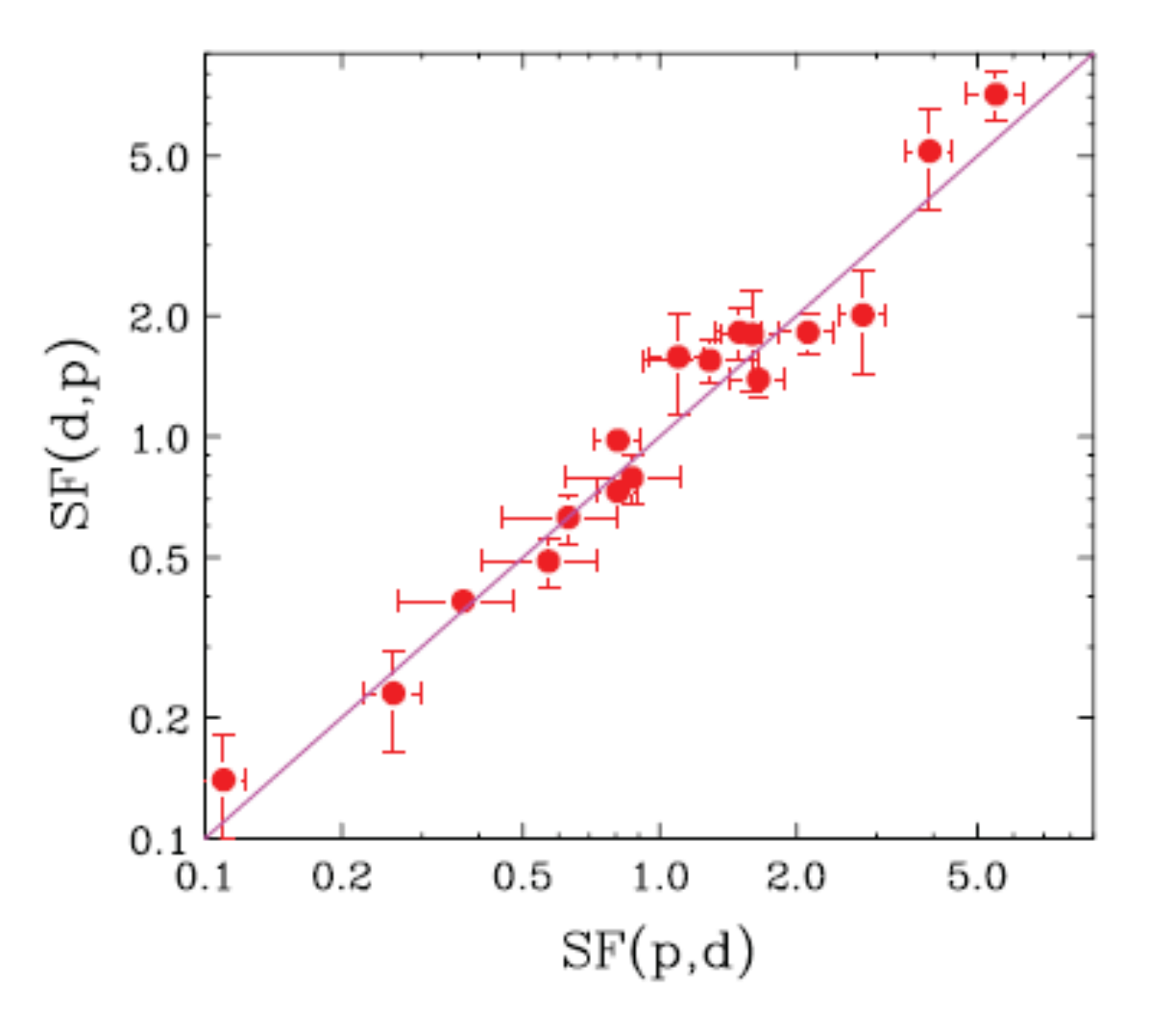}
\caption{Comparison of spectroscopic factors obtained from ($p$,$d$) and ($d$,$p$) reactions by Lee \textit{et al.} Perfect agreement is indicated by the line.  Reprinted figure with permission from \cite{Lee:2007} \textcopyright2007 by the American Physical Society }
\label{fig:Lee}
\end{center}
\end{figure}

The DWBA approximation fails for ($p$,$d$) and ($d$,$p$) reactions at energies of $\sim$20~MeV or higher as elastic scattering is no longer dominant and one must also take into account the breakup of the deuteron. In the Adiabatic Distorted Wave Approximation (ADWA) \cite{Johnson:1970} the ``deuteron'' distorted wave still describes the center-of-mass of a proton and neutron, but they are no longer propagating in the form of a bound deuteron. This distorted wave is calculated with the adiabatic OM potential 
\begin{eqnarray}
V(\bm{R}) &=& \frac{1}{D} \int d\bm{r} \left[ V_n(\bm{R}+\frac{\bm{r}}{2}) + V_p(\bm{R}-\frac{\bm{r}}{2}) \right] V_{np} \phi_{d}(r) d\bm{r}, \\
D &= & \int V_{np} \phi_{d}(r) d\bm{r},
\end{eqnarray} 
where $V_n(R)$ and $V_p(R)$ are neutron and proton optical-model potentials, $\phi_{d}(r)$ is the internal wavefunction of the deuteron with  $\bm{r}$ being is the relative proton-neutron position vector, and $V_{np}$ is the neutron-proton interaction. For energy-dependent potentials, $V_n$ and $V_{p}$ are evaluated at half the deuteron energy. Higher-order effects were subsequently included in Ref.~\cite{Johnson:1974}. In terms of input parameters, the main difference is that the DWBA requires protons and deuteron OM potentials while the AWBA requires proton and neutron OM potentials.

As transfer reactions are peripheral, they do not probe the magnitude of the overlap function in the center of the nucleus. As such, extracted parameters are sensitive to the radial extent of these function determined by the chosen radius and diffuseness parameters of the Wood-Saxon potential used. Pang \textit{et al.} have suggested a means of overcoming this by also measuring the asymptotic normalization coefficients in sub-barrier reactions \cite{Pang:2007}.

Finally we report on effects to include nonlocal potentials. Perey and Buck
prescribed a simple way to account for this by taking wavefunctions (sp overlap functions and distorted waves) obtained with a local potential and reducing the magnitude in the nuclear interior \cite{Perey:1961}. 
Titus and Nunes \cite{Titus:2014} compared this approach to one using exact solutions to the nonlocal potential \cite{Titus:2014}. The effect of nonlocality on the overlap function was more important than for the distorted waves with the net effect of a $\sim$20\% increase in cross section.  The use Perey-Buck correction factor was found to give results that differ by $\sim$10\% from the exact calculations. More recent studies have also considered the use of nonlocal dispersive-optical-model inputs for the distorted waves and overlap functions \cite{Ross:2015} (Sec.~\ref{sec:DOMknockout}). Efforts to make the AWBA more consistent with nonlocal potentials has also been made \cite{Johnson:2014,Timofeyuk:2016}.

\subsection{Heavy-Ion Knockout reactions}
\label{sec:knock}
Another type of knockout reaction is important at intermediate bombarding energies with radioactive beams. Here  a light target nucleus such as $^{12}$C or $^9$Be is typically used to knockout a nucleon, or two, from the projectile. An advantage of this method is its high sensitivity, the residual projectile fragment can be efficiently detected at small angles with a magnetic  spectrometer. Like other nucleon-removal reactions, the cross section is interpreted to be proportional to the spectroscopic factor and information on the angular momentum of the knocked-out nucleon (relative to the residual) can be obtained. In this case via the width of the longitudinal-momentum distribution of the residuals~\cite{Gade:2004}.

These knockout reactions have the cleanest interpretation when the residual nucleus is close to one of the drip lines and has no particle-bound excited states which can feed to the ground state. However if $\gamma$ rays are measured in coincidence, one can in principle disentangle the ground state and the feeding from a small number of excited states. The method can also be extended to residual nuclei beyond the drip line using the invariant-mass technique to identify the ground and excited states.

The total knockout cross section has two components; a stripping and elastic breakup or diffraction component. These are often calculated in the eikonal limit, \textit{i.e.} when the projectile and target interact, they have straight-line trajectories with a given impact parameter $\bm{b}$ and constant velocity. In addition, one makes use of the sudden approximation in that the nucleon is removed instantaneously and the remaining nucleons in the core are undisturbed. With these approximations, the sp stripping component is~\cite{Hussein:1985}
\begin{equation}
\sigma_{str} = \frac{1}{2I+1} \int \sum_{M} \braket{\phi_{IM} | (1-|S_{n}|^2) |S_c|^2 | \phi_{IM}} d\bm{b} 
\label{eq:stripping}
\end{equation}
where  $\phi_{IM}$ is the overlap function normalized to unity for the  removed nucleon with spin $I$ and projection $M$.  The $S$-matrices $S_n(b)$ and $S_c(b)$ are for the nucleon-target and core-target systems and can be expressed  in terms of the eikonal phase $\chi(b)$,
\begin{eqnarray}
 S(b) &=& e^{i\chi(b)} \\
 \chi(b) & = & -\frac{1}{\hbar v} \int_{-\infty}^{\infty} U_{opt}(r) dz.
\end{eqnarray}
Here $v$ is the relative velocity, $r=\sqrt{b^2+z^2}$, and $U_{opt}$ is the appropriate  optical-model potential. From Eq.~(\ref{eq:stripping}), the stripping cross section can understood from the product $(1-|S_n|^2) |S_c|^2$ where the first term corresponds to absorption of the removed nucleon by the imaginary  nucleon-target optical potential (non-elastic interaction between nucleon and target), while second term corresponds to an elastic interaction between the residual and target nucleus. 

 The second component of the breakup cross section, the diffractive or elastic-breakup component occurs when both the core and removed nucleon interact elastically with the target. If there are no particle-bound excited states, the sp cross section is \cite{Tostevin:2001} 
\begin{equation}
\sigma_{diff} = \frac{1}{2I+1} \int \sum_{M} \braket{\phi_{IM}|S_{c}S_{n}|\phi_{IM}} d\bm{b} .
\end{equation}
The spectroscopic factor can be obtained by scaling the calculated sp cross section $\sigma_{sp} =\sigma_{str} + \sigma_{diff}$ to the experimental value, or alternatively, if  theoretical shell-model spectroscopic factors are used, a theoretical cross section can be predicted. Tostevin and Gade \cite{Tostevin:2014}  have compiled  systematics of reduction factors of the experimental results from these expectations and found a very significant dependence on proton-neutron asymmery of the projectile. This result is in conflict with measurements obtained with  transfer \cite{Lee:2010,Flavigny:2018} and  ($p$,2$p$)  \cite{Atar:2018,pty011} reactions. More studies are needed to resolve this problem.

This eikonal approach breaks down if the bombarding energy is too low or the knockout nucleon is too deeply bound~\cite{Flavigny:2012} and an alternative theory, transfer to the continuum~\cite{Bonaccorso:1988,Bonaccorso:1991,Bonaccorso:2018} which includes conservation of energy and momentum is better able to describe the experimental distributions. This model also requires the core-target and nucleon-target $S$-matrices and the overlap function as input.

As the target nucleus most commonly used in these studies is $^{9}$Be, then 
optical-model potentials for scattering of nucleons and the residuals from $^{9}$Be are needed.  Note that as only the modulus squared of the two $S$-matrices are needed for the stripping component [Eq.~(\ref{eq:stripping})], then only the imaginary part of the eikonal phase, and hence, the imaginary optical potentials are needed.  However the real parts are needed for the diffractive part, but this component is typically much smaller~\cite{Bazin:2009}. 
All global nucleon optical-model potential are fitted for heavier targets typically $A>40$ (Sec.~\ref{sec:global}) so cannot be used.
 Most knockout studies have used theoretical OM parameters obtained with the folding technique (Sec.~\ref{sec:calc}). More recently, Bonaccorso and Charity have produced two empirical potentials by fitting experimental neutron and proton elastic-scattering angular distributions, reaction and total sections for energies  up to 136~MeV~\cite{Bonaccorso:2014}. Figure~\ref{fig:Bonaccorso} illustrates the quality of the fit to cross section data for $^{9}$Be for the two 
potentials (AB and DOM).

\begin{figure}[tpb]
\begin{center}
\includegraphics[scale=.6]{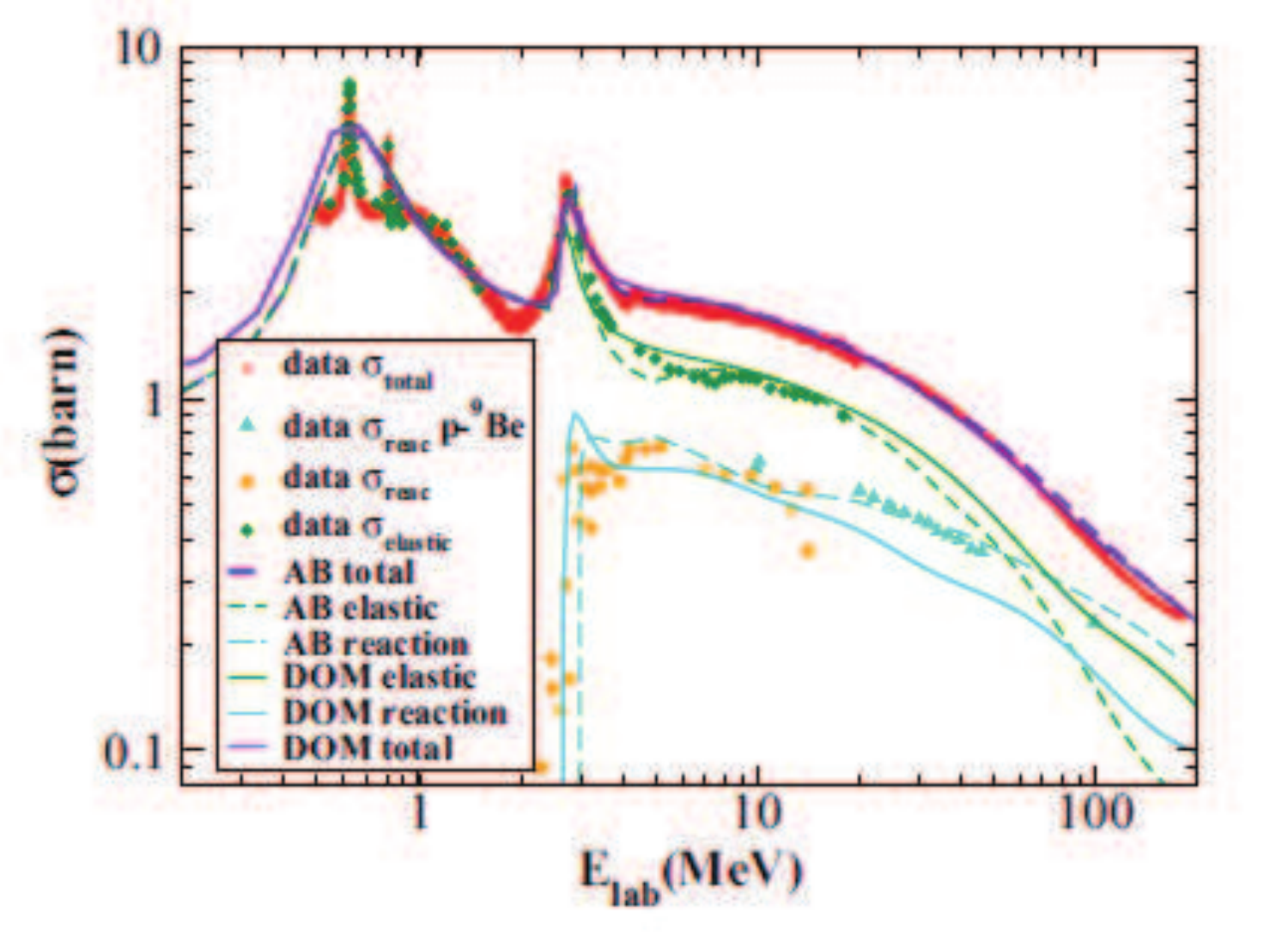}
\caption{Fits to experimental neutron total (diamonds at the top) and elastic (other diamonds), and reaction (circle) cross sections and proton reaction cross sections (triangles) for the two empirical potentials (AB and DOM) of Ref.~\cite{Bonaccorso:2014} for $^{9}$Be targets.  Reprinted figure with permission from \cite{Bonaccorso:2014} \textcopyright2014 by the American Physical Society.}
\label{fig:Bonaccorso}
\end{center}
\end{figure}

Empirical optical potentials for the core-target are not  available due to the non availability of elastic-scattering and absorption cross section data for most of these systems. Rather a double-folding procedure based on the theoretical density distributions for the core and target is often used. However, Bonaccorso \textit {et al.} have suggested that the single-folded procedure based on their empirical nucleon-$^{9}$Be OM potentials will be better \cite{Bonaccorso:2016,Bonaccorso:2016a}. They have compared total reactions cross sections calculated in this approach to experimental values for $^8$Li, $^8$B, and $^9$C projectiles on $^9$Be targets.  Good agreement is found especially at high energies where most of the knockout reactions are performed. At lower energies, an  additional small surface absorption component improves the reproduction of these experimental cross sections.  

\subsection{Nucleon knockout with electrons and proton beams}
\label{sec:pek}
 Both the transfer and heavy-ion knockout reactions are surface dominated and do not probe the overlap function in the interior region. Therefore more penetrating probes, such as high-energy electrons and protons, offer the possibility of studying knockout reactions with higher sensitivity.  At sufficiently high incident energies, both probes can be considered in the quasi-free limit. In the Distorted Wave Impulse approximation (DWIA), the differential cross section can be factorized as  \cite{Jacob:1966,Jacob:1973,Giusti:1988}
\begin{equation}
\frac{d^6\sigma}{d\Omega_{e}d\epsilon d\Omega_{N}dT_{N}} = K \left(\frac{d\sigma}{d\Omega}\right)_{free} \left|\Phi(\bm{Q})\right|^2
\end{equation}
where $\Omega_{e}$ and $\Omega$ are the solid angles of the outgoing probe and the knocked-out nucleon with energy and kinetic energies of $\epsilon$ and $T_{N}$ respectively. Also $K$ is a known  kinematical factor and $\left|\Phi(\bm{Q})\right|^2$ is the momentum distribution of the knockout nucleon in the original nucleus distorted by final-state interactions, and $(d\sigma/d\Omega)_{free}$ is the free differential cross section for the probe on a nucleon.

For ($p$,2$p$) and ($p$,$pn$) reactions, the transition matrix elements depends on one incoming and two outgoing distorted waves for the nucleons, while of course for ($e$,$e' p$) reactions one only considers an outgoing proton distorted wave. Of course in this case, Coulomb distortions of the electron waves should also be included \cite{Giusti:1988}. The standard global optical-model parametrizations (Sec.~\ref{sec:global}) are generally not useful for calculating these distorted waves as these were fit to low-energy data, whereas  ($e$,$e' p$) reactions are typical performed at outgoing proton energy of $\sim$100~MeV. The more recent Koning and Delaroche parametrization \cite{Koning:2003} does cover this energy-regime but we are not aware of its use for these reactions. An alternative high-energy parametrization [80-180~MeV, 24$\leq A \leq$ 208] by Schwandt \textit {et al.} \cite{Schwandt:1982} has been used extensively for DWIA calculations of ($e,e' p$) and some ($p$,2$p$) reactions \cite{denHerder:1986,Giusti:1987,Kramer:1989,Leuschner:1994,Branford:2000,Kramer:2001,Mazziotta:2002,Gaidarov:2002,Neveling:2002,Tabatabaei:2004,Giusti:2011}. As for transfer reactions, the extracted spectroscopic factor are sensitive to the choice of the optical-model parameters. A number of works have studied this dependence \cite{denHerder:1988,Cowley:1995,Leuschner:1994,Yasuda:2010} with variations in predicted cross section of up to 12\% found.
 At the high proton energies, some data have been analyzed within a relativistic 
DWIA which, for consistency, should use optical-model parameters fitted with the Dirac formalism. Global parameters obtained from fitting with a Dirac 
optical model used in ($e$,$e'p$) and ($p$,2$p$) reactions can be found in Refs.~\cite{Hama:1990,Cooper:1993}.

Recent studies with high-energy proton probes include the use of polarized proton beams at RCNP where cross sections and analyzing powers for ($\overrightarrow{p}$,2$p$) and ($\overrightarrow{p}$,$pn$) reactions \cite{Yasuda:2010, Wakasa:2017} were measured. While the extracted proton spectroscopic factor are similar to results obtained with ($e,e'p$) reactions, the neutron knockout results for $N=Z$ target are almost a factor of two larger than the corresponding ($e,e'p$) proton knockout values. Further efforts are needed to understand this. In addition, studies of ($p$,2$p$) reactions in inverse kinematics are being performed with exotic beam. At HIMAC (Heavy Ion Medical Accelerator in Chiba), they have performed measurements for $^{9-16}$C isotopes at 250 MeV/$A$ \cite{Kobayashi:2008} while at GSI Helmholtzzentrum f\"{u}r Schwerionenforschung,  measurements were performed for $^{13-23}$O isotopes at 310-451 MeV/$A$ \cite{Atar:2018}.


\subsection{Transfer and knockout reaction with the DOM}
\label{sec:DOMknockout}
While most calculations of the cross section for transfer and the various types of knockout reactions use either theoretical or empirical OM parameters to calculate the distorted waves, the radial dependencies of  overlap functions come from a separate source. Typically, a sp wavefunction is solved for a Wood-Saxon potential whose depth is adjusted to match the experimental binding energy.

With the dispersive-optical-model fits to both positive scattering and negative, bound-state data, both the nucleon distorted wave, the overlap function, and its normalization (spectroscopic factor) are provided by a single model. Thus transfer and knockout data can be used to test the validity of the DOM fit or even further constrain its parametrization.  Reproduction of transfer and knockout data in addition to the elastic-scattering, reaction and bound-state data brings greater confidence in extracted DOM quantities.  In ($d$,$p$), ($d$,$n$) and ($p$,$d$) transfer reactions, if the ADWA formalism is used, then both the proton and neutron distorted waves can be provided by the DOM. However in DWBA calculations of the same reactions, the deuteron distorted waves need to be obtained from a separate source.

In 2011,  Nguyen \textit{et al.} \cite{Nguyen:2011} used the local DOM fit to closed-shell nuclei by the St.-Louis group \cite{Charity:2006,Charity:2007,Mueller:2011} with the ADWA formalism to look at ($d$,$p$) transfer reactions. They found that calculations with their DOM input were able to describe the transfer angular distributions as well as or better than global parametrizations. In addition, spectroscopic factors extracted with DOM ingredients show less dependence on the beam energy and their magnitudes are more in line with the ($e,e'p$) values than those from a standard analysis. For an example, a comparison of the reproduction of  experiment data is shown for a $^{48}$Ca target in Fig.~\ref{fig:Nguyen}.

 Ross \textit{et al.} used the non-local DOM potential fit to $^{40}$Ca by the St.-Louis group\cite{Mahzoon:2014} to study the ($p$,$d$) reaction in the DWBA \cite{Ross:2015}. This study was focused on the importance of non-locality which produced opposing effects on the predicted cross section from the distorted waves and the overlap function.

\begin{figure}[tpb]
\begin{center}
\includegraphics[scale=.6]{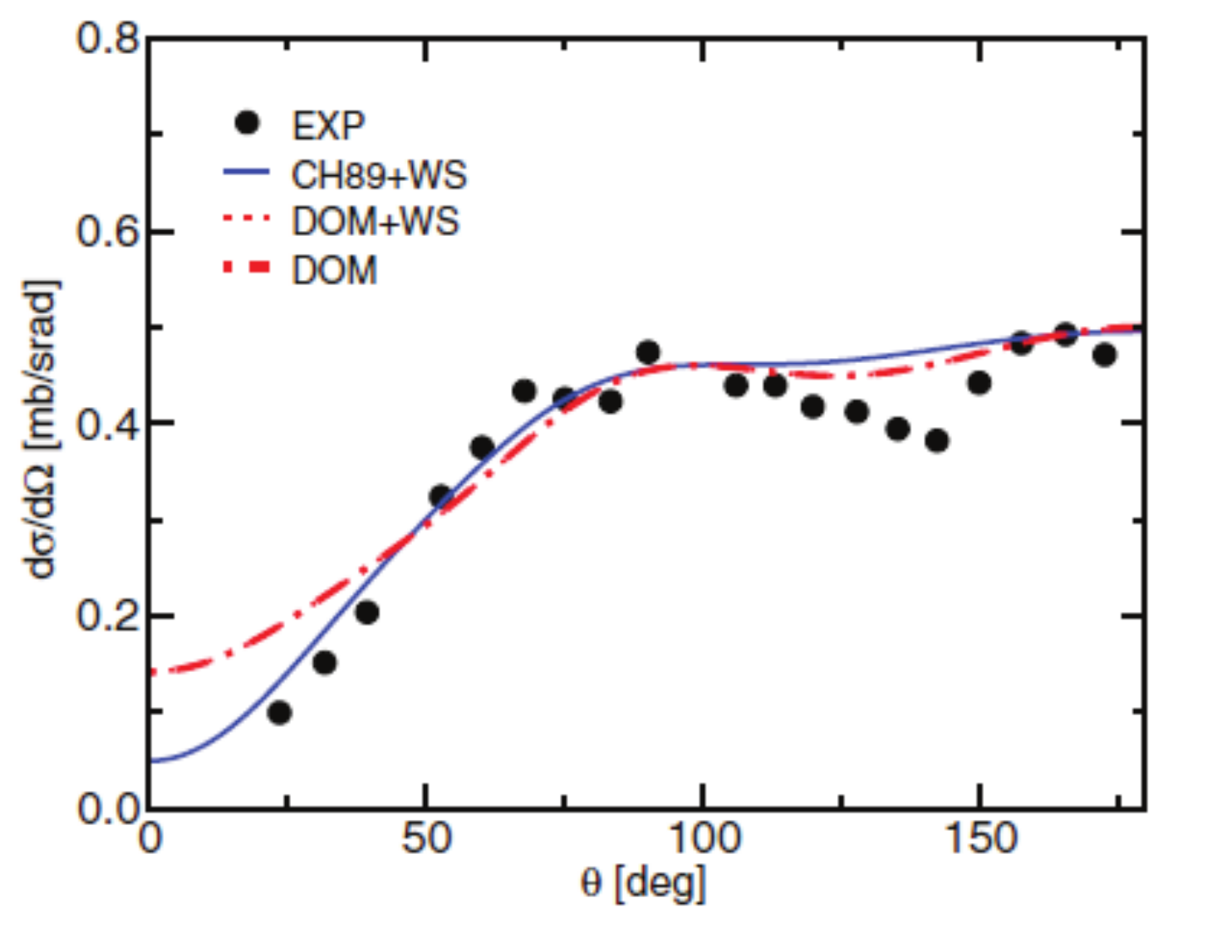}
\caption{Comparison of experimental $^{48}$Ca($d$,$p$)$^{49}$Ca ($E_d$=2~MeV) angular distribution to AWDA calculations with  inputs from the CH89 potential for the distorted wave and a Wood-Saxon potential for the overlap function, the DOM potential for the distorted waves and the same Wood-Saxon potential for the overlap function, and both the overlap function and distorted waves from the DOM potential.  All calculations have been normalized to the data at backward angles. The latter two calculations overlap.  Reprinted figure with permission from \cite{Nguyen:2011} \textcopyright2011 by the American Physical Society.} 
\label{fig:Nguyen}
\end{center}
\end{figure}

The current status of the theoretical treatment of inclusive $(d, p)$ reactions was recently reviewed in Ref.~\cite{Potel2017} and applications to Ca isotopes illustrated.
The required neutron-Ca propagator was described within the local DOM framework, and the interplay between elastic breakup and non-elastic breakup was studied for three Ca isotopes. 
The accuracy of the description of different reaction observables was favorably assessed by comparing with experimental data of $(d, p)$ on ${}^{40,48}$Ca.
Predictions of the model for the extreme case of ${}^{60}$Ca which has recently been identified \cite{Tarasov:2018}.
The capability of the DOM optical potential to describe negative-energy physics was contrasted with the use of the empirical potential of Ref.~\cite{Koning:2003}.

Quite recently, an updated version of the St.-Louis nonlocal DOM parametrization for $^{40}$Ca was used to calculate the $^{40}$Ca($e,e'p$)$^{39}$K reactions with the DWIA~\cite{Atkinson:2018}. In this case the proton distorted wave, the radial-dependence of the overlap function, it normalization (spectroscopic factor) were calculated from the DOM parametrization. There was no fit parameters used in these DWIA calculations.  Agreement with data obtained for the knockout of the 0$d_{3/2}$ and 1$s_{1/2}$ orbitals with outgoing proton energy to 100 MeV were as good as, or better than, previous descriptions employing the Schwand \textit{et al.}'s local global OM potential~\cite{Schwandt:1982} and overlap functions from Wood-Saxon potentials where the radius and  normalization factor are adjusted to fit the knockout data. The net effect of nonlocality, was in the opposite direction to the transfer study of Ross \textit{et al.}~\cite{Ross:2015}, and  lead to larger spectroscopic factors.  Presumably this is related to larger importance of knockout from the interior of the nucleus in ($e,e'p$) case.
Fig.~\ref{fig:mack5} shows the agreement of these DWIA calculation with the distorted spectral functions obtained from the ($e,e'p$) raw data.
A slight increase of the spectroscopic factors obtained from the DOM description compared to the values of the standard Nikhef analysis~\cite{Lapikas:1993} is obtained otherwise validating the procedure of the Nikhef group~\cite{Atkinson:2018}.

\begin{figure}[tpb]
\begin{center}
\includegraphics[scale=.6]{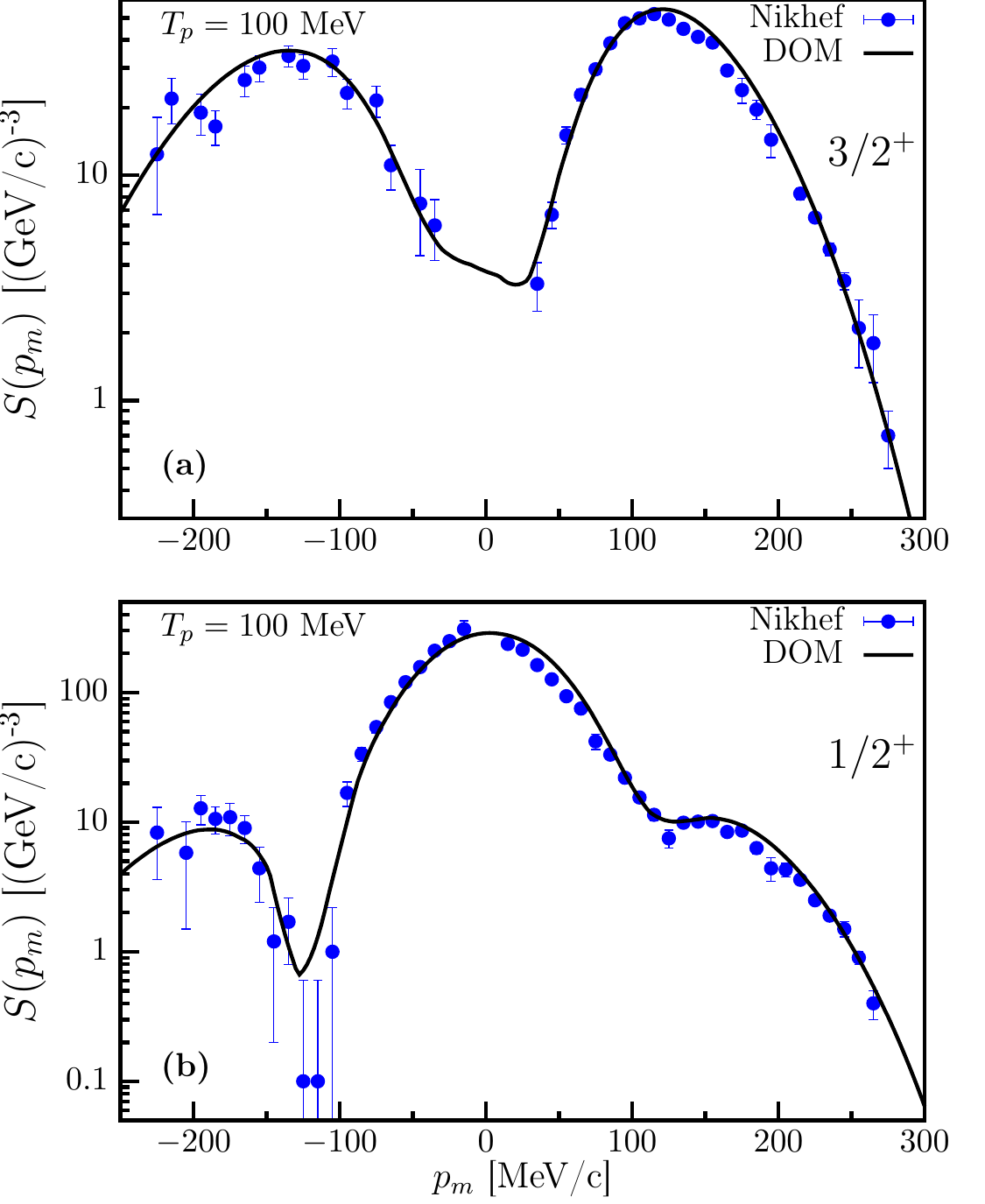}
\caption{Comparison of the spectral functions measured at Nikhef for outgoing proton energies of 100~MeV to DWIA calculations using the proton distorted waves, overlap function and its normalization from a non-local DOM parametrization.
Results are shown for the knockout of a 0$d_{3/2}$ and 1$s_{1/2}$ proton in the upper and lower panels respectively.
Reprinted figure with permission from \cite{Atkinson:2018} \textcopyright2018 by the American Physical Society.} 
\label{fig:mack5}
\end{center}
\end{figure}

\section{Conclusions and Outlook}
\label{sec:conc}
The topics presented in this review clearly indicate that the optical model is still of great interest and importance to the nuclear physics community. Not only is the optical model used to describe elastic scattering, it is also an  ingredient in theories for other reaction types. As such the desire to know the potential not only for beta-stable nuclei but for exotic proton and neutron rich systems has driven both experimental and theoretical programs. 

Of particular note is the variant known as the dispersive optical model which
enforces causality through dispersion relations between the real and imaginary potentials. 
When nonlocality is also included, the optical potential can be considered as describing the nucleon self-energy in the Green's-function formalism. 
This allows for a consistent description of both positive-energy scattering data and negative-energy bound-state properties. With both these regimes determined from a single optical potential, this produces all of the  ingredients  needed for sp knockout and pick-up reactions. Some initial studies along these lines were discussed in this review, but the expectation is that the DOM will be used more widespread for such purposes in the future.  

With the available of newer data, especially neutron scattering and cross section data, there have more refined global nucleon optical-model parametrizations which cover wider regions of bombarding energy. 
In addition, better parametrizations for composite particles have also been produced. The $N-Z$ asymmetry dependence of the imaginary potential for neutron scattering is still not well determined. The data that exist at present suggest that it does not follow the functional formed assumed in present global parametrizations. Further experimental studies on separated isotopes are needed to resolve this issue and allow for better global parametrizations. This is especially important if one wants to extrapolate to nuclei close to the drip lines.
Within the DOM approach this will also require to incorporate feature associated with pairing correlations as it can be done in the \textit{ab initio} Gorkov-Green's function method.

Such extrapolations can also be addressed theoretically. We have discussed some recent advances in the microscopic calculation of optical potentials or nucleon self-energies. Various methods include the short and long-range correlations to various degrees.
Such microscopic calculations provide valuable insights into the functional forms that are expected to be relevant for empirical potentials.
Methods that rely on local-density approximations typically require further fine tuning but have found global application.
Multiple-scattering approaches have recently been extended to include a consistent description of one-body densities and interactions that originate from the same underlying NN interaction.
Such methods are reasonably successful at high energy but are not easy to improve and also tend to have difficulty with polarization data.
The use of chiral interactions both in nuclear-matter and finite-nucleus calculations for the optical potential has been illustrated.
Some shortcomings may be due to the softness of such interactions as the coupled-cluster method also exhibits difficulties in generating accurate results, in particular for absorptive potentials.

The authors therefore conclude that all many-body methods at present do not provide an unambiguous path to an accurate description of the optical potential.
Important future developments are therefore necessary.
 As an intermediate between theory and experimental data, we therefore advocate the use of the dispersive optical model.
Is has shown to be able to predict ground-state information related to neutrons and has been used successfully to assess the distorted-wave impulse approximation description of the $(e,e'p)$ reaction. 
Its application to transfer reactions has already been useful in its local form and is currently extended to nonlocal potentials.

\section{Acknowledgements}
   This work was supported by the U.S. Department of Energy, Division of
   Nuclear Physics under grant No. DE-FG02-87ER-40316 and by the U.S. National Science Foundation under grant PHY-1613362.

\bibliographystyle{elsarticle-num}
\bibliography{opt}
\end{document}